\documentclass[12pt]{article}
\usepackage{appendix}
\usepackage{anysize}
\usepackage{graphics}
\usepackage{color}
\usepackage{amssymb}
\usepackage{graphicx}
\usepackage{epstopdf}
\usepackage[hang,small,bf]{caption}
\usepackage{hyperref}
\usepackage{amsmath}
\usepackage{latexsym}
\usepackage{setspace}
\usepackage{sectsty}
\usepackage[square,comma,numbers,sort&compress,merge]{natbib}
\usepackage{breqn}
\usepackage{color}

\sectionfont{\large\bf}
\subsectionfont{\normalsize\bf}
\subsubsectionfont{\small\bf}
\def \beq{\begin{equation}}
\def \eeq{\end{equation}}
\def \bea{\begin{align}}
\def \eea{\end{align}}

\def\met{\mbox{${\hbox{$E$\kern-0.6em\lower-.1ex\hbox{/}}}_T$}}
\def\4vol{{\int d^4x \sqrt{-g}}}

\def\simlt{\stackrel{<}{{}_\sim}}
\def\simgt{\stackrel{>}{{}_\sim}}

\title{
\vspace*{-1.3cm}
\begin{flushright}
\normalsize{
ANL-HEP-PR-12-26\\
EFI-12-06\\
FERMILAB-PUB-12-132-PPD-T}
\end{flushright}
\vspace{0.5cm}
\Large
\textbf{The pMSSM Interpretation of LHC Results Using
Rernormalization Group Invariants}
\author{\textbf{Marcela Carena$^{a,b,c}$, Joseph Lykken$^{a}$, Sezen Sekmen$^c$}\\
\textbf{Nausheen R.~Shah$^{a}$, and Carlos E.~M.~Wagner$^{b,d,e}$}\\
[1.5cm]
\normalsize\emph{$^a$~Fermi National Accelerator Laboratory, P.~O.~Box 500, Batavia, IL 60510, USA\footnote{http://theory.fnal.gov}}\\
\normalsize\emph{$^b$~Enrico Fermi Institute and $^d$~Kavli Institute for Cosmological Physics,}\\
\normalsize\emph{University of Chicago, Chicago, IL 60637, USA} \\
\normalsize\emph{$^c$~Physics Department, CERN, CH 1211 Geneva 23, Switzerland}\\
\normalsize\emph{$^e$~HEP Division, Argonne National Laboratory, 9700 Cass Ave., Argonne, IL 60439, USA}}}
\begin{document}
\nocite{*}
\setcounter{page}{0}
\maketitle
\thispagestyle{empty}
\begin{abstract}
The LHC has started to constrain supersymmetry-breaking parameters by setting bounds on possible colored particles at the weak scale. Moreover, constraints from Higgs physics, flavor physics, the anomalous magnetic moment of the muon, as well as from searches at LEP and the Tevatron have set additional bounds on these parameters. Renormalization Group Invariants (RGIs) provide a very useful way of representing the allowed parameter space by making direct connection with the values of these parameters at the messenger scale. Using a general approach, based on the pMSSM parametrization of the soft supersymmetry-breaking parameters, we analyze the current experimental constraints to determine the probability distributions for the RGIs. As examples of their application, we use these distributions to analyze the question of Gaugino Mass Unification and to probabilistically determine the parameters of General and Minimal Gauge Mediation with arbitrary Higgs mass parameters at the Messenger Scale.
\end{abstract}
\thispagestyle{empty}

\newpage

\setcounter{page}{1}

\onehalfspacing


\section{Introduction}
The Standard Model~(SM) provides an excellent description of all experimentally measured observables at present. Mass generation relies on the Higgs mechanism, which is based on the introduction of an elementary scalar field transforming in the fundamental representation of the $SU(2)_L$ group. The vacuum expectation value~(vev) of this scalar field sets the weak scale, which is then proportional to the magnitude of the square root of the negative squared mass parameter in the scalar Higgs potential~\cite{Higgs:1964pi},\cite{Higgs:1966ev}. The SM provides no explanation for the magnitude of this mass parameter, which is sensitive via radiative corrections to new physics at high scales.

The Minimal Supersymmetric Extension of the Standard Model~(MSSM) has most of the virtues of the SM~\cite{PRPLC.110.1}--\cite{Martin:1997ns}. Apart from a loop factor, the magnitude of the Higgs mass parameter is determined by the size of the supersymmetry-breaking parameters of the third generation squarks. These also determine the value of the SM-like Higgs mass at the loop level. Values of the third generation squark masses of about 1~TeV lead to SM-like Higgs masses in the 115--130~GeV range~\cite{Okada:1990vk}--~\cite{Martin:2002wn}. Hence, recent hints of a Higgs mass of about 125~GeV are consistent with MSSM predictions~\cite{Carena:2011aa}.

The supersymmetry-breaking mass parameters depend on the unknown mechanism of supersymmetry breaking and on the messenger scale, at which supersymmetry breaking is transmitted to the observable sector. Recent experimental bounds from the LHC set strong constraints on colored particles at the TeV scale, and therefore on the parameters of minimal models of supersymmetry breaking.

Several works have studied the relationship of the supersymmetric mass parameters between the messenger scale and the weak scale~\cite{Lafaye:2004cn}--\cite{Carena:1996km}. It would be very useful to have a method that allowed us to set bounds on the supersymmetry-breaking parameters at the messenger scale, independent of the unknown supersymmetry-breaking scheme and of the unknown value of the messenger scale. Renormalization Group Invariants (RGIs)~\cite{Kane:2006hd}--\cite{Carena:2010wv} provide such a method. Determination of the value of the RGIs at the TeV scale sets their values at the messenger scale. One can then use the information provided by the RGIs to set constraints on general classes of models~\cite{Carena:2010gr},\cite{Carena:2010wv}. An exhaustive analysis of the RGIs for different supersymmetry-breaking scenarios is performed in Ref.~\cite{Hetzel:2012bk}. 

The effects of various preLHC and LHC results on the phenomenological MSSM~(pMSSM) parameter space have been studied in detail in Refs.~\cite{Berger:2008cq}--\cite{Arbey:2011aa}. In this article, we use the pMSSM parametrization of the soft supersymmetry-breaking parameters~\cite{Berger:2008cq} to determine the current probability distribution of the RGIs at the TeV scale. We shall compare the situation before and after constraints from the LHC are imposed.   

To illustrate the power of this framework, we will use the pMSSM RGI probability distributions to analyze three particular issues: 
\begin{itemize}
\item{Possible scale of Gaugino Mass Unification.}
\item{ Messenger scale parameters in a realization of General Gauge Mediation.}
\item{ Messenger scale parameters associated with  Minimal Gauge Mediation.}
\end{itemize}
The probabilistic interpretation of the RGIs  can be applied to other quantities of interest in the MSSM using for example the analysis presented in Ref.~\cite{Hetzel:2012bk}.

In section 2 we list the RGIs to be used in this paper, outlining the methodology to be used in our analyses. We then compute the  RGI probability distributions obtained by imposing current experimental constraints. In section 3 we study the question of Gaugino Mass Unification and the consistency of the scale of this Gaugino Mass Unification with experimental constraints. In section 4 we look at General Gauge Mediation and determine the probability distribution of the relevant parameters of this model. Section 5 discusses the probability distributions for the Minimal Gauge Mediated parameters. We reserve section 6 for our conclusions. Details about our probability analysis are given in Appendix A. Appendix B gives the specific definition of the pMSSM. Appendix C gives the inverted relationships between the soft masses of the pMSSM and the RGIs. Appendix D lists these in the case of flavor-blind models.

\section{RG~Invariants: Probabilistic Interpretation}

\subsection{RGI-pMSSM Basis }
\label{RGIpMSSM}

There are 14 relevant RGIs, analyzed in Ref.~\cite{Carena:2010gr},\cite{Carena:2010wv}, involving the soft supersymmetry-breaking parameters, which we will use as the basis of our current work. These  are summarized at one-loop accuracy in Table~\ref{table.Inv}; two-loop corrections were studied in Ref.~\cite{Carena:2010gr} and shown to be of order of a few percent or less. Moreover, there are 2 RGIs relating only the gauge couplings~($I_{g_2}$ and $I_{g_3}$), which we can use to redefine the other 12 RGIs in terms of just the soft masses and the scale. These soft masses, ignoring  possible small flavor dependence of the sfermion and Higgs mass parameters, are given by a total of 17 scalar masses plus 3 gaugino masses. One can make the additional well-motivated assumption of degeneracy for the first and second generation sfermion mass parameters. In such a case, one is left with 12 scalar masses. Therefore, the 12 RGIs, which are linearly independent, can be inverted to give 12 soft supersymmetry-breaking masses in terms of these RGIs as a function of 3 given soft masses.

\begin{table}[!ptbh]
\begin{center}
\caption{1-Loop RG Invariants in the MSSM }
\scriptsize
\begin{tabular}{@{}| c | c | c | c | c | @{}}
\hline
\hline
&&&&\\
\;\;RGI\;\; & Definition in Terms of Soft Masses & MGM($M$) & GGM($M$) & CMSSM+NUHM($M$)\\ [0.8ex]
 & & & &\\
\hline
\hline
&&&&\\
$D_{B_{13}}$&$2(m_{\tilde{Q}_1}^2-m_{\tilde{Q}_3}^2)-m_{\tilde{u}_1}^2+m_{\tilde{u}_3}^2-m_{\tilde{d}_1}^2+m_{\tilde{d}_3}^2$&0&0&0\\ [3ex]
\hline
&&&&\\
$D_{L_{13}}$&$2(m_{\tilde{L}_1}^2-m_{\tilde{L}_3}^2)-m_{\tilde{e}_1}^2+m_{\tilde{e}_3}^2$&0&0&0\\ [3ex]
\hline
&&&&\\
$D_{\chi_1}$&$3(3m_{\tilde{d}_1}^2-2(m_{\tilde{Q}_1}^2-m_{\tilde{L}_1}^2)-m_{\tilde{u}_1}^2)-m_{\tilde{e}_1}^2$&0&0& $5m_0^2$ \\ [3ex]
\hline&&&&\\
$D_{Y_{13H}}$&$\begin{array}{c} m^2_{\tilde{Q}_1}-2m^2_{\tilde{u}_1}+m^2_{\tilde{d}_1}-m^2_{\tilde{L}_1}+m^2_{\tilde{e}_1}\\-\frac{10}{13}\left(m^2_{\tilde{Q}_3}-2m^2_{\tilde{u}_3}+m^2_{\tilde{d}_3}-m^2_{\tilde{L}_3}+m^2_{\tilde{e}_3}+m^2_{H_u}-m^2_{H_d}\right)\end{array}$&$-\frac{10}{13}(\delta_u-\delta_d)$&$-\frac{10}{13}(\delta_u-\delta_d)$&$-\frac{10}{13}(\delta_u-\delta_d)$\\ [6ex]
\hline
&&&&\\
$D_{Z}$&$3(m_{\tilde{d}_3}^2-m_{\tilde{d}_1}^2)+2(m_{\tilde{L}_3}^2-m_{H_d}^2)$&$-2\delta_d$&$-2\delta_d$&$-2\delta_d$\\ [3ex]
\hline
&&&&\\
$I_{Y\alpha}$&$\left(m^2_{H_u}-m^2_{H_d}+\sum_{gen}(m^2_{\tilde{Q}}-2m^2_{\tilde{u}}+m^2_{\tilde{d}}-m^2_{\tilde{L}}+m^2_{\tilde{e}})\right)/g_1^2$&$\left(\delta_u-\delta_d\right)/g_1^2$&$\left(\delta_u-\delta_d\right)/g_1^2$&$\left(\delta_u-\delta_d\right)/g_1^2$\\ [3ex]
\hline
&&&&\\
$I_{B_i}$&$M_i/\theta_i^2$&$B$&$B_i$& $m_{1/2}/\theta_i^2$\\ [3ex]
\hline
&&&&\\
$I_{M_1}$&$M_1^2-\frac{33}{8}(m_{\tilde{d}_1}^2-m_{\tilde{u}_1}^2-m_{\tilde{e}_1}^2)$&$\frac{38}{5}g_1^4B^2$&$g_1^4\left(B_1^2+\frac{33}{10}A_1\right)$& $m_{1/2}^2+\frac{33}{8}m_0^2$\\ [3ex]
\hline
&&&&\\
$I_{M_2}$&$M_2^2+\frac{1}{24}\left(9(m_{\tilde{d}_1}^2-m_{\tilde{u}_1}^2)+16m_{\tilde{L}_1}^2-m_{\tilde{e}_1}^2\right)$&$2g_2^4B^2$&$g_2^4\left(B_2^2+\frac{1}{2}A_2\right)$& $m_{1/2}^2+\frac{5}{8}m_0^2$\\ [3ex]
\hline
&&&&\\
$I_{M_3}$&$M_3^2-\frac{3}{16}(5m_{\tilde{d}_1}^2+m_{\tilde{u}_1}^2-m_{\tilde{e}_1}^2)$&$-2g_3^4B^2$&$g_3^4\left(B_3^2-\frac{3}{2}A_3\right)$& $m_{1/2}^2-\frac{15}{16}m_0^2$\\ [3ex]
\hline
&&&&\\
$I_{g_2}$&$ 1/g_1^2-33/(5g_2^2)$&$\approx -10.9$&$\approx -10.9$&$\approx -10.9$\\ [3ex]
\hline
&&&&\\
$I_{g_3}$&$ 1/g_1^2+11/(5g_3^2)$&$\approx 6.2$&$\approx 6.2$&$\approx 6.2$\\ [3ex]
\hline
\end{tabular}
\label{table.Inv}
\end{center}
\end{table}
In the pMSSM, apart from the 15 soft supersymmetry-breaking parameters discussed above, there are 3 soft supersymmetry-breaking parameters, $A_f$ ($f=t,b,\tau$), denoting the mixing of the left- and right-handed third generation sfermions, and $\tan\beta$, the ratio of the Higgs vevs~(or equivalently the soft supersymmetry-breaking parameter $B_\mu$). These 19 parameters then give the complete basis for the pMSSM.

The Higgs soft supersymmetry-breaking parameters may be determined, up to loop corrections, as a function of the Higgsino mass parameter, $\mu$, the CP-odd Higgs mass, $m_A$, and $\tan\beta$. The tree-level expressions for $\mu$ and $m_A$ in terms of $m_{H_u}$ and $m_{H_d}$ are given by:
\begin{dmath}
\mu ^2=\frac{m_{H_u}^2\tan^2\beta-m_{H_d}^2}{ \left(1- \tan^2\beta\right) } - \frac{1}{2}m_Z^2\;,
\end{dmath}
\begin{dmath}
m_A^2=\frac{(m_{H_u}^2-m_{H_d}^2)}{\cos 2 \beta} -m_Z^2\;.
\end{dmath}
These can be inverted to give the Higgs soft supersymmetry-breaking mass parameters:
\begin{dmath}
m_{H_u}^2= \frac{1}{2} \left[(1+\cos 2 \beta)m_A^2+m_Z^2 \cos 2 \beta -2 \mu ^2\right],
\end{dmath}
\begin{dmath}
m_{H_d}^2= \frac{1}{2} \left[(1-\cos 2 \beta )m_A^2-m_Z^2 \cos 2 \beta-2 \mu ^2\right]\;.
\end{dmath}

Using the above relations, one can define a 1-to-1 correspondence between the 19 pMSSM parameters and the 12 RGIs, complemented by $\mu$, $m_A$, $\tan\beta$, the 3 mixing parameters $A_t$, $A_b$ and $A_{\tau}$ and one third generation squark mass parameter, for instance $m_{Q_3}$. The expressions for the soft masses in terms of the RGIs are given in Appendix C.

\subsection{Methodology}

In Ref.~\cite{Sekmen:2011cz}, the probability distributions of the 19 pMSSM parameters were computed, analyzing the differences between the results for these distributions considering some preLHC measurements (listed in Table~\ref{tab:preLHCobs}) and after including various 1~fb$^{-1}$ LHC results (listed in Table~\ref{tab:LHCobs}). We refer the reader to  Ref.~\cite{Sekmen:2011cz} for specific details about the likelihood analysis.  We shall use the set of pMSSM points and their corresponding preLHC and LHC likelihoods from Ref.~\cite{Sekmen:2011cz} and obtain the probability distributions for the RGIs in Table~\ref{table.Inv} projected to 5~fb$^{-1}$ of LHC data.

In Ref.~\cite{Sekmen:2011cz}, a flat prior for the all the soft supersymmetry-breaking parameters was used. The scalar masses were varied between 0 and 3 TeV. The gaugino masses and the $\mu$ parameter were scanned between -3 and 3 TeV, and the mixing parameters, $A_f$, were scanned from -7 to 7 TeV. The range of $\tan \beta$ considered was 2 to 60.

\begin{table}[t]
\begin{center}
\caption{The pre-LHC experimental results that are the basis of our pMSSM parameter scan using Markov Chain Monte Carlo (MCMC) sampling. We re-weight a~posteriori with the limit $BR(B_s \rightarrow \mu \mu) \le 1.08 \times 10^{-8}$ at 95\% CL \cite{newbsmm}. However, this hardly has any effect. In evaluating the Higgs mass limit, we apply a Gauss-distributed theoretical uncertainty with $\sigma=1.5$~GeV to the $m_h$ computed by {\tt SoftSUSY}. }
\begin{tabular}{|c|c|c|c|c|}
\hline
$i$ 	& Observable 	& Experimental result 	& Likelihood function \\
 	& $\mu_i$		& $D_i$				&  $L(D_i|\mu_i)$ \\
\hline\hline
1 & $BR(b \rightarrow s\gamma)$  
   & $(3.55 \pm 0.34)\times 10^{-4}$~\cite{Asner:2010qj,Misiak:2006zs}
   & Gaussian \\
\hline
2 & $BR(B_s \rightarrow \mu \mu)$
   & $\le 4.7 \times 10^{-8}$~\cite{Nakamura:2010zzi} 
   & $1/\big(1+{\mathrm{exp}}{(\frac{\mu_2-D_2}{0.01D_2})}\big)$ \\
\hline
3 & $R(B_u \rightarrow \tau \nu)$  
   & $1.66\pm 0.54$~\cite{Nakamura:2010zzi}
   & Gaussian \\
\hline
4 & $\Delta a_\mu$ 
   & $(28.7\pm8.0)\times 10^{-10}$ $[e^+e^-]$ ~\cite{Davier:2010nc}& Weighted Gaussian average  \\
   & & $(19.5\pm8.3)\times 10^{-10}$ [$\tau^+\tau^-$] ~\cite{Davier:2010nc}&   \\
\hline
5 & $m_t$ 
   & $173.3\pm1.1$~GeV~\cite{tev:2009ec}
   & Gaussian \\
\hline
6 & $m_b(m_b)$  
   & $4.19^{+0.18}_{-0.06}$~GeV\cite{Nakamura:2010zzi}
   & Two-sided Gaussian  \\
\hline
7 & $\alpha_s(M_Z)$ 
   & $0.1176\pm 0.002$~\cite{Amsler:2008zzb}
   & Gaussian  \\
\hline\hline
 &  & LEP\&Tevatron &  $m_h$ sampled from $Gauss(m_h, 1.5)$ \\
8  &         $m_h$    & (\texttt{HiggsBounds} \cite{Bechtle:2011sb})    & $L_8=1$ if allowed. \\
& & &  $L_8=10^{-9}$ if excluded. 
\\
\hline
9 & sparticle & LEP & $L_9=1$ if allowed  \\
   & masses  & Neutral LSP (\texttt{MicrOMEGAs} \cite{Belanger:2001fz}) & $L_9=10^{-9}$ if excluded  \\
\hline
\end{tabular}\label{tab:preLHCobs}
\end{center}
\end{table}

\begin{table}[t]
\caption{LHC measurements used in the current study. The $\alpha_T$ variable is effective in suppressing background from light-quark QCD. SS $2\ell$, and OS $2\ell$ denote same-sign and opposite-sign dileptons, respectively. The $\alpha_T$~\cite{Collaboration:2011zy}, SS~\cite{CMS:SS}, and OS~\cite{CMS:OS} results were published by the CMS Collaboration. We re-weight a~posteriori with the limit $BR(B_s \rightarrow \mu \mu) \le 4.5 \times 10^{-9}$ at 95\% CL \cite{LHCb}. This has an effect only on $D_Z$, which depends on the Higgs mass parameter $m_{H_d}^2$. We also update the Higgs bounds, imposing the constraints from the CMS di-photon searches~\cite{Chatrchyan:2012tw}, which do not have a strong effect on the probability distribution of the soft supersymmetry-breaking parameters.}
\begin{center}
\begin{tabular}{|c|c|c|c|c|}
\hline
$j$&~Analysis and search region & Observed event count   & Data-driven SM \\
         &~ (values in GeV)              & $(N_j) $ & BG estimate      \\
&             &        & $(B_j \pm \delta B_j)$ \\
\hline\hline
1&~$\alpha_T$ hadronic, $275 \le H_T < 325$~ & $782$ & $787.4^{+31.5}_{-22.3}$ \\
2&~$\alpha_T$ hadronic, $325 \le H_T < 375$~ & $321$ & $310.4^{+8.4}_{-12.4}$ \\
3&~$\alpha_T$ hadronic, $375 \le H_T < 475$~ & $196$ & $202.1^{+8.6}_{-9.4}$ \\
4&~$\alpha_T$ hadronic, $475 \le H_T < 575$~ & $62$ & $60.4^{+4.2}_{-3.0}$ \\
5&~$\alpha_T$ hadronic, $575 \le H_T < 675$~ & $21$ & $20.3^{+1.8}_{-1.1}$ \\
6&~$\alpha_T$ hadronic, $675 \le H_T < 775$~ & $6$ & $7.7^{+0.8}_{-0.5}$ \\
7&~$\alpha_T$ hadronic, $775 \le H_T < 875$~ & $3$ & $3.2^{+0.4}_{-0.2}$ \\
8&~$\alpha_T$ hadronic, $875 \le H_T$ & 1 & $2.8^{+0.4}_{-0.2}$ \\
\hline
9&~SS $2\ell$, $H_T>400$, $\met >120$ & $1$ & $2.3 \pm 1.2$ \\
\hline
10&~OS $2\ell$, $H_T>300$, $\met >275$ & $8$ & $4.2 \pm 1.3$ \\
\hline
\hline
$$ 	& Observable 	& Experimental result 	& Likelihood function \\
\hline\hline
11 & $BR(B_s \rightarrow \mu \mu)$ 
   & $\le 4.5 \times 10^{-9}$ \cite{LHCb}
   & $1/\big(1+{\mathrm{exp}}{(\frac{\mu_{11}-D_{11}}{0.01D_{11}})}\big)$ \\
\hline
12 & $m_h$ & $\frac{\sigma(H\rightarrow\gamma\gamma)}{\sigma(H\rightarrow\gamma\gamma)_{SM}}$ \cite{Chatrchyan:2012tw} &  $L_{12}=1$ if allowed. \\
& & &  $L_{12}=10^{-9}$ if excluded.
\\
\hline
\end{tabular}\label{tab:LHCobs}
\end{center}
\end{table}

The RGIs, are functions of the soft mass parameters and therefore a flat prior for the later does not imply a flat prior for the RGIs. In particular, even in the simplest cases, the fact that the parameters have boundary values imply that certain regions are preferred. As an example, consider the subtraction of two mass-squared parameters,
\begin{equation}
f(a,b) = m_a^2 - m_b^2.
\end{equation}
If both $m_{a}$ and $m_{b}$ have a uniform distribution between 0 and 3~TeV, it is clear that the probability of $f(a,b) \simeq \pm (3~{\rm TeV})^2$ will be much smaller than the probability of  $f(a,b) \simeq 0$. This is because in the former case one of the mass parameters has to be equal to 0 while the other is 3~TeV, while the later situation comprises of all cases in which $m_a \simeq m_b$, independent of their value. Therefore, in order to determine the probability distributions of the invariants (and other functions considered later), one should normalize them such that they can be compared to a flat prior for the functions under consideration and not for the masses. In order to do this, we have re-scaled the experimentally weighted probability distributions of the RGIs by the probability distributions for these quantities obtained by varying the mass parameters with a flat uniform distribution in the region originally scanned. We will refer to the later distributions as the ``Flat Un-weighted'' distributions. The details of the exact procedure are given in Appendix A.

The results are shown in Figs.~\ref{Invariants1} and \ref{Invariants2}. The shaded green region represents the flat un-weighted distribution for the RGI being considered. For the $I_{B_i}$s, which depend only on, and are linearly proportional to the gaugino masses, this distribution is flat (apart from a small variation with the gauge couplings). However, for the other RGIs, these distributions acquire definite features. The green line represents the probability distribution obtained considering only the pre-LHC constraints listed in Table~\ref{tab:preLHCobs}. These depend heavily on the constraints on the weak eigenstates coming from LEP and $(g_\mu-2)$, the bounds on the CP-odd and charged Higgs masses and third generation masses coming from the $BR(b \to s \gamma)$ and the LEP/Tevatron Higgs results. The red line, instead, represents the probability distributions obtained after the LHC results are considered~(Table~\ref{tab:LHCobs}).

The details on the computation of the final resultant distribution we label as ``$p(\theta| \rm Exp.)$ Reweighted'' are given explicitly in Appendix A.  However, the method can be simply understood by noting that, as discussed in Appendix A, the ratio of the difference of any 2 probabilities with the flat distribution, $(p_1-p^f)/(p_2-p^f)$,  is preserved when the scan range on the original masses is changed. Therefore, we first subtract the probability distribution determined after the LHC measurements (red line) from the one obtained with the flat masses prior (shaded green) and then shift this distribution by the minimum, setting the minimum at zero. Once this is done, we re-scale the distribution such that the total probability is 1.  This then gives the dashed black line denoting our final resultant distribution. This distribution is flat in regions not scanned or impacted by experiment and  enhances and reflects the actual effect of the experiments on the RGIs. Larger values of this distribution highlight the regions where experimental input has increased the likelihood and values less than the flat probability show regions where experiments disfavor model space. 

Due to this method of re-scaling and subtracting the probabilities, one has to be careful when using these distributions to calculate resultant quantities of interest~(as will be done, for example, when calculating product probabilities). One needs to convert the distribution back to a true probability via the scale factor $SF$~(labeled ``SF'' in plots) defined in Appendix A:
\begin{equation}\label{SF}
p(\theta | \mbox{post-LHC})=p(\theta | Flat )+\frac{1}{SF}\left[p(\theta | \mbox {Exp.) Reweighted}-p(\theta | Flat )\right],
\end{equation}
where $p(\theta | Flat )=1/({\rm no. \; bins})$. Unless otherwise noted, ${\rm no. \; bins}=100$ in all plots, implying $p(\theta | Flat )=0.01$. 

\subsection{Results}

\begin{figure}
\begin{center}
\begin{tabular}{c c}
\includegraphics[width=0.45\textwidth]{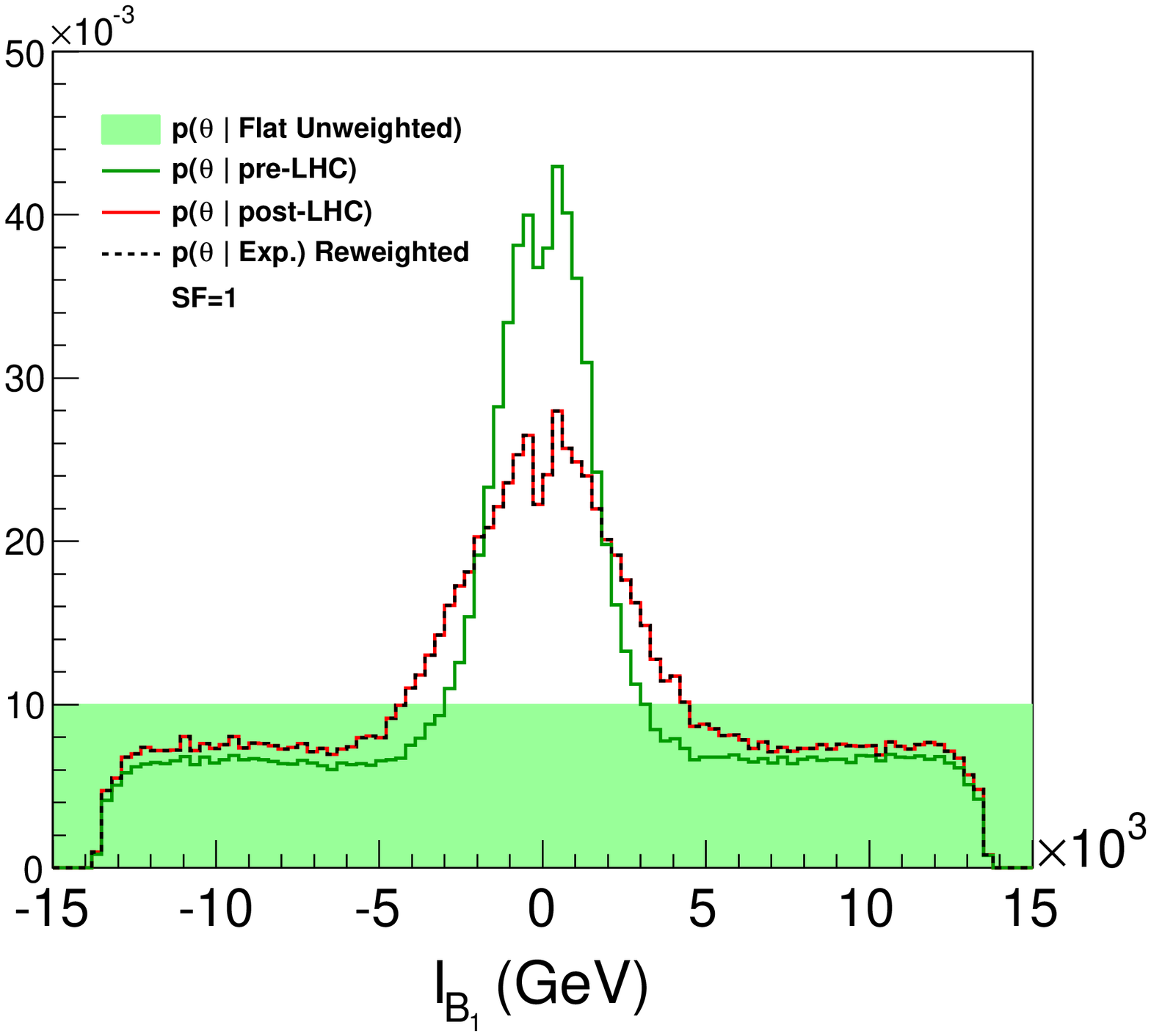}  &
\includegraphics[width=0.45\textwidth]{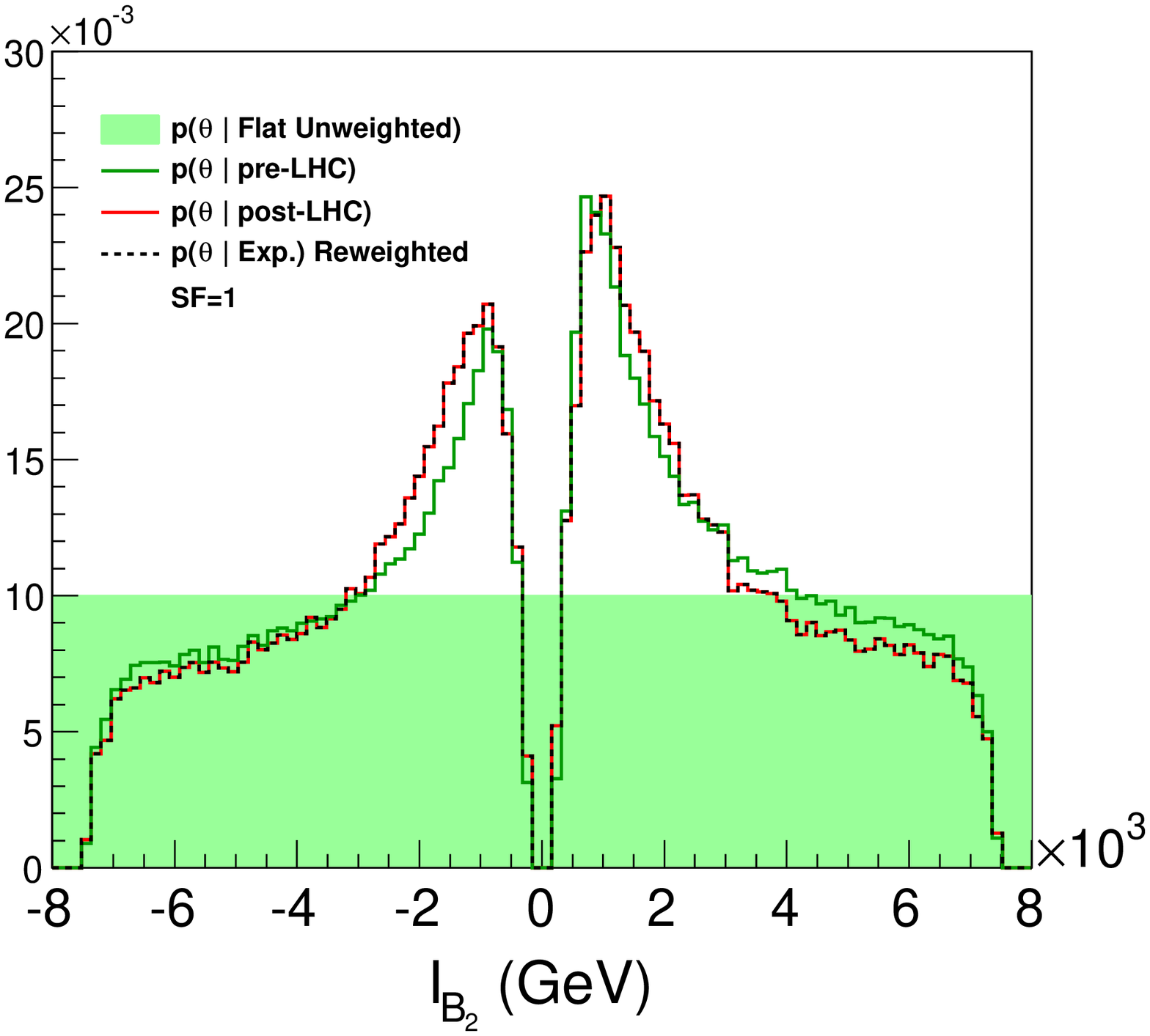}  \\
\includegraphics[width=0.45\textwidth]{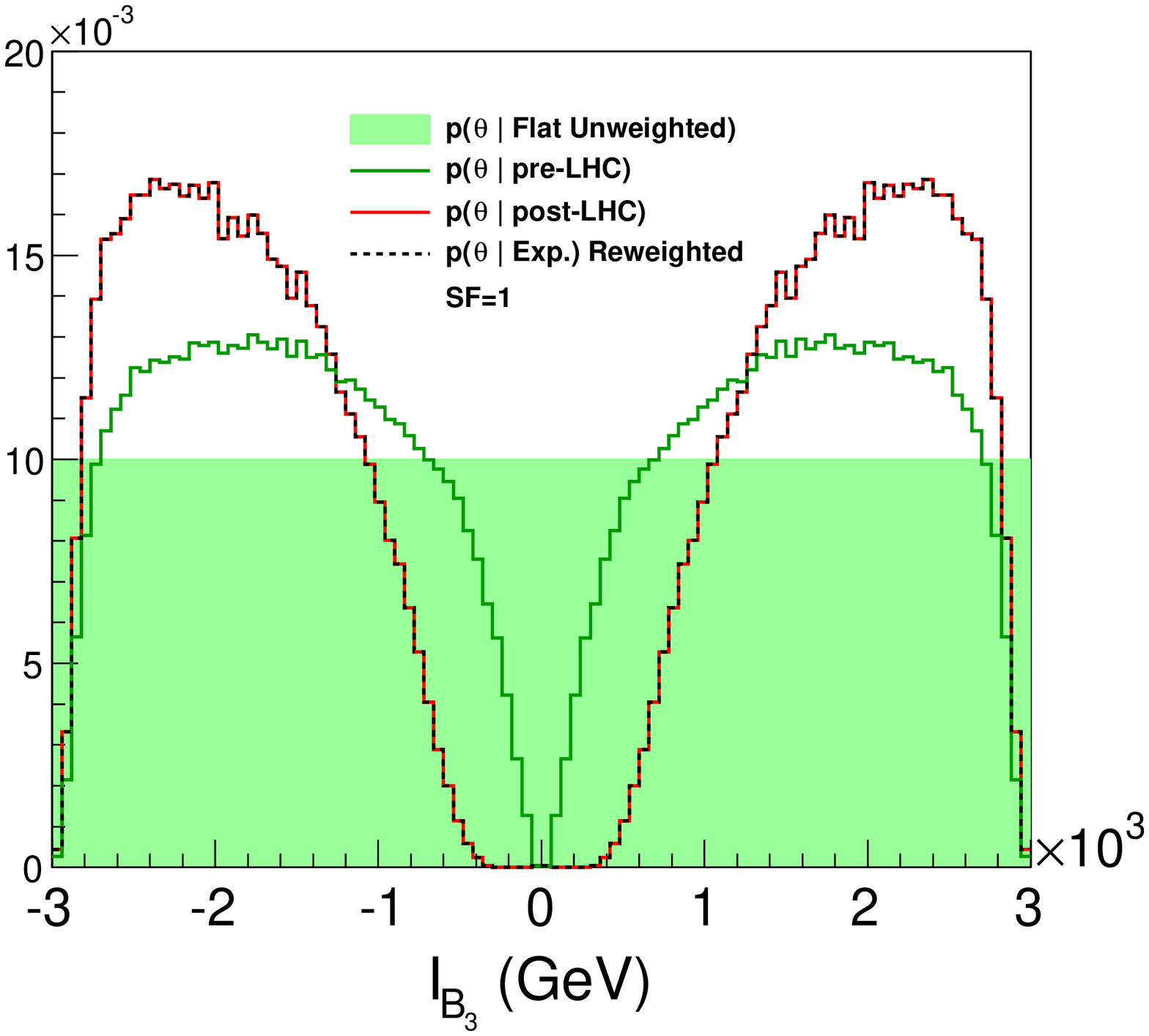}  &
\includegraphics[width=0.45\textwidth]{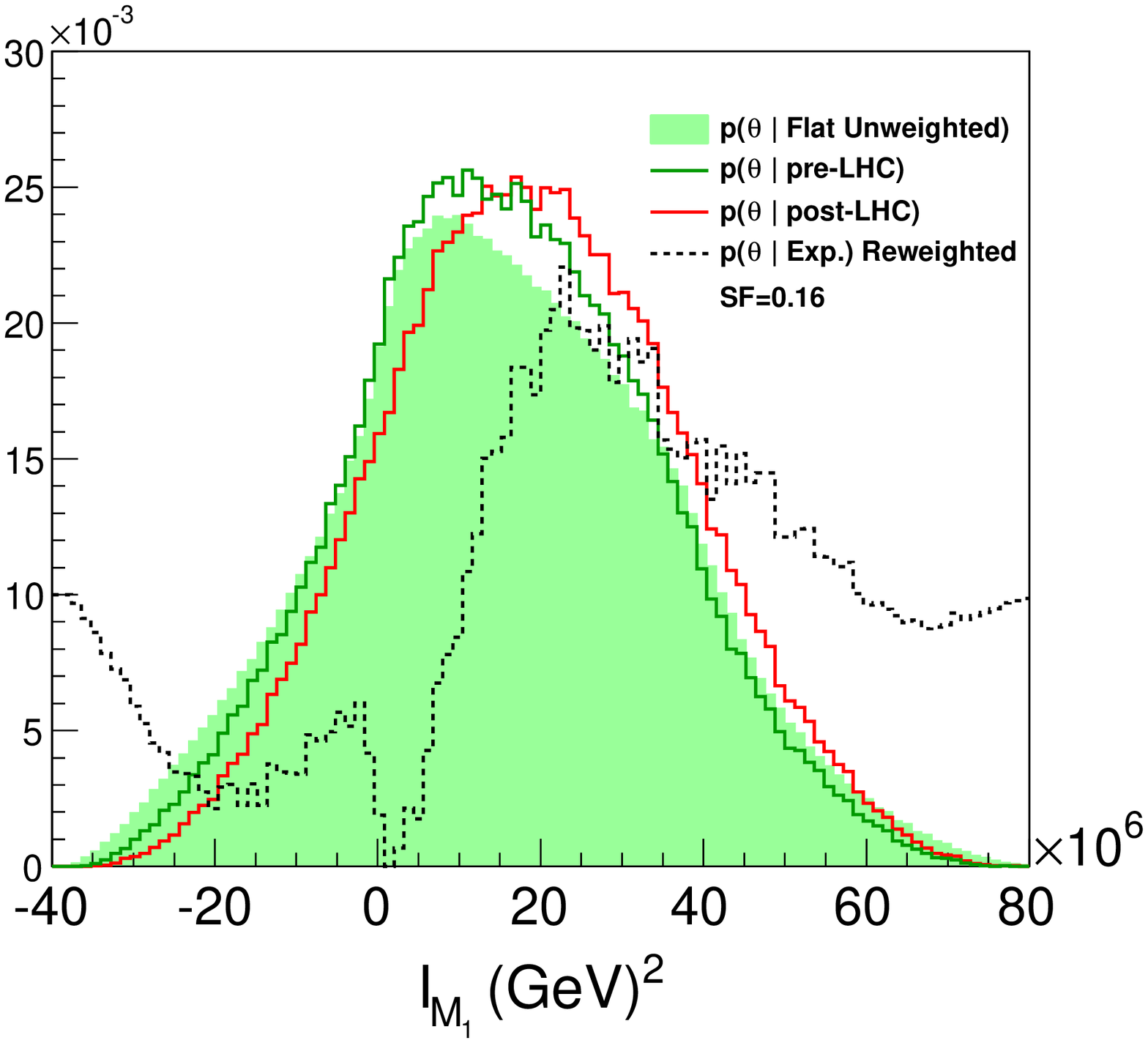}  \\
\includegraphics[width=0.45\textwidth]{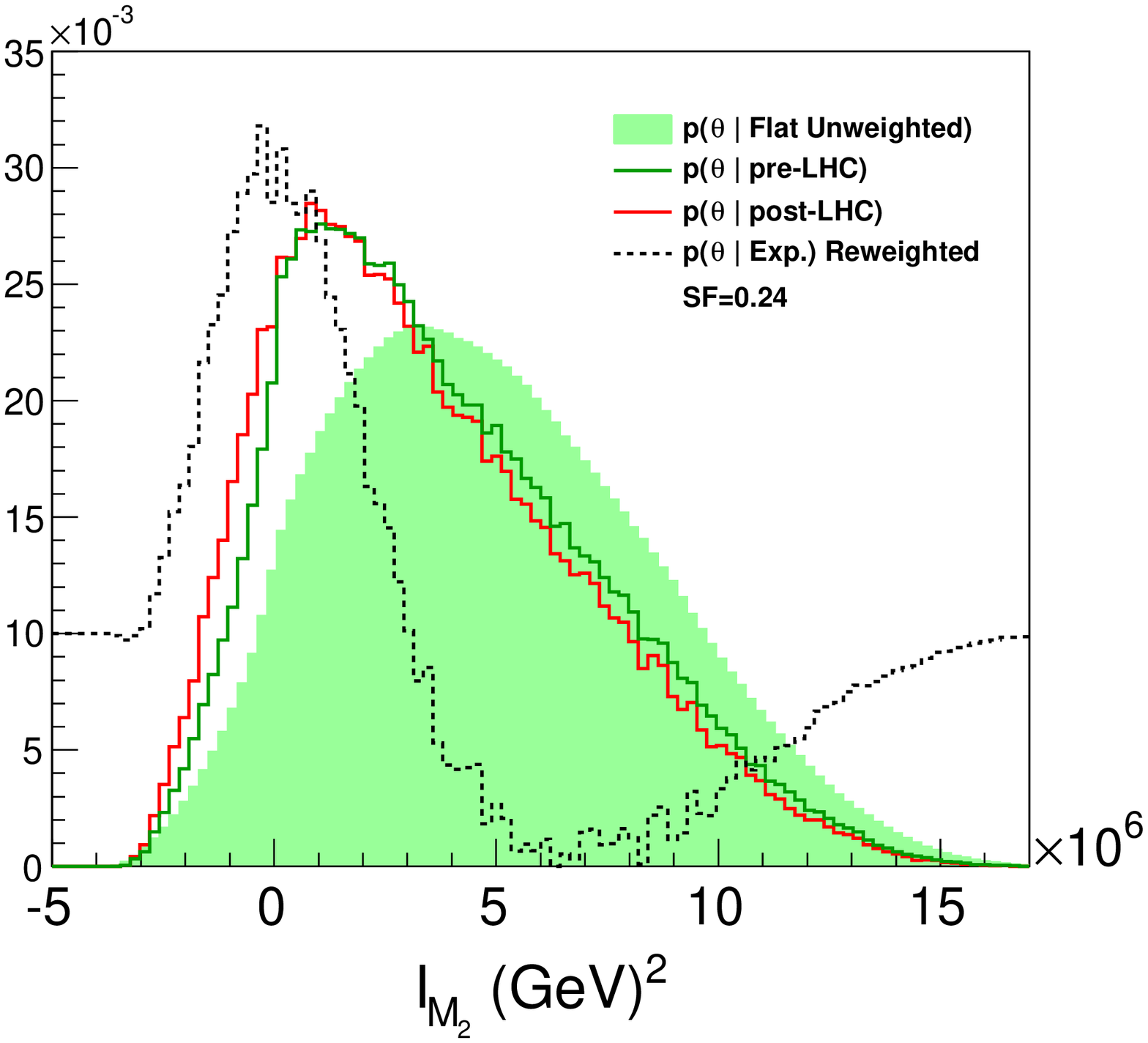}  &
\includegraphics[width=0.45\textwidth]{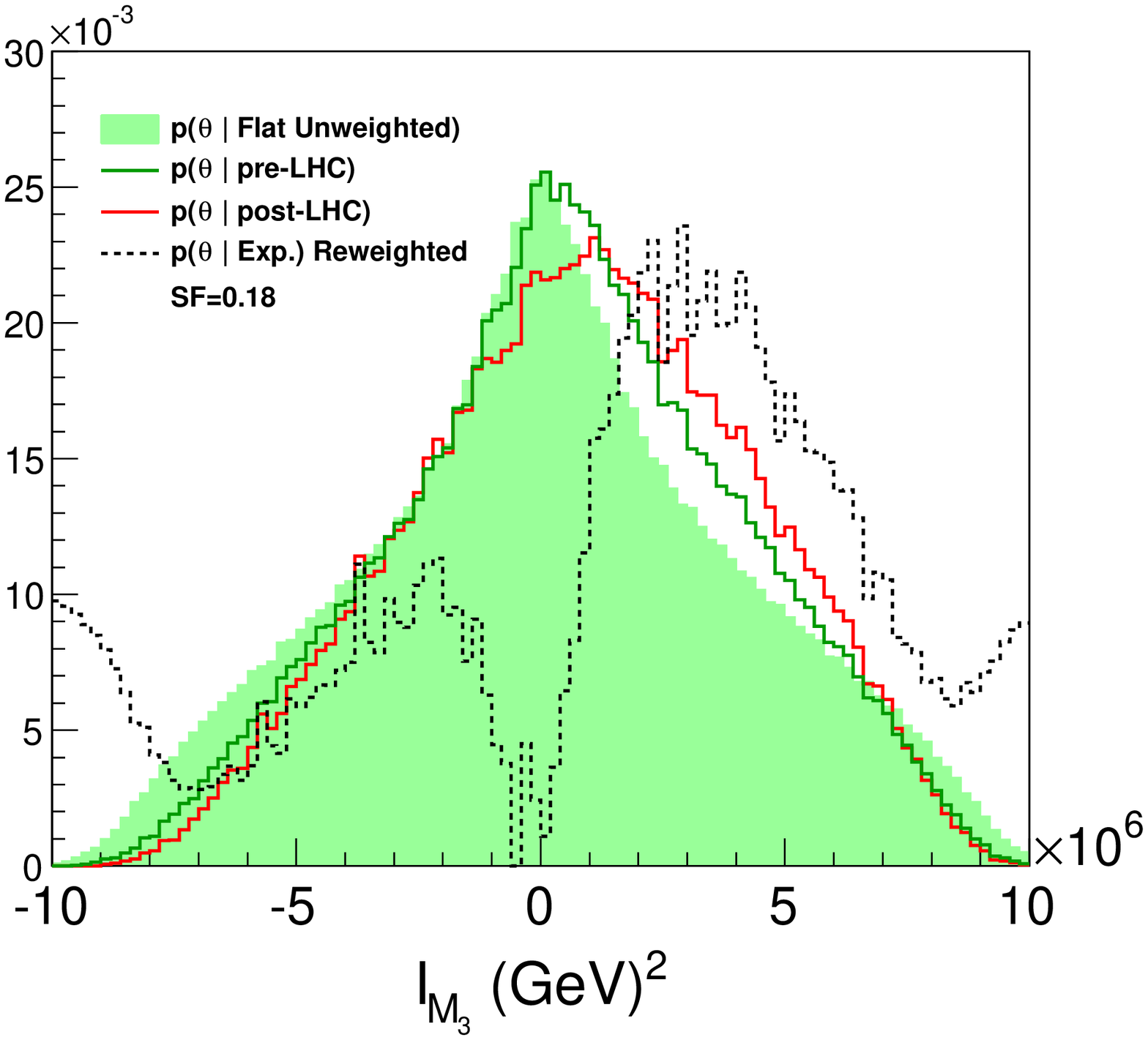}  \\
\end{tabular}
\end{center}
\caption{Distribution of the RGIs before and after the LHC constraints are added (green and red lines), flat distribution (shaded green) and subtracted probability distribution (dashed black line). }
\label{Invariants1}
\end{figure}
\begin{figure}
\begin{center}
\begin{tabular}{c c}
\includegraphics[width=0.45\textwidth]{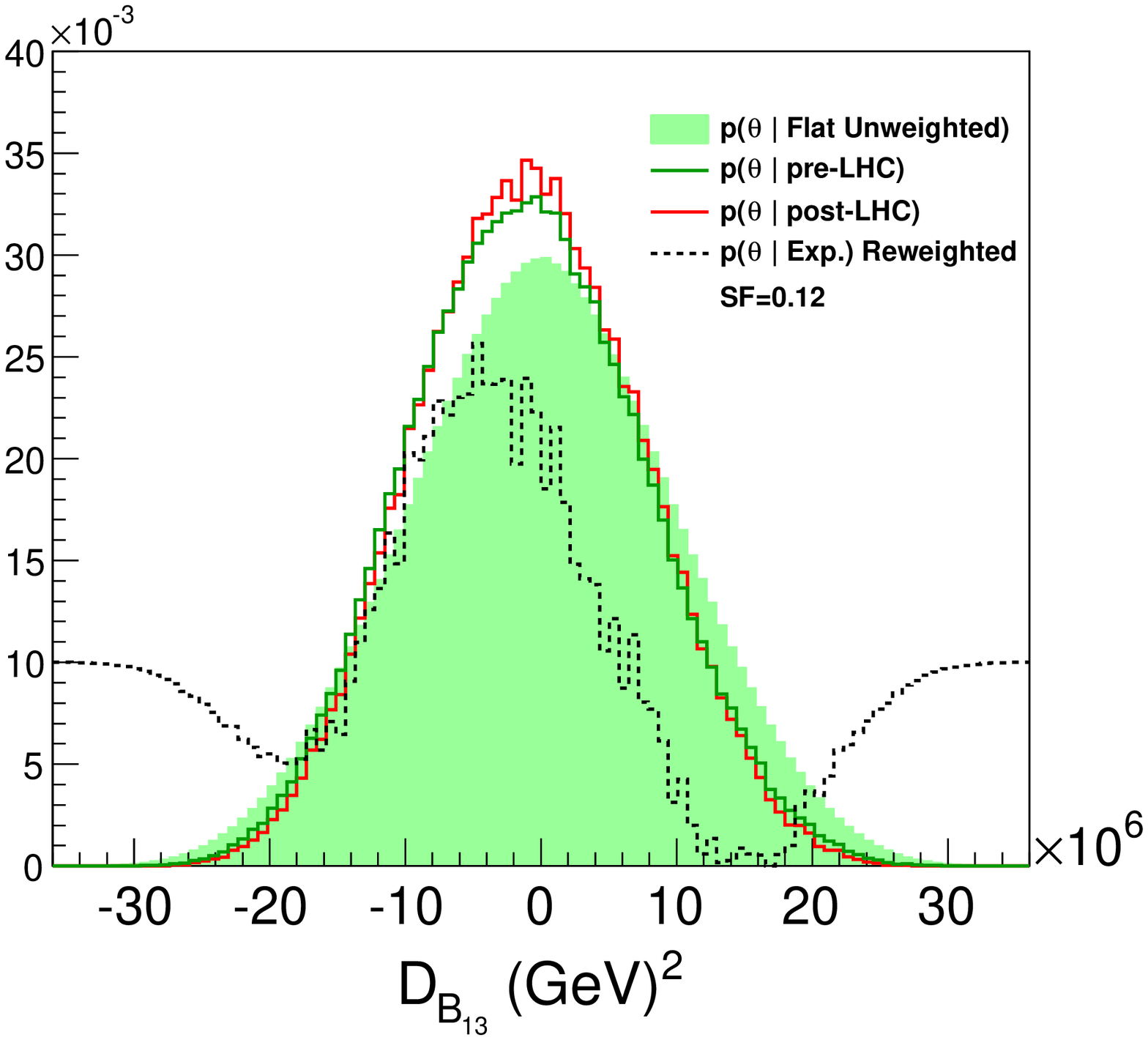}  &
\includegraphics[width=0.45\textwidth]{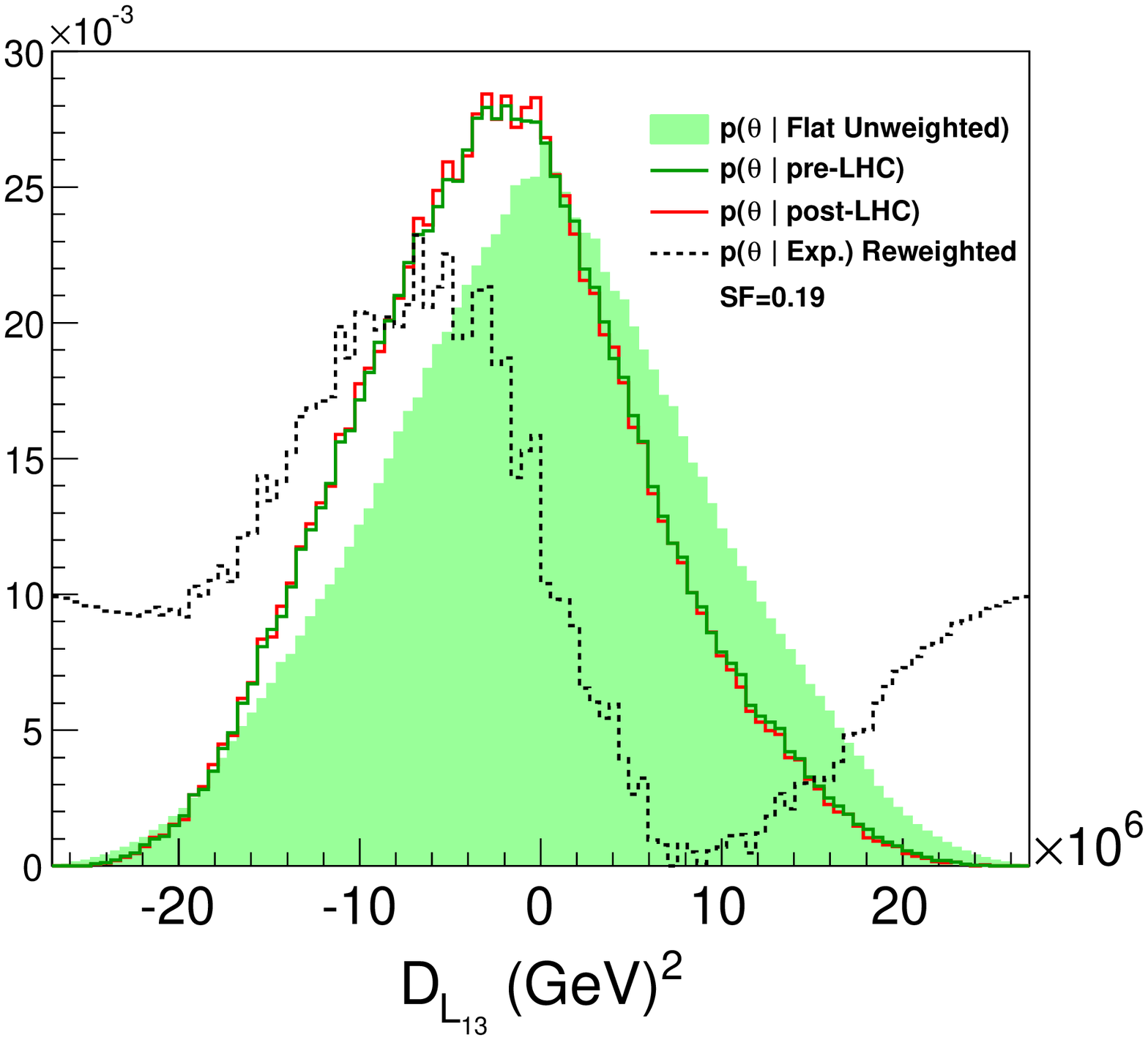}  \\
\includegraphics[width=0.45\textwidth]{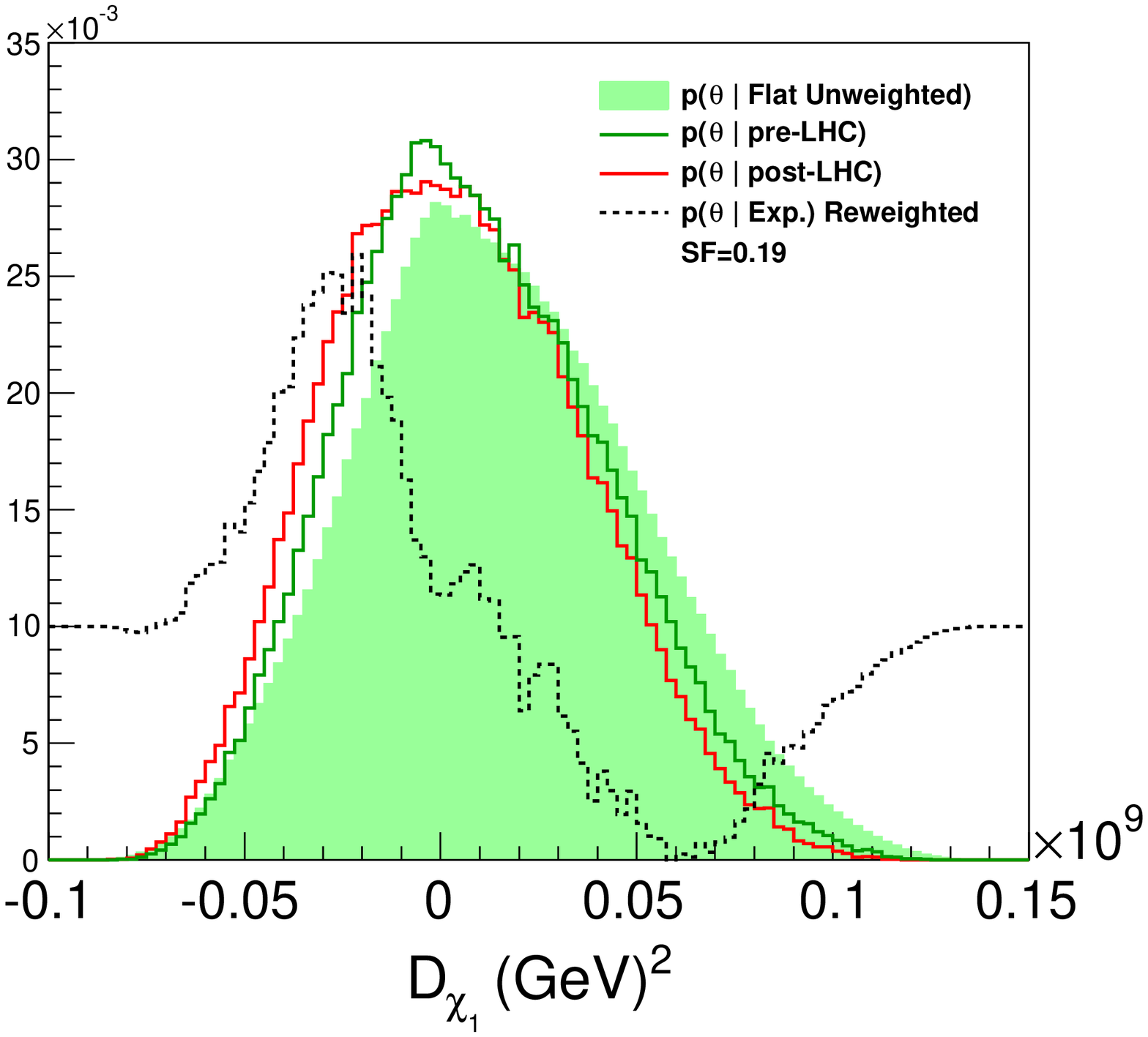}  &
\includegraphics[width=0.45\textwidth]{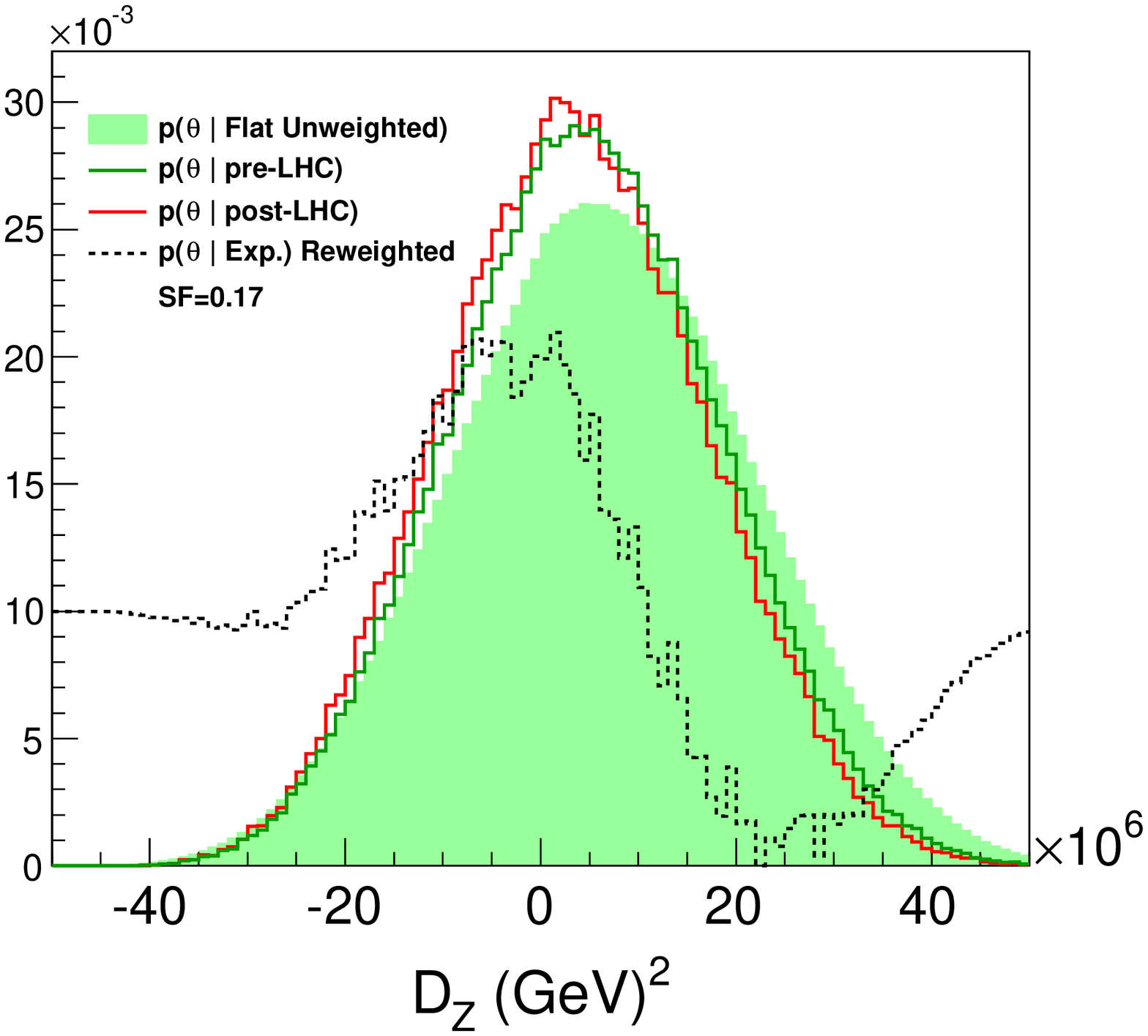}  \\
\end{tabular}
\end{center}
\caption{Distribution of the RGIs before and after the LHC constraints are added (green and red lines), flat distribution (shaded green) and subtracted probability distribution (dashed black line). }
\label{Invariants2}
\end{figure}

Here we will discuss the probability distributions presented in Figs.~\ref{Invariants1} and~\ref{Invariants2} for the different RG Invariants. We don't analyze $D_{Y_{13H}}$ and $D_{Y_\alpha}$ since these two depend on almost all the soft masses and the current experimental bounds on these combinations are too weak to show an effect on the probability distributions. 

The first three distributions displayed in Fig.~\ref{Invariants1} are for the three  $I_{B_i}$ which are equal to the gaugino masses divided by the square of the corresponding gauge coupling, $M_i/g_i^2$.  The LEP constraints on charginos, sleptons  and gluinos, together with the requirement of a neutral particle to be the lightest supersymmetric one, lead to a preference towards low values of the bino mass, $M_1$, increasing the probability for small $I_{B_1}$. The LHC modifies this distribution indirectly, through the updated bounds on the gluino and squark masses.  This is due to the requirement of having a neutral particle as the LSP:  When at least one of the squarks and/or the gluino is light, one neutralino or a sneutrino is forced to be even lighter. Of all the neutral particles, the bino is the only one that is not related to the mass of other charged particles and therefore is not pushed to larger values. Hence, the bino can be very light increasing its  probability of being the LSP.  For heavier squarks and gluinos, the neutral particles can be heavier and consequently the bino mass probability distribution moves to larger values. Regarding $I_{B_2}$, the LEP constraints on chargino masses, together with the bounds on $(g_{\mu}$$-$$2)$ restrict small values of $M_2$, while leading to a preference for values of $M_2$ near the weak scale. The LHC does not significantly modify this constraint. Since we have not implemented the Tevatron bounds on the squark and gluino masses, the pre-LHC constraints on $I_{B_3}$ are dominated by the indirect effect of requiring that  a gluino cannot be the LSP, which therefore disfavor the region of small values of this quantity. The LHC SUSY searches further constrain values of $M_3$ up to $\sim$1~TeV, which is clearly shown in the $I_{B_3}$ distribution. 

Looking at $I_{M_1}$, also displayed in Fig.~\ref{Invariants1}, the lower bound on the slepton and squark masses from LEP disfavor the lowest values of this RGI. The LHC further strengthens this trend by increasing the bounds on the squark masses. $I_{M_2}$ is strongly shifted to lower values by $(g_{\mu}$$-$$2)$, which leads to a preference towards small values of $M_2$ and the left-handed slepton masses. The LHC data does not modify this trend in any significant way. Finally, for $I_{M_3}$, the previously discussed bounds on the  gluino (and similar bounds on the squark) masses, push them to larger values, leading to a preference towards non-zero values of this RGI. The LHC, with significantly larger bounds on the gluino mass, leads to a further preference towards larger values of this RGI.  The asymmetry between positive and negative values comes from the fact that the gluino has a larger cross section and is constrained to be heavy even in the case in which only the third generation squarks are lighter than the gluinos.  Individual squark species, on the other hand have lower cross sections and therefore have a higher probability of being lighter in the pMSSM~\cite{Sekmen:2011cz}.

In order to understand the behavior of the other four invariants displayed in Fig.~\ref{Invariants2}, it is convenient to analyze the results of Ref.~\cite{Sekmen:2011cz}. The bound on the Higgs mass leads to a preference for larger values of the third generation masses, beyond the bounds on the first and second generation masses obtained at the Tevatron. Since the Higgs mass bounds are approximately symmetric in their dependence on $m_{Q_3}$ and $m_{u_3}$ and the negative weight of $m_{Q_3}$ on $D_{B_{13}}$ is twice as large as the positive one of $m_{u_3}$, we see a preference towards negative values of $D_{B_{13}}$. The LHC SUSY searches have not yet changed this tendency in a significant way. Regarding $D_{L_{13}}$, the preference towards small values of $(g_\mu$$-$$2)$ lead to a preference towards small left-handed second generation sleptons, beyond the LEP constraints, and therefore towards lower values of this RGI. The LHC, again, does not have a strong impact on this distribution. Lower values of the left-handed slepton masses also affect the pre-LHC distribution of $D_{\chi_1}$, pushing it to lower values. At the LHC, there is a somewhat stronger constraint on the left-handed squarks with respect to the right-handed ones, leading to slightly lower values of $D_{\chi_1}$. Finally, $D_Z$ is strongly dominated by the bounds on the CP-odd Higgs mass coming from $B_s \to \mu\mu$, which push $m_{H_d}^2$ to larger values and $D_Z$ to lower ones.

\section{Gaugino Mass Unification}
Gaugino Mass unification is a common feature of models in which supersymmetry breaking occurs at scales larger than the GUT scale. In such a case, up to threshold corrections, one should expect that due to the extended gauge structure, the gaugino masses associated with the $SU(3)_c$, $SU(2)_L$ and $U(1)_Y$ unify at the GUT scale.

At scales lower than the GUT scale, however, threshold corrections can be large and could lead to quite different values of the three gaugino masses. This happens, in particular, if the gaugino masses receive large contributions induced by gravitational interactions governed by the scale anomaly. These contributions are proportional to the $\beta$ function coefficients of the respective gauge couplings.

Further contributions to the gaugino masses may come from gauge mediation, induced by messengers charged under the SM gauge groups and with direct coupling to the supersymmetry-breaking sector. In minimal models, the gauge-mediated gaugino mass contributions at the messenger mass scale are proportional to the value of the square of the gauge couplings at the same scale.

The simplest contributions to the gaugino masses at the messenger scale are then given by

\begin{equation}
M_i =  A \;  \beta_i \; g_i^2   +  B \;  g_i^2  \; \theta(M_{\rm mess}^2 - Q^2) +  \frac{M_{1/2}}{g_{\rm GUT}^2}  \; g_i^2
\label{m_gaugino}
\end{equation}
where the coefficients $A$, $B$ and $M_{1/2}$ parameterize the anomaly-mediated, minimal gauge-mediated and minimal SUGRA-mediated contributions to the gaugino masses. We have inserted a $\theta$ function to denote the fact that the gauge mediated contribution is only relevant at energy scales, $Q$, below the messenger mass scale, $M_{\rm mess}$. 

The condition of gaugino mass unification can be written in terms of RGIs. Indeed, assuming that the gaugino masses unify at some scale, $M_{\rm unif}$,
\begin{eqnarray}
\frac{\beta_2 I_{B_1} - \beta_1 I_{B_2}}{\frac{\beta_2}{g_1^2} - \frac{\beta_1}{g_2^2}} = M_{g}\;,
\nonumber\\
&\nonumber\\
\frac{\beta_3 I_{B_1} - \beta_1 I_{B_3}}{\frac{\beta_3}{g_1^2 }- \frac{\beta_1}{g_3^2}} = M_{g}\;,
\end{eqnarray}
where $\beta_i = \{33/5,1,-3\}$ for $i =\{ 1,2,3\}$, $g_i$ are the gauge couplings at the gaugino mass unification scale and $M_g$ is the common gaugino mass. The denominators in the above equation are nothing but $I_{g_2}$ and $-3 I_{g_3}$, respectively. The value of the gauge couplings at the gaugino unification scale may be obtained by just dividing the above expressions by the corresponding $I_{B_i}$ invariant. In particular ~\cite{Carena:2010gr},
\begin{eqnarray}
g_1^2(M_{\rm unif})  = \frac{\beta_2  - \beta_1 I_{B_2}/I_{B_1}}{I_{g_2}} \simeq  \frac{\beta_2 - \beta_1 I_{B_2}/I_{B_1}}{2 (\beta_2 - \beta_1)}
\nonumber\\
g_1^2(M_{\rm unif})  = \frac{\beta_3  - \beta_1 I_{B_3}/ I_{B_1}}{-3 I_{g_3}}  \simeq  \frac{\beta_3 - \beta_1 I_{B_3}/I_{B_1}}{2(\beta_3 - \beta_1)}
\label{gauginounif}
\end{eqnarray}
where we have used the fact that $g^2_{GUT} \simeq 1/2$. From the equality of the first and second line in Eq.~(\ref{gauginounif}), one can see that gaugino mass unification requires that
\begin{equation}
(\beta_3 - \beta_2) I_{B_1} - (\beta_3 - \beta_1) I_{B_2} + (\beta_2 - \beta_1) I_{B_3} = 0,
\label{condition123}
\end{equation}
or, inserting the numerical values of the $\beta_i$ coefficients~\cite{Carena:2010gr},
\begin{equation}
12 I_{B_2} - 5 I_{B_1} - 7 I_{B_3} = 0.
\end{equation}

Using the expression for the gaugino mass, Eq.~\ref{m_gaugino},  we get that at the weak scale,
\begin{equation}\label{IB_beta}
I_{B_i} = A \;  \beta_i + \left( B + M_{1/2}/g_{\rm GUT}^2 \right)  =  A \; \beta_i  + C
\end{equation}
where we have joined the scale and gauge mediated contributions, $C\equiv \left( B + M_{1/2}/g_{\rm GUT}^2 \right)$, since they cannot be distinguished at low energies. Note that this is similar to the case of Mirage mediation~\cite{Choi:2004sx}~\cite{Choi:2005uz}~\cite{Endo:2005uy}. Observe that the condition given in Eq.~(\ref{condition123}) is always satisfied in this case.

Interestingly enough, inserting Eq.~\ref{IB_beta} in both the expressions for $g_1^2(M_{\rm unif})$ in Eq.~(\ref{gauginounif}), the same equation is obtained, giving the necessary condition for the unification of gaugino masses at some scale, 
\begin{eqnarray}
g_1^2(M_{\rm unif}) \simeq \frac{1}{2} \times \frac{C}{C + A \beta_1}.
\label{g12gaugino}
\end{eqnarray}
In addition, the above expression is independent of $\beta_{2,3}$.

In general, for positive values of $A$ and $C$, depending on which supersymmetry-breaking mechanism is dominant, the apparent gaugino unification scale can vary from the infrared to the GUT scale. In order for gaugino mass unification to take place at a physical scale, however, we need that $0.5 \geq  g_1^2(M_{\rm unif}) \geq 0.2$, which sets interesting constraints on the values of $A$ and $C$.  For $A=0$, one gets that unification is at the GUT scale. For $C =0$, instead, one obtains that unification occurs for $g_1^2(M_{\rm unif}) \simeq 0$, which is an un-physical value, and for which $I_{B_1}$ diverges unless the gaugino mass also vanishes. Let us remark, however, that the unification of gaugino masses obtained by extrapolating the RG evolution into un-physical scale values, at which the physical spectrum is not the MSSM one, could still say something relevant about the supersymmetry-breaking mechanism.  For example, the unphysical $g_1^2(M_{\rm unif}) \simeq 0$ for $C=0$ is the expected apparent unification value in anomaly mediation scenarios. It is therefore very interesting to use the above expressions, Eq.~(\ref{gauginounif}), to check the consistency of gaugino mass unification, and to obtain information about the scale at which it may occur.

Let us elaborate further on the above point. Although we computed the gaugino unification scale by using $g_1^2$, we could have used any other gauge coupling, and the condition in Eq.~(\ref{g12gaugino}) would be the same, with $g_1^2$ and $\beta_1$ replaced by the corresponding $g_i^2$ and $\beta_i$. The fact that the conditions one obtains are consistent with each other can be obtained by rewriting Eq.~(\ref{g12gaugino}) for any $g_i^2$,  in the following way
\begin{equation}
\frac{1}{g_i^2 (M_{\rm unif})} = 2\left(1  + \; \frac{A}{C}\; \beta_i\right).
\label{gi2}
\end{equation}
This has the correct form for the evolution equation for the gauge couplings from the scale $M_{GUT}$, where $1/g_i^2 = 2$, to other energies, provided we interpret
\begin {equation}
\log\left(\frac{M_{GUT}}{M_{\rm unif}} \right) = 16 \pi^2 \frac{A}{C}\;,
\end{equation}
or, equivalently
\begin{eqnarray}
M_{\rm unif}  & = & M_{GUT} \; \exp\left(-16 \pi^2 \frac{A}{C}\right) ,
 \nonumber \\
 & \simeq & M_{GUT} \; \exp\left[ - \frac{8 \pi^2}{\beta_i}  \left(   \frac{1}{g_i^2(M_{\rm unif})} - 2 \right) \right] .
 \label{Munifgunif}
\end{eqnarray}
Since $\beta_3$ is negative, for large values of $A/C$, the effective scale, $Q_{\rm unif}$, may be below the QCD Landau pole and therefore $g_3^2$, from Eq.~(\ref{gi2}) becomes negative, and so clearly un-physical. Negative values of $g_{1,2}^2$ may also be obtained for negative values of $A$ or $C$. 

\begin{figure}
\begin{center}
\begin{tabular}{c c}
\includegraphics[width=0.45\textwidth]{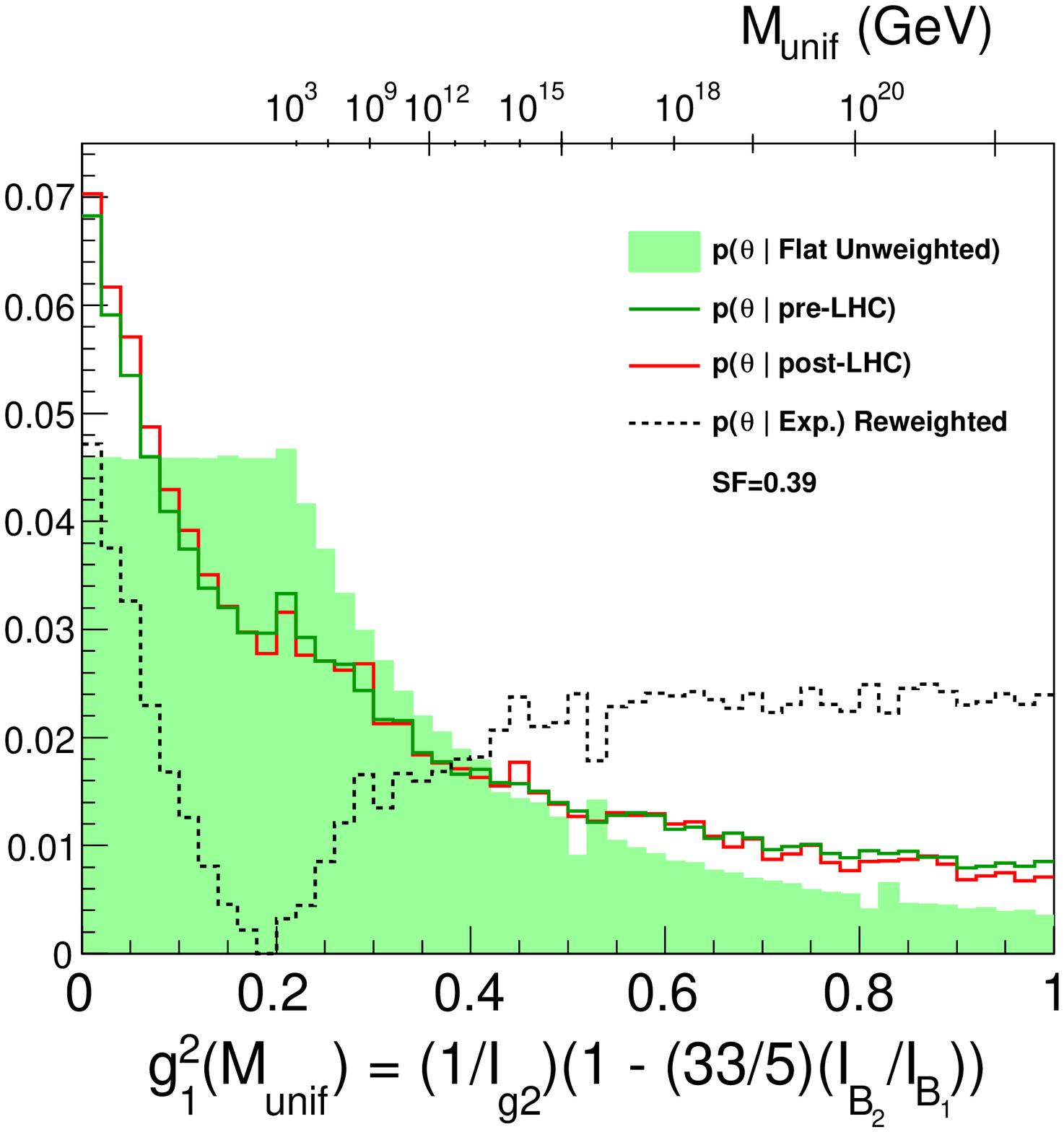}  &
\includegraphics[width=0.45\textwidth]{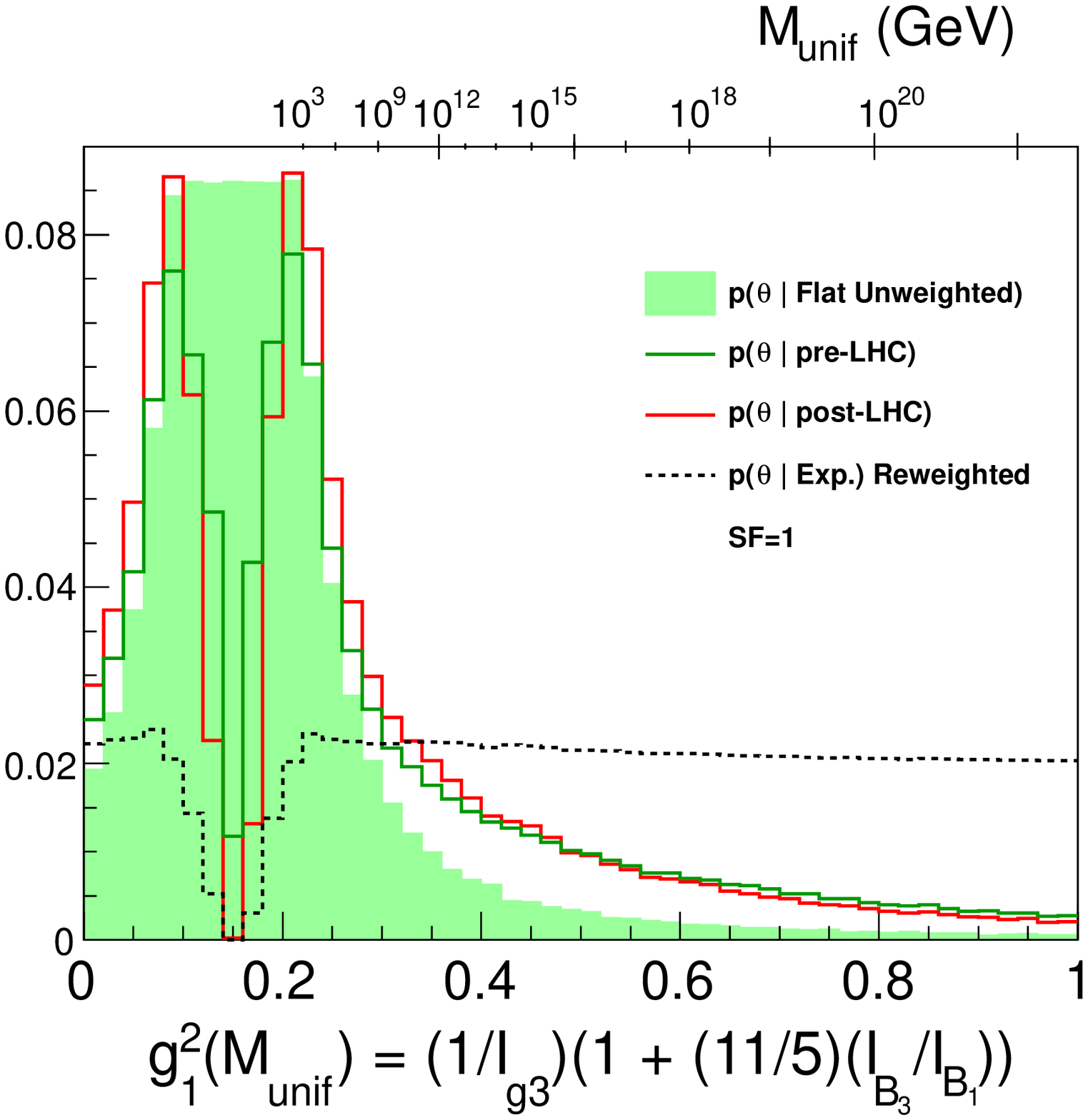}  \\
\end{tabular}
\end{center}
\caption{Distributions of the value of the gauge coupling $g_1^2$ at the Gaugino Mass unification scale before and after LHC constraints (green and red lines), flat distribution (shaded green) and subtracted probability distribution (dashed black line). $\rm no.\; bins=50$ for these plots, implying $p(\theta|Flat)=0.02$.}
\label{GauginoUnif}
\end{figure}

\begin{figure}
\begin{center}
\begin{tabular}{c c}
\includegraphics[width=0.5\textwidth]{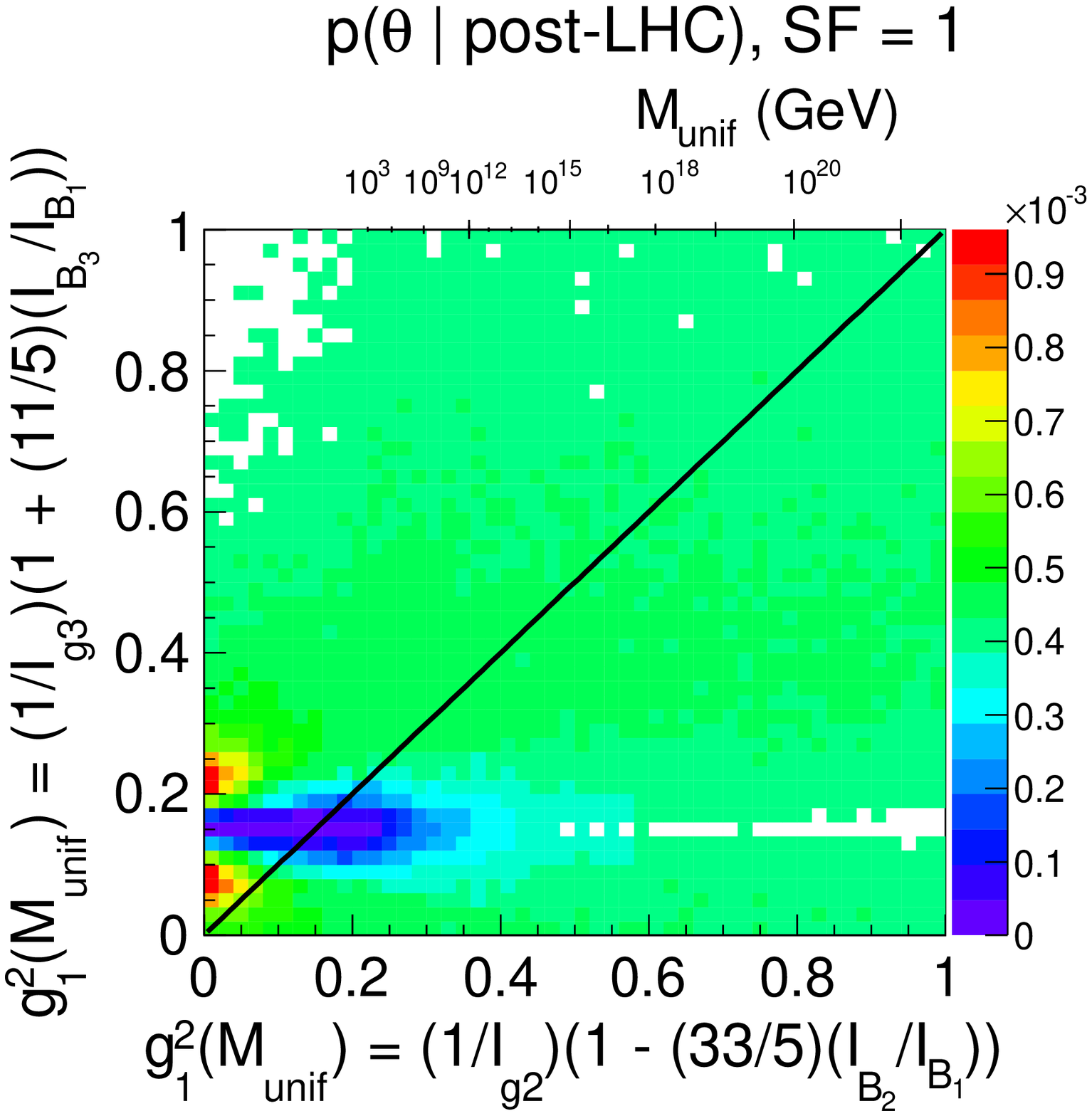}  &
\includegraphics[width=0.5\textwidth]{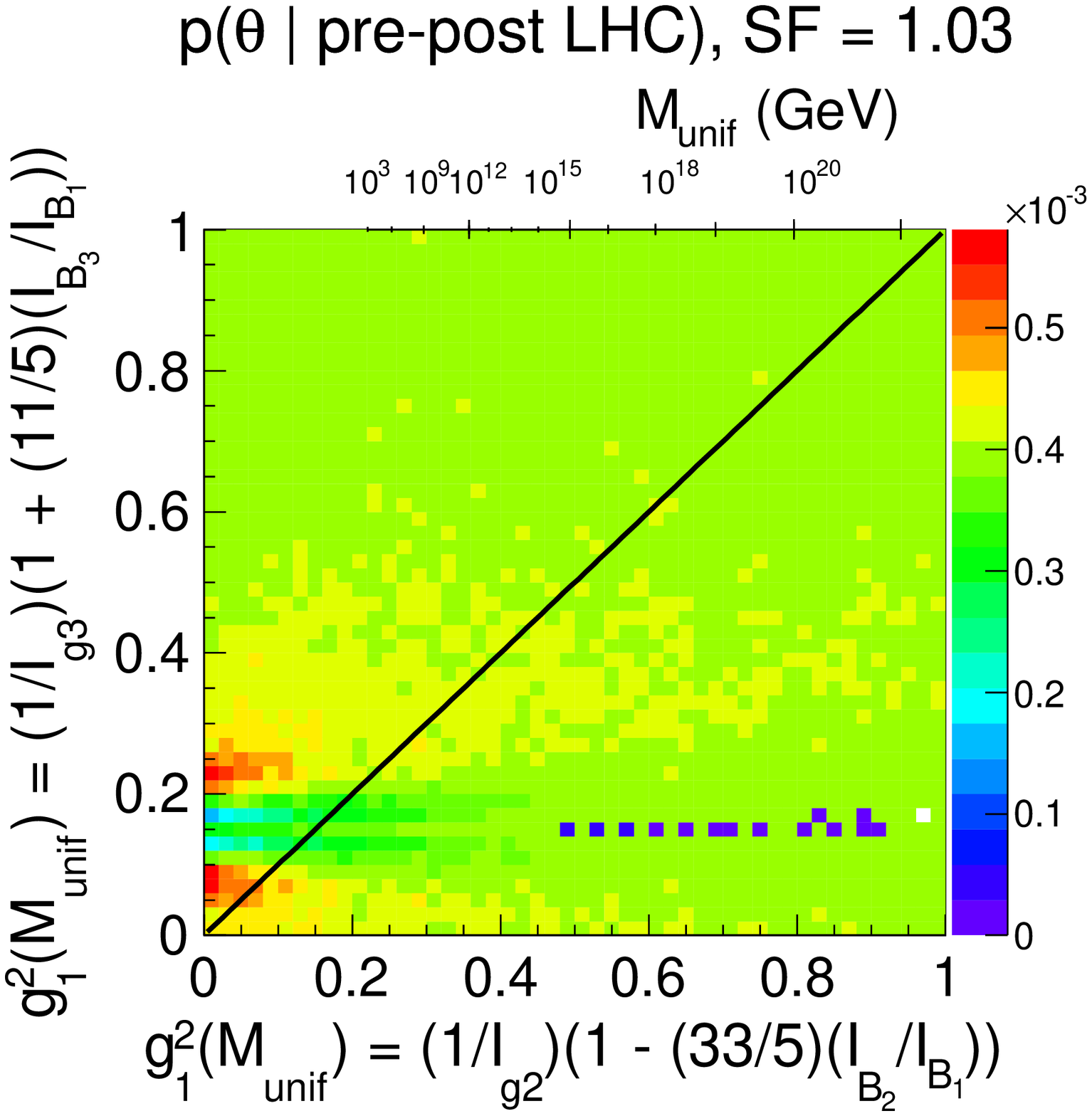}  \\
\end{tabular}
\end{center}
\caption{Probability Distributions for the gauge coupling, $g_1^2$, at the Gaugino Mass unification scale. \textit{Left:} After LHC constraints. \textit{Right:} Difference between pre and post LHC probabilities.  There are 50 bins for each axis in these plots, implying $p(\theta|Flat)=0.4\times 10^{-3}$.}
\label{GauginoUnif1223}
\end{figure}

\begin{figure}
\begin{center}
\includegraphics[width=0.85\textwidth]{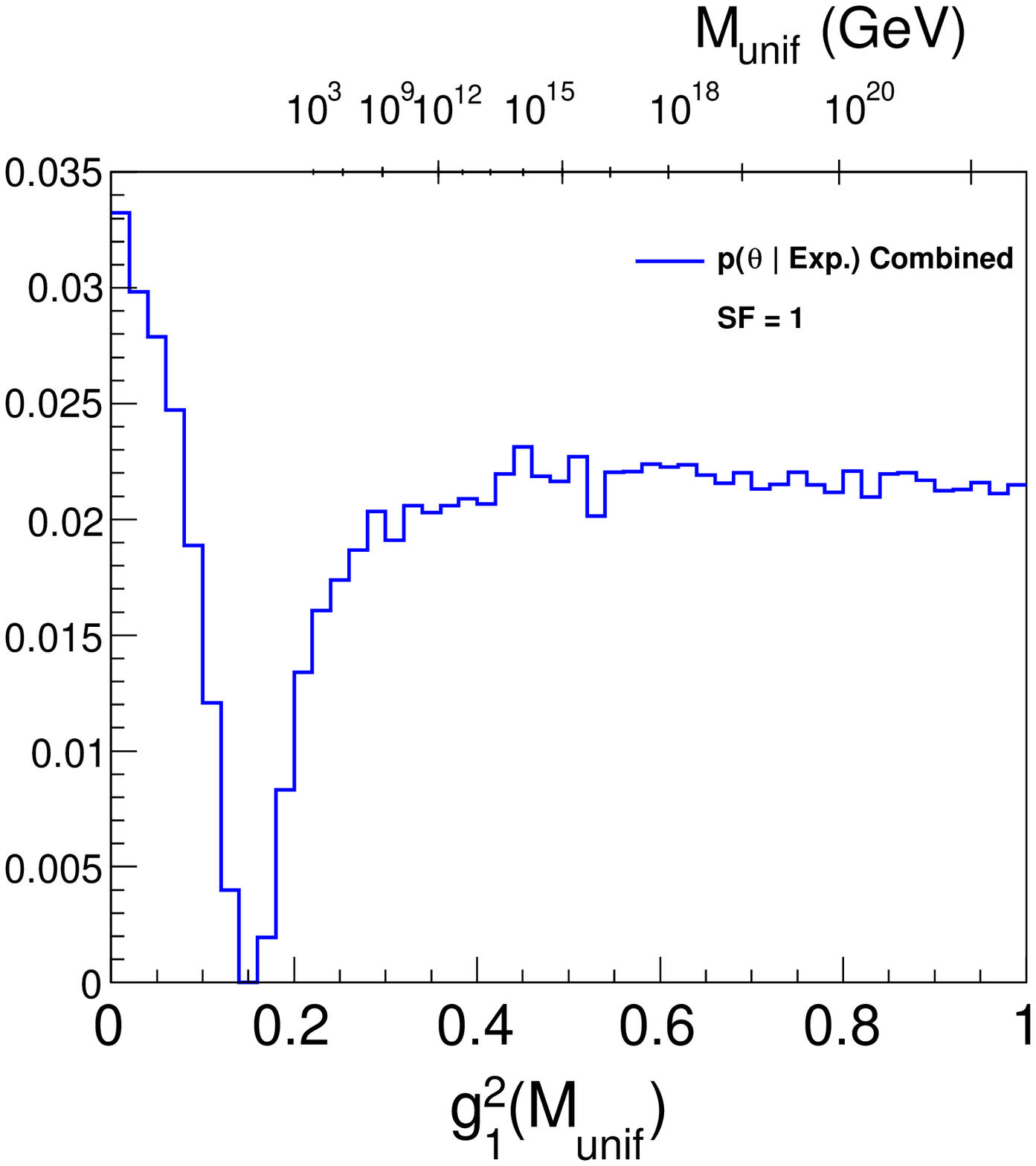}
\end{center}
\caption{Final product probability distributions of the value of the gauge coupling $g_1^2$ at the Gaugino Mass unification scale, demanding that both determinations of $g_1^2$ shown in Fig.~\ref{GauginoUnif} agree and apparent Gaugino Mass Unification takes place at the scale $M_{\rm{unif}}$.  The number of bins is taken to be 50 for this plot, implying $p(\theta|Flat)=0.02$.}
\label{GauginoUnifComb}
\end{figure}

In order to analyze the experimental impact on the scale of Gaugino Mass Unification, we have studied the two possible independent determinations of $g_1^2(M_{\rm unif})$ coming from the ratios $I_{B_2}/I_{B_1}$ and $I_{B_3}/I_{B_1}$, Eq.~(\ref{gauginounif}), respectively. If gaugino masses unify at a certain scale, those two determinations should lead to the same value of $g_1^2(M_{\rm unif})$. Since the unification scale is not necessarily the messenger scale, we have only restricted the value of the gauge coupling, $g_1^2(M_{\rm unif})$, to lie between 0 and 1.

Fig.~\ref{GauginoUnif} represents the probability distributions of $g_1^2(M_{\rm unif})$ obtained by the two ways described above (Eq.~\ref{gauginounif}). The green, red and black curves and the shaded green area have the same interpretation as the one in Figs.~\ref{Invariants1} and~\ref{Invariants2}. The results are very  interesting, since values of the gauge coupling of about its weak scale value $g_1^2(M_{\rm unif}) \simeq 0.2$, are clearly disfavored, while unification at the GUT scale $g_1^2(M_{\rm unif}) \simeq 0.5$ or at values consistent with anomaly mediation $g_1^2(M_{\rm unif})\simeq 0$ are somewhat preferred.

Fig.~\ref{GauginoUnif1223} shows a two-dimensional representation of these results, comparing the results for $g_1^2(M_{\rm unif})$ obtained by the two different equations. The left panel shows the results after LHC constraints are used and the right panel shows the difference between pre- and post-LHC. Dark (light) blue is strongly (weakly) disfavored, while the green regions provide an acceptable description. Yellow (red ) is weakly (strongly) preferred. The black diagonal line shows the (equal) values that should be obtained for Gaugino Mass Unification to be realized. From the right panel in this figure, we see that the LHC has had a pretty significant effect on the expectation of $M_{\rm{unif}}$. We clearly see that scales of order $10^8 - 10^{15}$ GeV have become more favored, whereas there is a clear depletion of probability near the weak scale. 

Fig.~\ref{GauginoUnifComb} shows the final product probability distribution for $g_1^2(M_{\rm unif})$ of the two distributions in Fig.~\ref{GauginoUnif}, demanding that both expression in Eq.~\ref{gauginounif} agree. The result, not surprisingly, leads to a current preference towards small values of $g_1^2(M_{\rm unif})$ or values of $g_1^2(M_{\rm unif}) \simgt 0.4$.

\section{General Gauge Mediation}

Gauge mediated SUSY-breaking is generically defined as a model in which supersymmetry breaking is transmitted to the observable sector via gauge interactions, leading therefore to flavor-blind parameters~\cite{Giudice:1998bp}--\cite{McGarrie:2010kh}. In Ref.~\cite{Meade:2008wd} General Gauge Mediation~(GGM) was defined as any theory in which all SUSY-breaking effects decouple from the MSSM in the limit of vanishing MSSM gauge couplings.

The soft sfermion masses in GGM can be parameterized by a set of three parameters, $A_i$,
\begin{equation}\label{softfm}
m_{\tilde{f}}^2=\sum_{i=1}^3g_i^4C_i(f)A_i\;,
\end{equation}
where the sum runs over the gauge groups of the MSSM. Here $C_i(f)$  is the quadratic Casimir of the superfield $f$ under the gauge group  $i$, which, for a fundamental representation of $SU(N)$ takes the value $C_i(f) = (N^2 - 1)/2N$, while for $U(1)$, $C_1(f) = Y^2/4$ . Observe that we are implicitly working with a normalization of the gauge couplings consistent with their unification at the GUT scale, so $g_1^2 = 5 g_1^2(SM)/3$, and $Y^2/4 = 3/5 \times (Q -T_3)^2$.

The gaugino masses are expressed in terms of three more parameters, $B_i$, given by
\begin{equation}\label{softMm}
M_i=g_i^2 B_i\;.
\end{equation}

In order to generate a Higgsino mass parameter, $\mu$, and soft term, $B_{\mu}$, of the correct order, gauge mediation may need to be supplemented by additional SUSY-breaking contributions in the Higgs sector. Therefore, we assume that in the case of the soft Higgs masses, the expression given in Eq.~(\ref{softfm}) may be modified,
\begin{align}
m_{H_u}^2&=m_{\tilde{L}_3}^2+\delta_u\;,\nonumber\\
m_{H_d}^2&=m_{\tilde{L}_3}^2+\delta_d\;.
\label{deltas}
\end{align}.

Due to flavor independence, $D_{B_{13}}$ and $D_{L_{13}}$ vanish in GGM. Moreover, the RGI $D_{\chi_1}$ also vanishes, as can be easily checked using its definition in Table~\ref{table.Inv}. The invariant $D_Z$ presents a simple dependence on the mass parameters and provides information on $\delta_d$. Therefore, the probability distribution for $\delta_d$ in GGM can be read directly from the one of $D_Z$ presented before in Fig.~\ref{Invariants2}. Of the other RGIs, there are six that probe the high-scale mass parameters of pure GGM, namely the $I_{B_i}$s and the $I_{M_i}$s. We shall mostly concentrate on those invariants in this section. 

As mentioned in Section 2.3, we exclude the other two invariants, $I_{Y_{\alpha}}$ and $D_{Y_{13H}}$ as they depend on too many parameters and currently it is difficult to obtain meaningful information from them. Observe, that, eventually, the invariants $D_{Y_{13H}}$ and $I_{Y\alpha}$ can be used to determine the gauge couplings at the high scale and also probe possible non-universal corrections to the Higgs soft masses~\cite{Carena:2010gr}.

As emphasize above, we shall concentrate on  the RGIs with explicit dependence on the gaugino mass parameters to extract information about the $A_i$ and $B_i$. From the $I_{B_i}$ we immediately obtain
\begin{equation}
B_i=I_{B_i}\;,
\label{Br}
\end{equation}
and these distributions can be seen in Fig.~\ref{Invariants1}.

In order to obtain information on the $A_i$, both $I_{B_i}$ and $I_{M_i}$ must be used:
\begin{align}
A_1&=\;\frac{10}{33}\left(\frac{I_{M_1}}{g_1^4}-I_{B_1}^2\right)\;,\nonumber\\
A_2&=\quad2\left(\frac{I_{M_2}}{g_2^4}-I_{B_2}^2\right)\;,\nonumber\\
A_3&=-\frac{2}{3}\left(\frac{I_{M_3}}{g_3^4}-I_{B_3}^2\right)\;,
\label{Ar}
\end{align}
where the $g_i$ are the gauge couplings at the messenger scale, $M$.

\begin{figure}
\begin{center}
\begin{tabular}{c c c}
\includegraphics[width=0.34\textwidth]{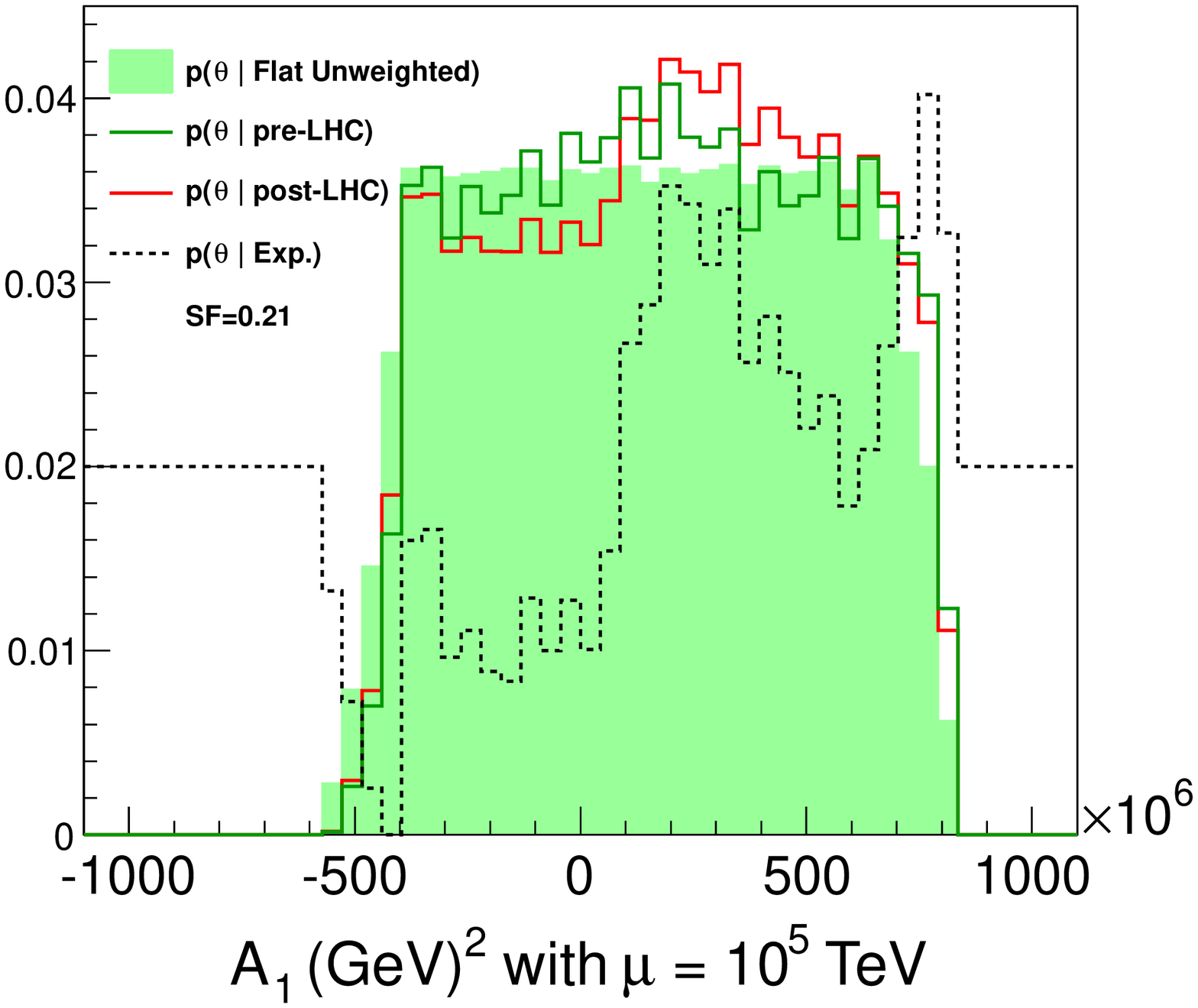}  &
\includegraphics[width=0.34\textwidth]{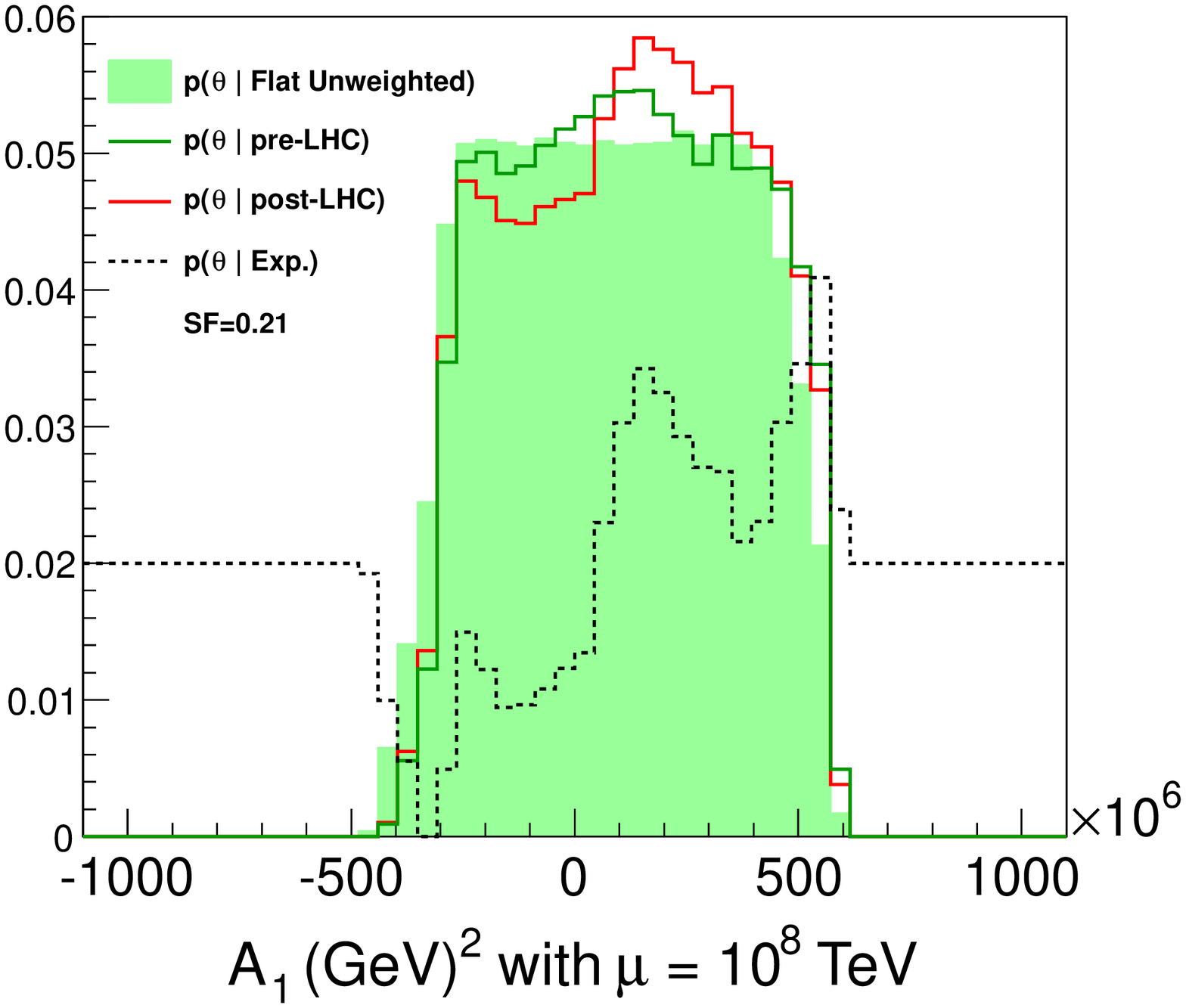}  &
\includegraphics[width=0.34\textwidth]{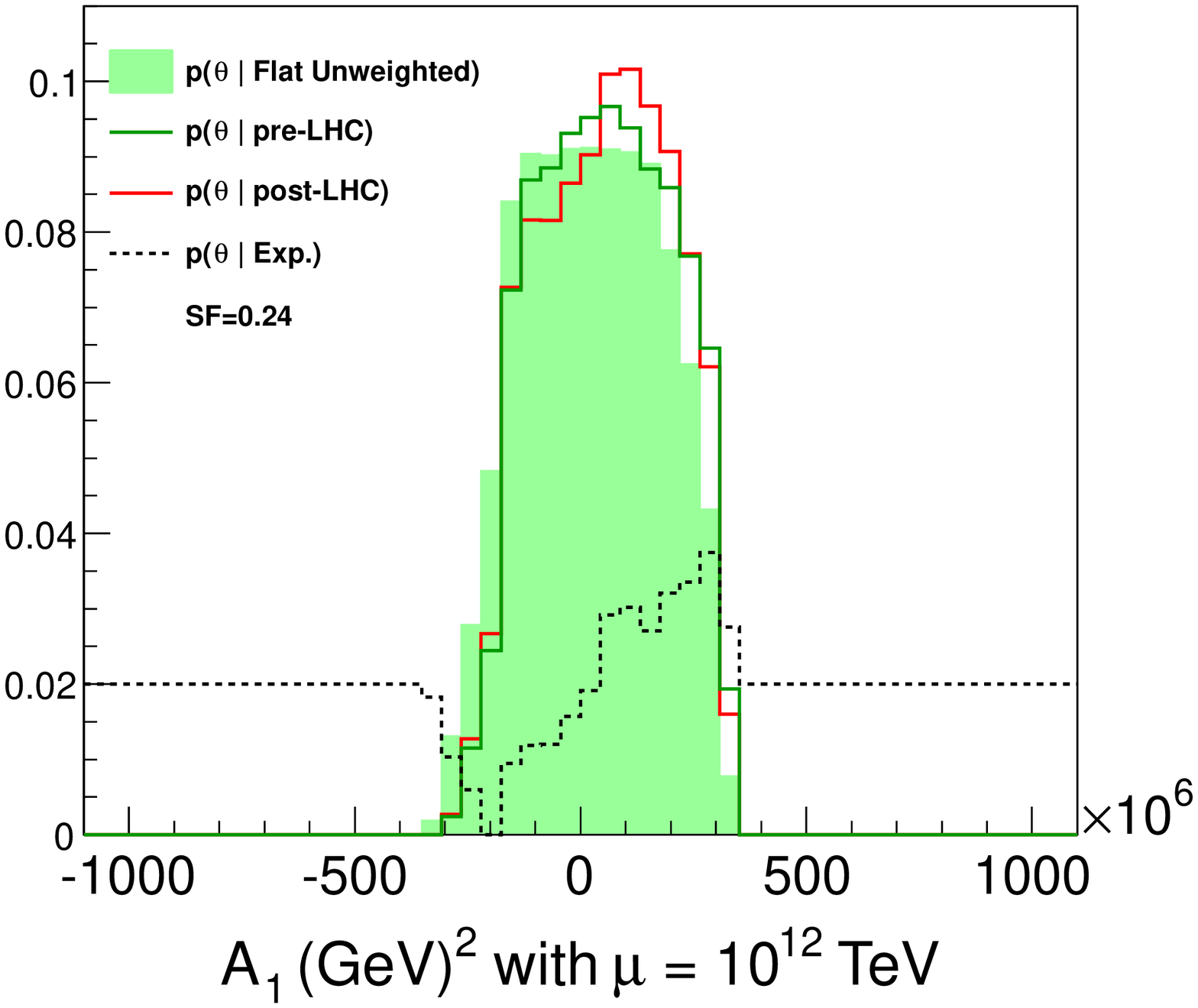}  \\
\includegraphics[width=0.34\textwidth]{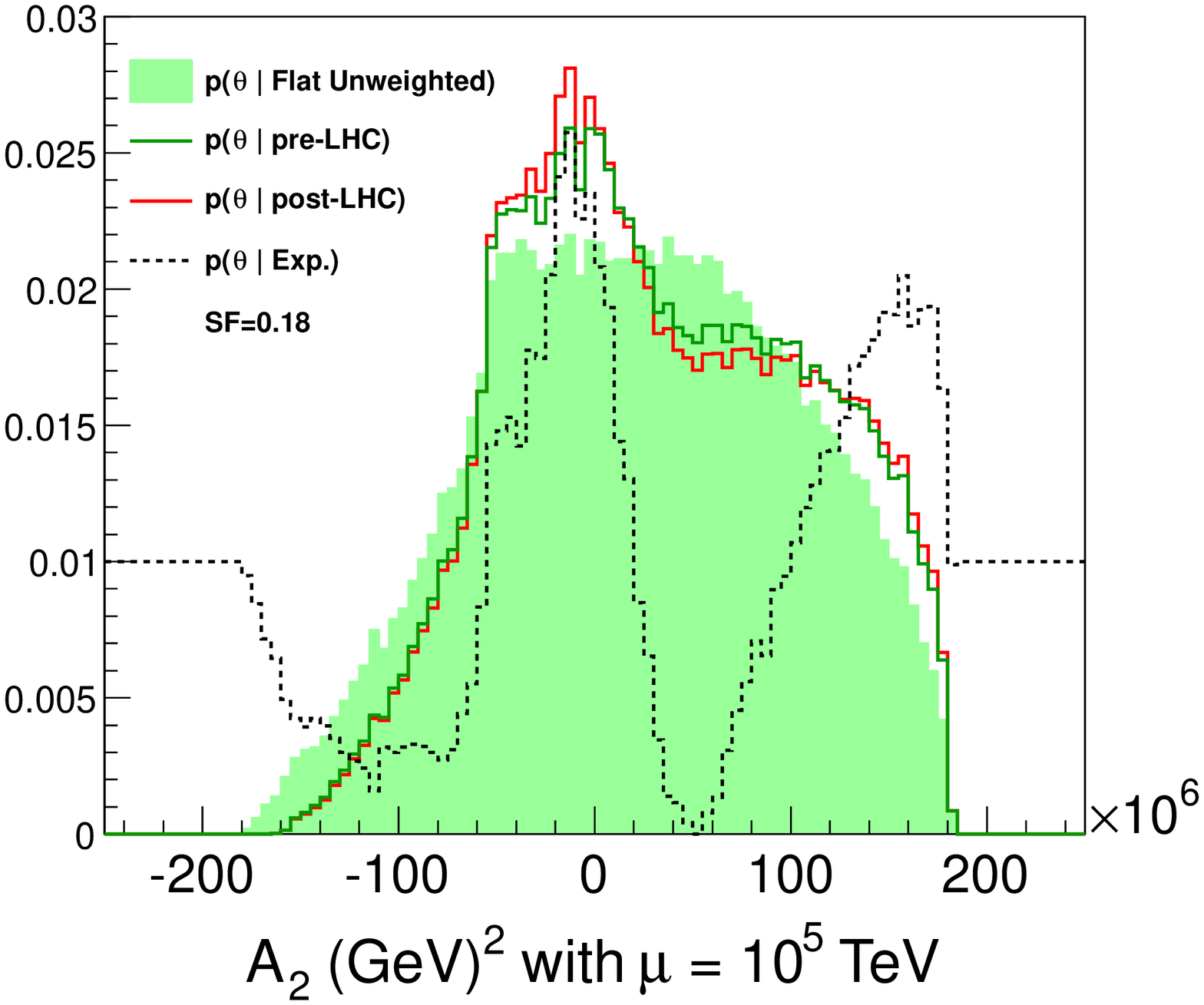}  &
\includegraphics[width=0.34\textwidth]{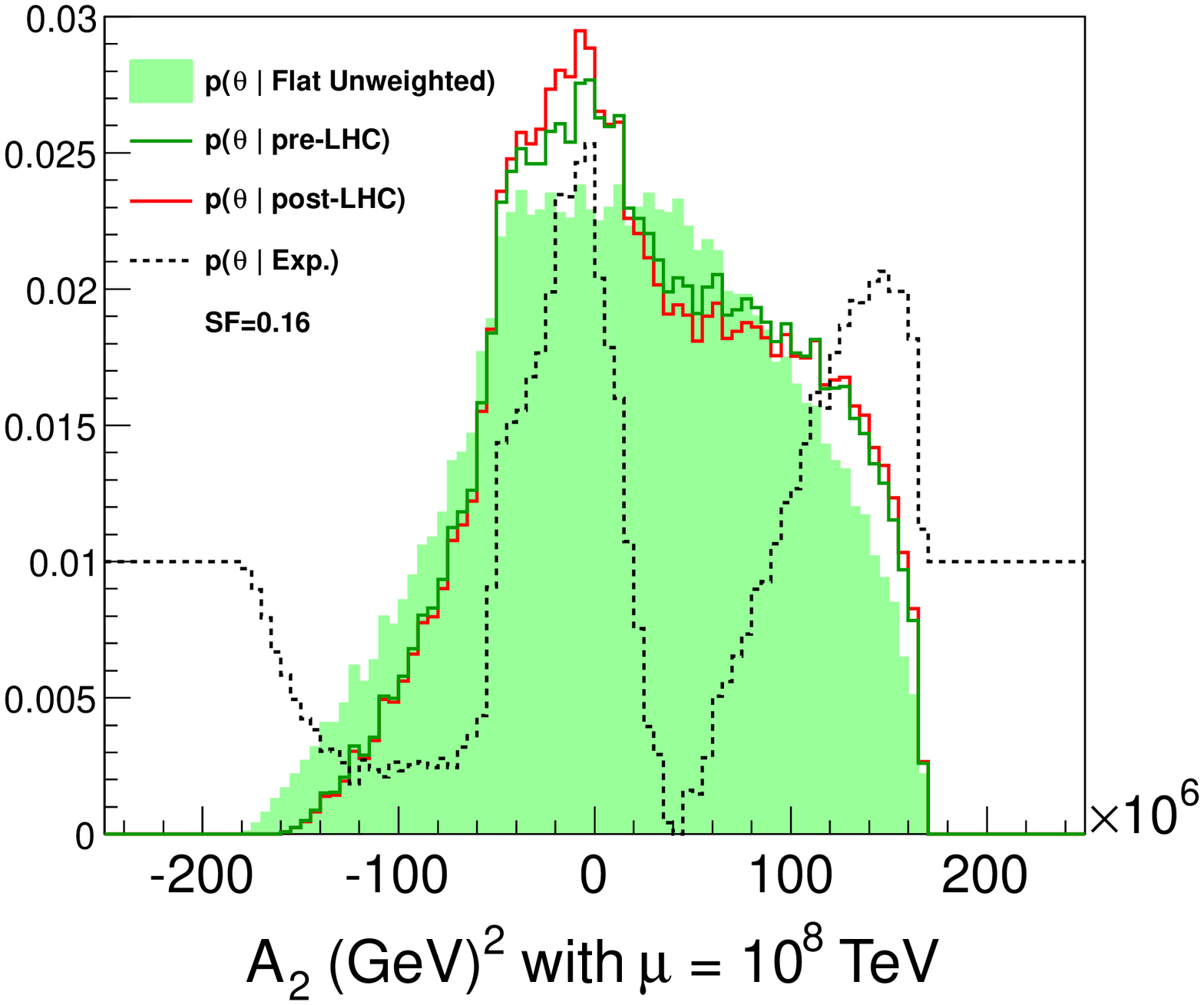}  &
\includegraphics[width=0.34\textwidth]{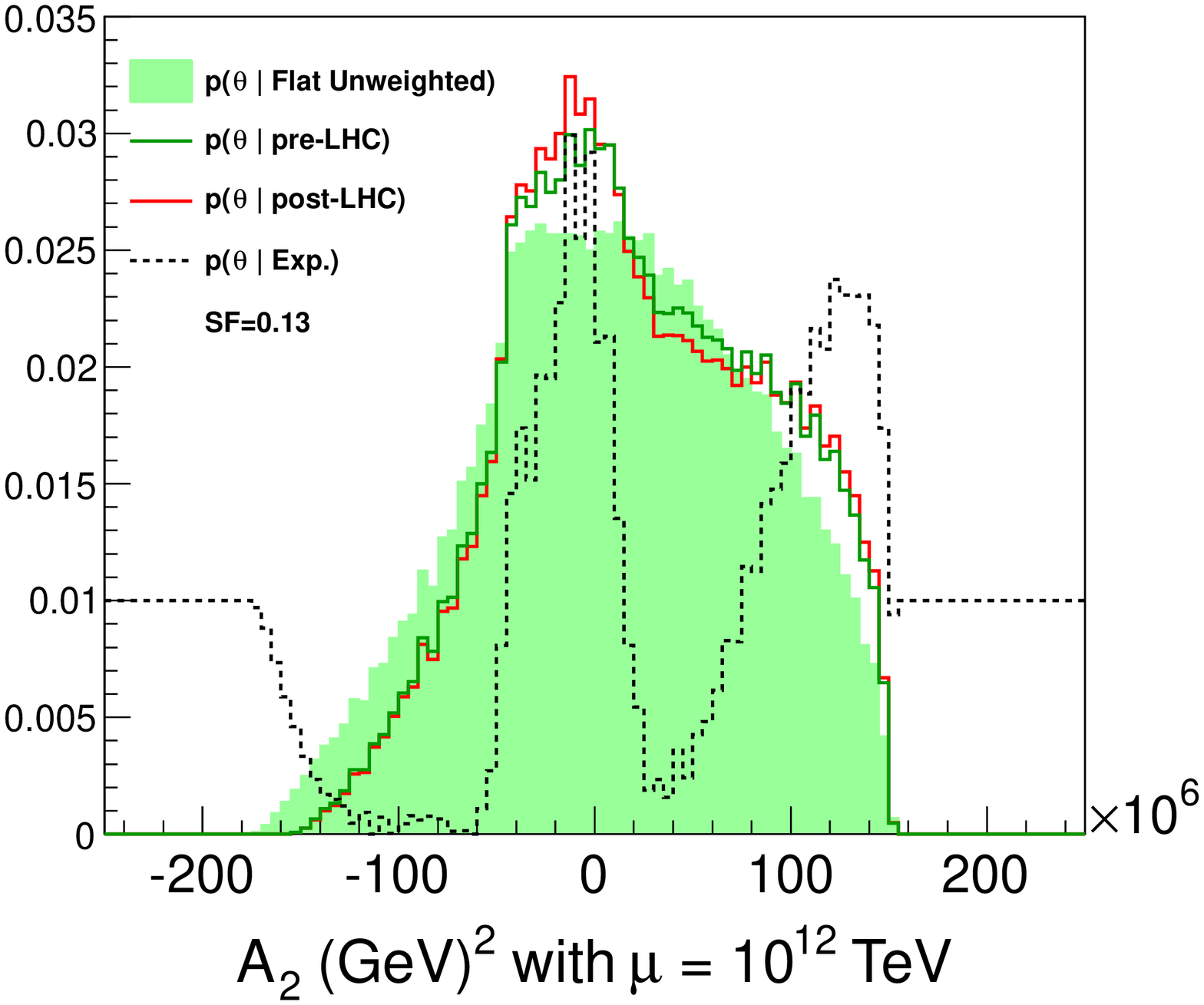}  \\
\includegraphics[width=0.34\textwidth]{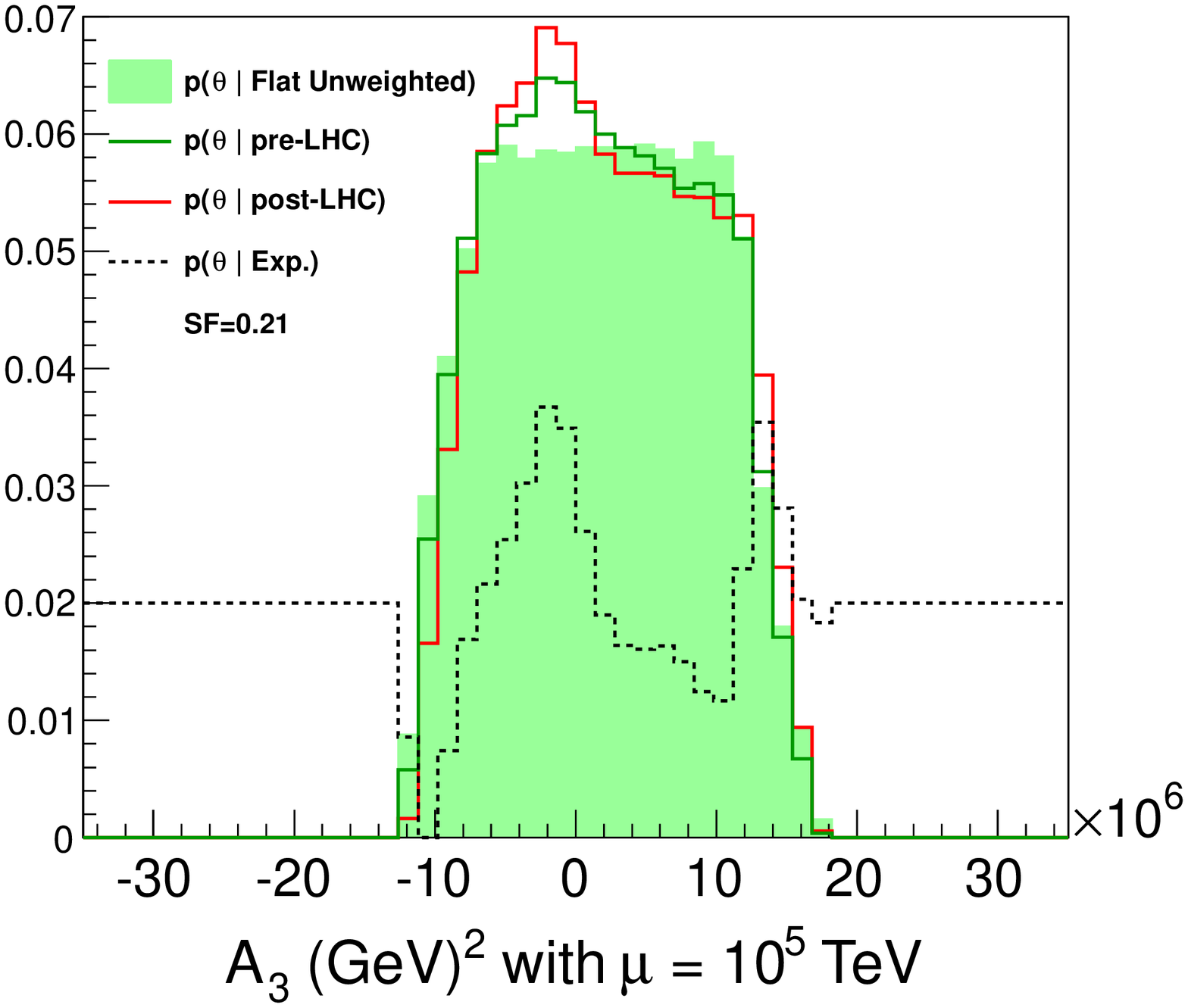}  &
\includegraphics[width=0.34\textwidth]{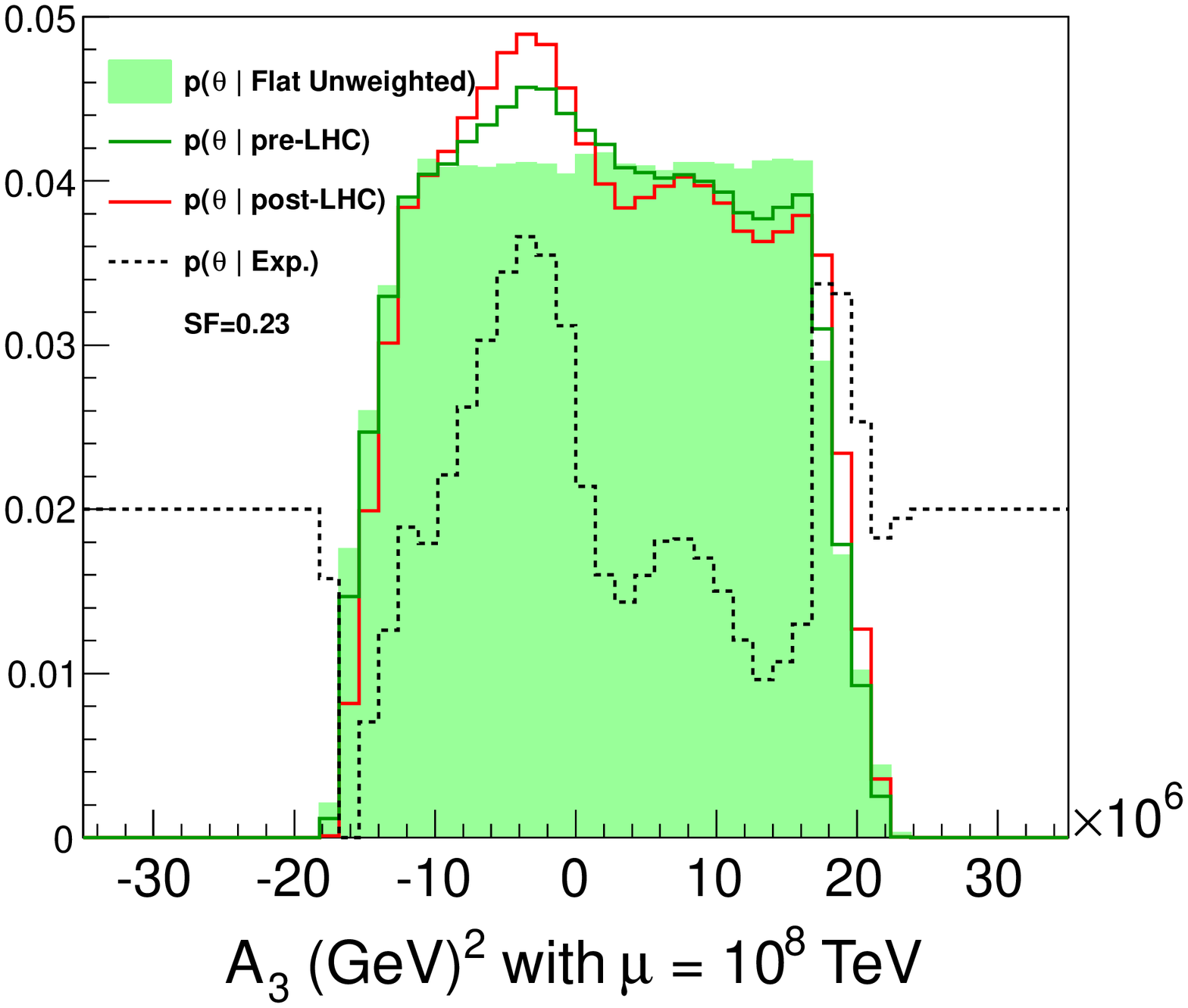}  &
\includegraphics[width=0.34\textwidth]{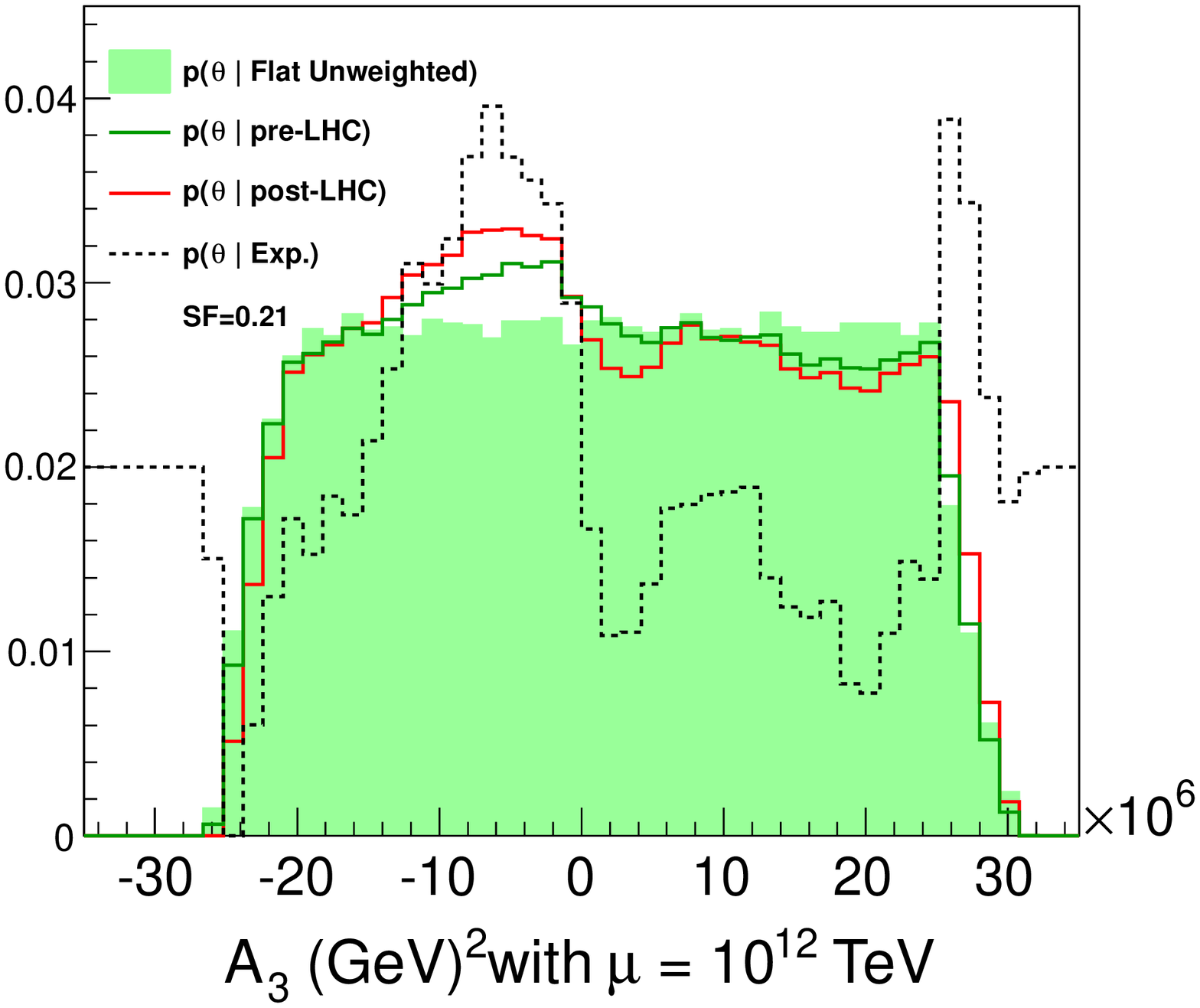}  \\
\end{tabular}
\end{center}
\caption{Distribution of the GGM parameters $A_i$ before and after the LHC constraints are added (green and red lines), flat distribution (shaded green) and subtracted probability distribution (dashed black line). The three set of values are associated with different values of the messenger scale, $M=10^5, 10^8$ and $M=10^{12}$ TeV. For the first and last row, $\rm no.\; bins =50$ giving $p(\theta|Flat)=0.02$.}
\label{GGMParameters}
\end{figure}

The probability distributions for the $A_i$ for three different values of the messenger scale are given in Fig.~\ref{GGMParameters}. The shaded green area, as well as the green, red and black lines have the same interpretation as in Figs.~\ref{Invariants1} and \ref{Invariants2}.

From Fig.~\ref{GGMParameters}, it is clear that  positive and sizable values of $A_1$ are preferred, independent of the messenger scale. On the other hand, values of $A_2$ close to zero are somewhat favored, although sizable values are equally likely. Finally, $A_3$ can be small and negative, but positive and sizable values are equally or more likely than the negative ones. One can check, that if the values of the $A_3$ are the ones associated with the regions of maximal likelihood, which correspond to negative values of this parameter, and assume $A_2$ to be small, the boundary condition for the left-handed squarks square mass parameters would be negative, unless the largest values of $A_1$ at each messenger scale are selected. These boundary conditions, together with the ones for the $B_i$, would then lead to somewhat light left-handed sfermions compared to the right-handed ones.

\section{Minimal Gauge Mediation}

Minimal Gauge Mediation~(MGM) is a particular gauge mediated model in which the soft supersymmetry-breaking parameters are obtained through the interaction of messenger particles that transform under the $5$ + $\bar{5}$ representation of $SU(5)$. The general assumption is that these messengers acquire mass via their interaction with a singlet superfield. When this superfield acquires a vev, it fixes the messenger scale, and its F-term, $F_S$, fixes the scale of soft supersymmetry-breaking parameters through the identity
\begin{equation}
M_i = \frac{g_i^2}{16 \pi^2} \frac{F_S}{S}\;.
\end{equation}
Hence in minimal gauge mediated models
\begin{equation}
B_i = B=\frac{F_S}{ 16 \pi^2 S}\;,
\end{equation}
with $i = {1,2,3}$. Moreover, the scalar masses at the messenger scale are obtained at the two-loop level and acquire the value
\begin{equation}\label{softfmMGM}
m_{\tilde{f}}^2=\sum_{i=1}^3 2 \frac{g_i^4(M)}{(16\pi^2)^2} C_i(f) \frac{F_S^2}{S^2}\;.
\end{equation}
Hence, in MGM
\begin{equation}
A_i = A=\frac{2 F_S^2}{(16 \pi^2 S)^2} = 2 B^2\;.
\end{equation}

Therefore, minimal gauge mediation is a model with just two parameters, $B $ and the messenger scale, $M$ ($g_2$ and $g_3$ can be written in terms of $g_1$ through the RGIs $I_{g_2}$ and $I_{g_3}$, which is then just a function of $M$).

In order to determine the probability distribution for the parameters $B$ and $g_1^2(M)$, we use the ones of the $I_{M_i}$s and the $I_{B_i}$s. From Table~\ref{table.Inv}, we see that the $I_{M_i}$ must fulfill the following relations
\begin{eqnarray}
I_{M_1} - \frac{38 g_1^4(M) B^2}{5} = 0,
\nonumber\\
I_{M_2}  - 2 g_2^4(M) B^2 = 0,
\nonumber \\
I_{M_3} + 2 g_3^4(M) B^2 = 0,
\label{IMiB}
\end{eqnarray}
which together with the equations $I_{B_i} = B$ define a system of 6 equations with only 2 unknowns. For every value of $I_{B_i}$ one can obtain a value of $B$ that leads, from the 3 equations in Eq.~(\ref{IMiB}), to 9 independent values of $g_1^2(M)$. In addition, these 3 equation lead to 3 different sets of simultaneous equations that can be solved for $B$ and $g_1^2(M)$ independently. This leads to another 6 solutions  for $B$ and 3 for $g_1^2(M)$, leading to a total of 9 solutions for $B$ and 12 for $g_1^2(M)$.

The set of 9 solutions that we use to compute the probability distributions for $B$ are:

\begin{eqnarray}
\label{MGM_Bis}
B_1&=&I_{B_1},\\
B_2&=&I_{B_2},\\
B_3&=&I_{B_3},\\
B_{4,5}&=&\mp\frac{5 I_{g_2} \left(33 I_{M_1} \sqrt{I_{M_2}}+\sqrt{95} \sqrt{I_{M_1}} I_{M_2}\right)}{\sqrt{2} (1089 I_{M_1}-95 I_{M_2})},\\
B_{6,7}&=&\mp\frac{5 I_{g_3} \left(11I_{M_1} \sqrt{-I_{M_3}}+\sqrt{95} \sqrt{I_{M_1}} I_{M_3}\right)}{\sqrt{2} (121 I_{M_1}+95 I_{M_3})},\\
B_{8,9}&=&\mp\frac{5 (I_{g_2}-I_{g_3}) I_{M_2} \sqrt{-I_{M_3}}}{11 \sqrt{2} \left(I_{M_2}+3\sqrt{I_{M_2}} \sqrt{-I_{M_3}}\right)}
\end{eqnarray}

The corresponding set of 12 solutions for the gauge coupling at the messenger scale, $g_1^2(M)$ are instead given by:

\begin{eqnarray}
g_{1_{1}}^2&=&\frac{\left(-33 \sqrt{I_{M_1}}+\sqrt{95} \sqrt{I_{M_2}}\right)}{I_{g_2}\sqrt{95 I_{M_2}}},\\
g_{1_{2}}^2&=&\frac{\left(-11 \sqrt{I_{M_1}}-\sqrt{95} \sqrt{-I_{M_3}}\right)}{I_{g_3}\sqrt{-95  I_{M_3}}},\\
g_{1_{3}}^2&=&\frac{\left(\sqrt{I_{M_2}}+3 \sqrt{-I_{M_3}}\right)}{\left(I_{g_2} \sqrt{I_{M_2}}+ 3I_{g_3} \sqrt{-I_{M_3}}\right)},\\
g_{1_4}^2&=&\sqrt{\frac{5 I_{M_1}}{38 I_{B_1}^2}},\\
g_{1_5}^2&=&\sqrt{\frac{5 I_{M_1}}{38 I_{B_2}^2}},\\
g_{1_6}^2&=&\sqrt{\frac{5 I_{M_1}}{38 I_{B_3}^2}},\\
g_{1_{7}}^2&=&\frac{5 \sqrt{I_{M_2}}}{\left(33 \sqrt{2} I_{B_1}+5 I_{g_2} \sqrt{I_{M_2}}\right)},\\
g_{1_{8}}^2&=&\frac{5 \sqrt{I_{M_2}}}{\left(33 \sqrt{2} I_{B_2}+5 I_{g_2} \sqrt{I_{M_2}}\right)},\\
g_{1_{9}}^2&=&\frac{5 \sqrt{I_{M_2}}}{\left(33 \sqrt{2} I_{B_3}+5 I_{g_2} \sqrt{I_{M_2}}\right)},\\
\end{eqnarray}
\begin{eqnarray}
g_{1_{10}}^2&=&\frac{5 \sqrt{-I_{M_3}}}{\left(11\sqrt{2} I_{B_1}- 5 I_{g_3} \sqrt{-I_{M_3}}\right)},\\
g_{1_{11}}^2&=&\frac{5 \sqrt{-I_{M_3}}}{\left(11 \sqrt{2} I_{B_2}-5 I_{g_3} \sqrt{-I_{M_3}}\right)},\\
g_{1_{12}}^2&=&\frac{5 \sqrt{-I_{M_3}}}{\left(11\sqrt{2} I_{B_3}-5 I_{g_3} \sqrt{-I_{M_3}}\right)}\;.
\label{MGM_gis}
\end{eqnarray}

\begin{figure}
\begin{center}
\begin{tabular}{c }
\includegraphics[width=0.7\textwidth]{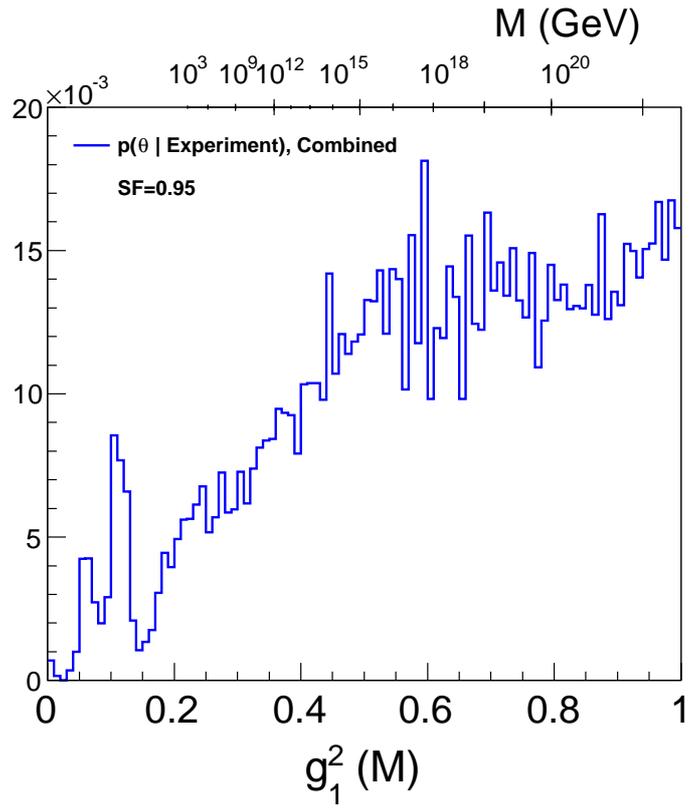}  \\
\includegraphics[width=0.7\textwidth]{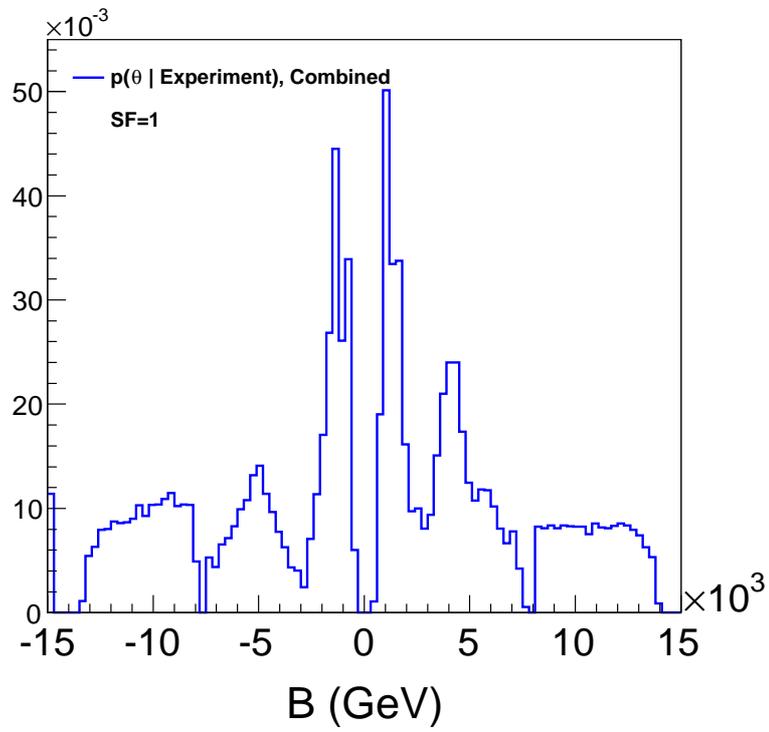}  \\
\end{tabular}
\end{center}
\caption{Final product distribution of the MGM parameters $B$ and $g_1^2(M)$.}
\label{MGMpardist}
\end{figure}

\begin{figure}
\begin{center}
\begin{tabular}{c c c}
\includegraphics[width=0.34\textwidth]{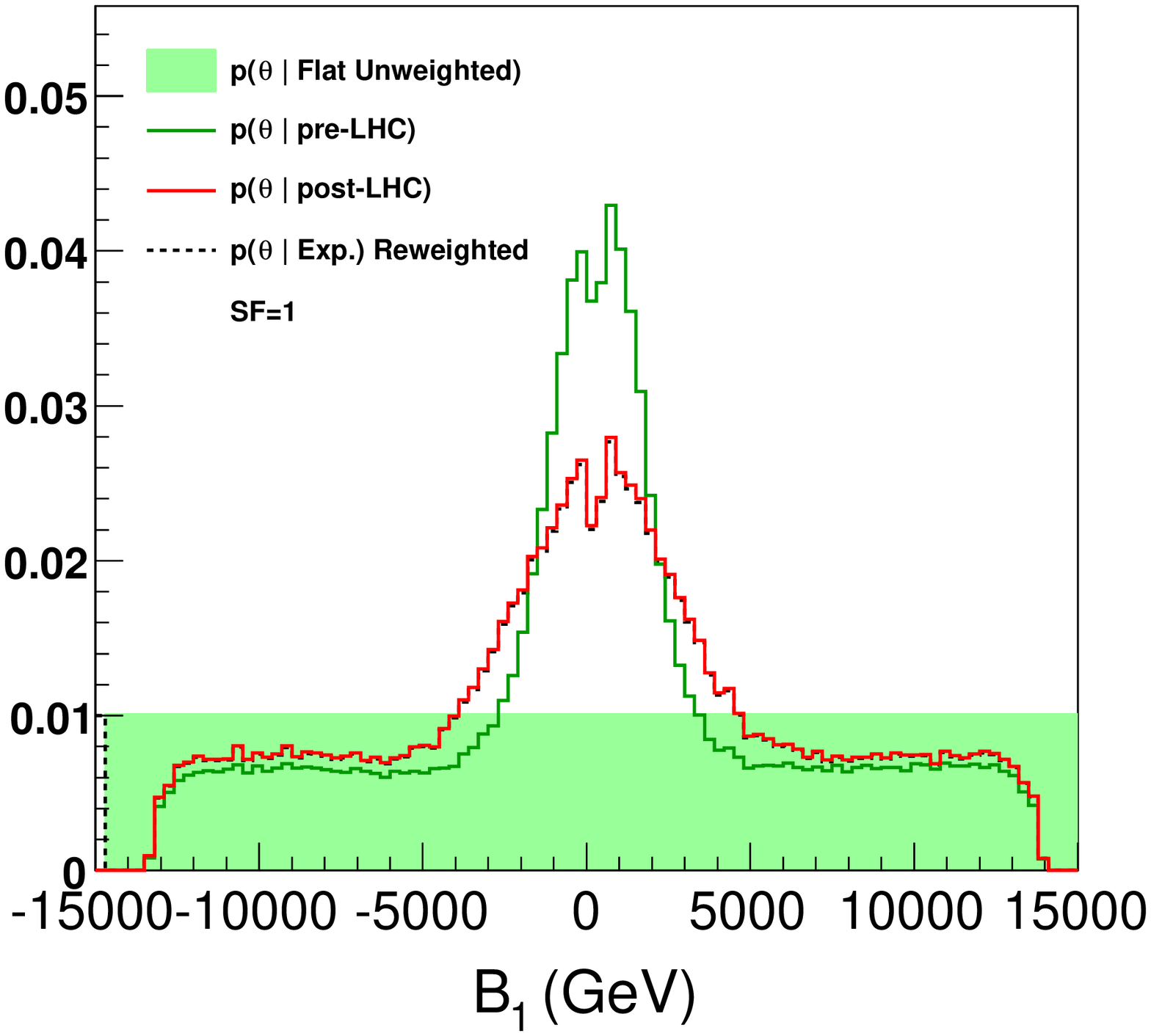}  &
\includegraphics[width=0.34\textwidth]{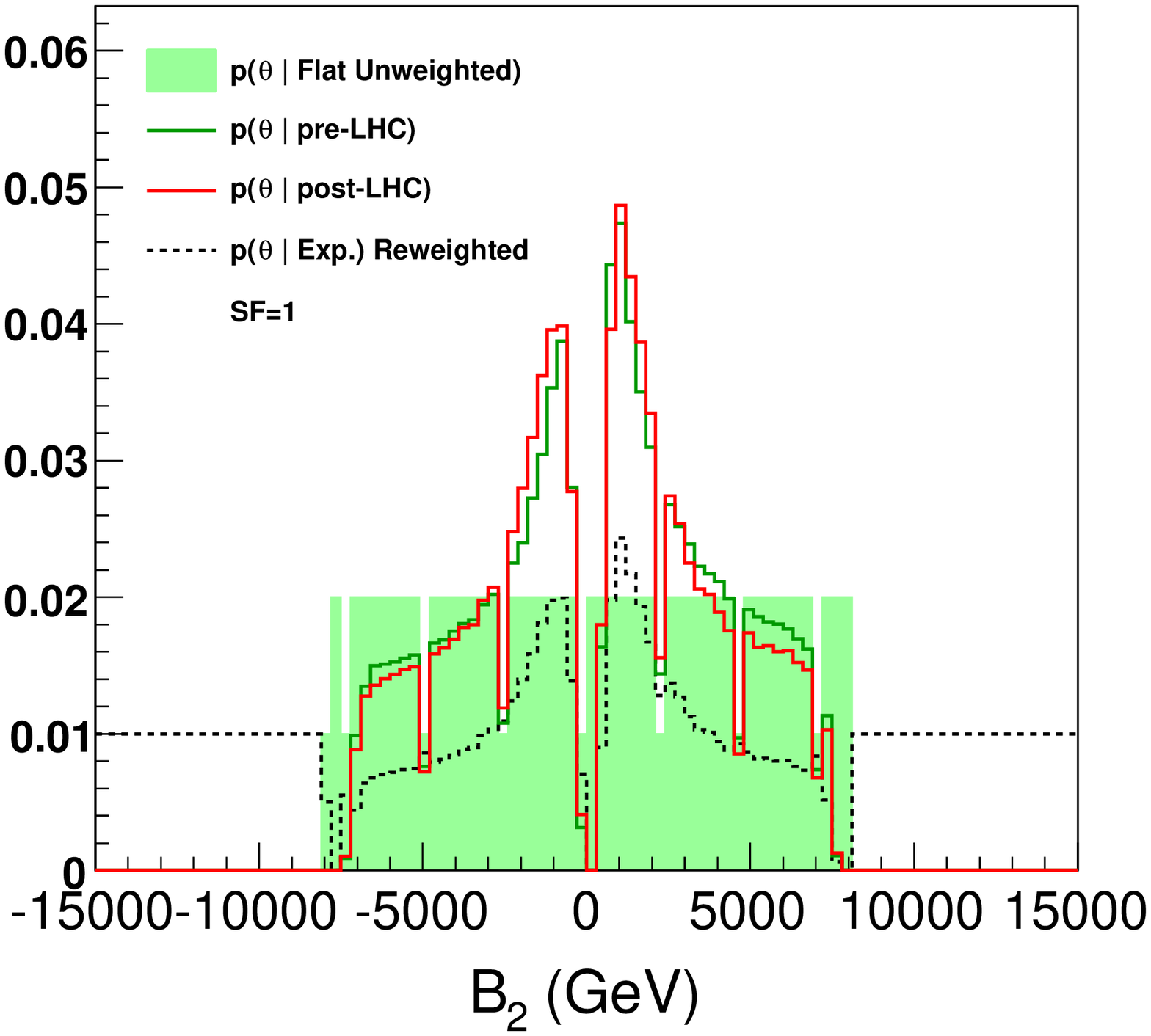}  &
\includegraphics[width=0.34\textwidth]{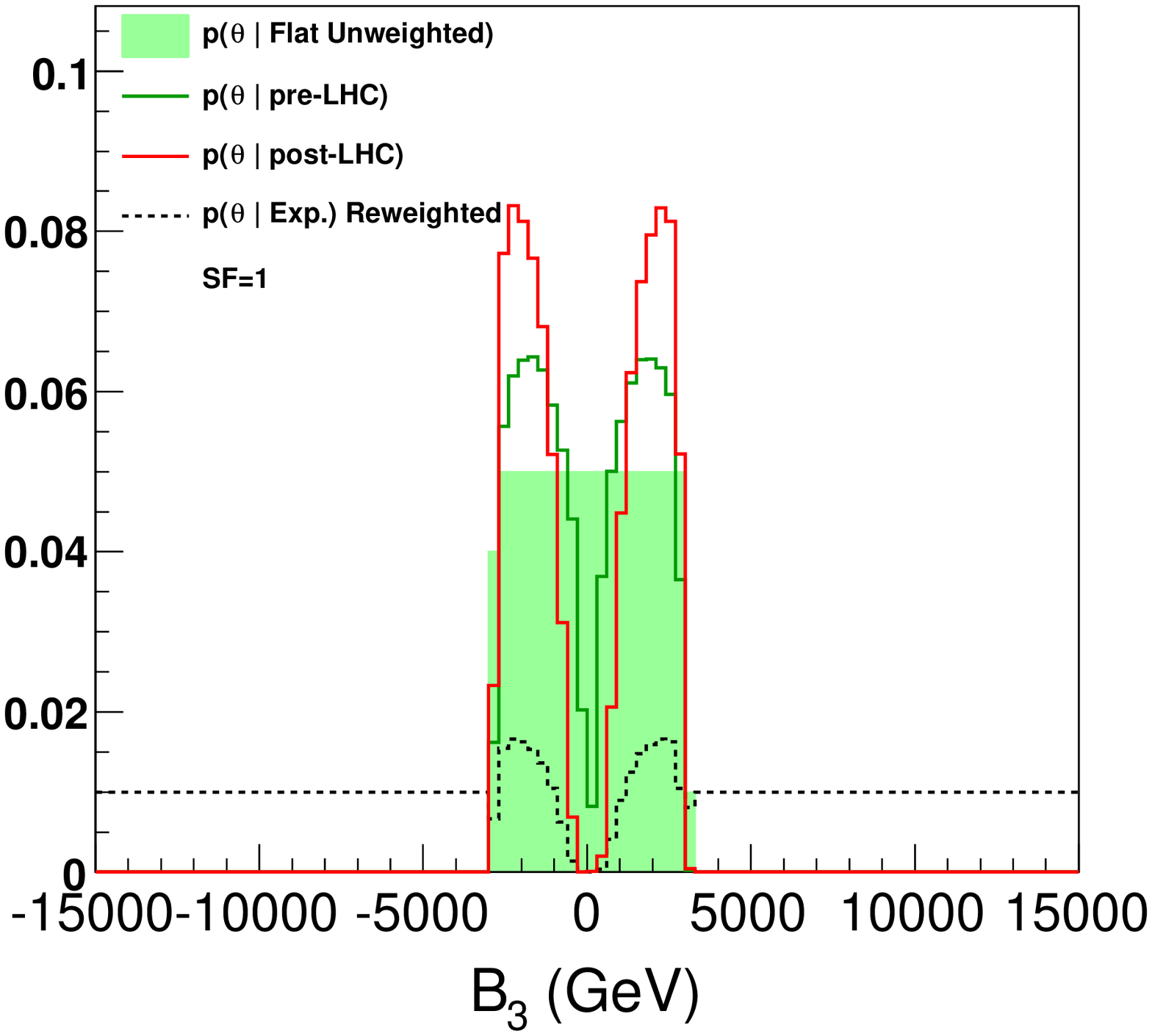}  \\
\includegraphics[width=0.34\textwidth]{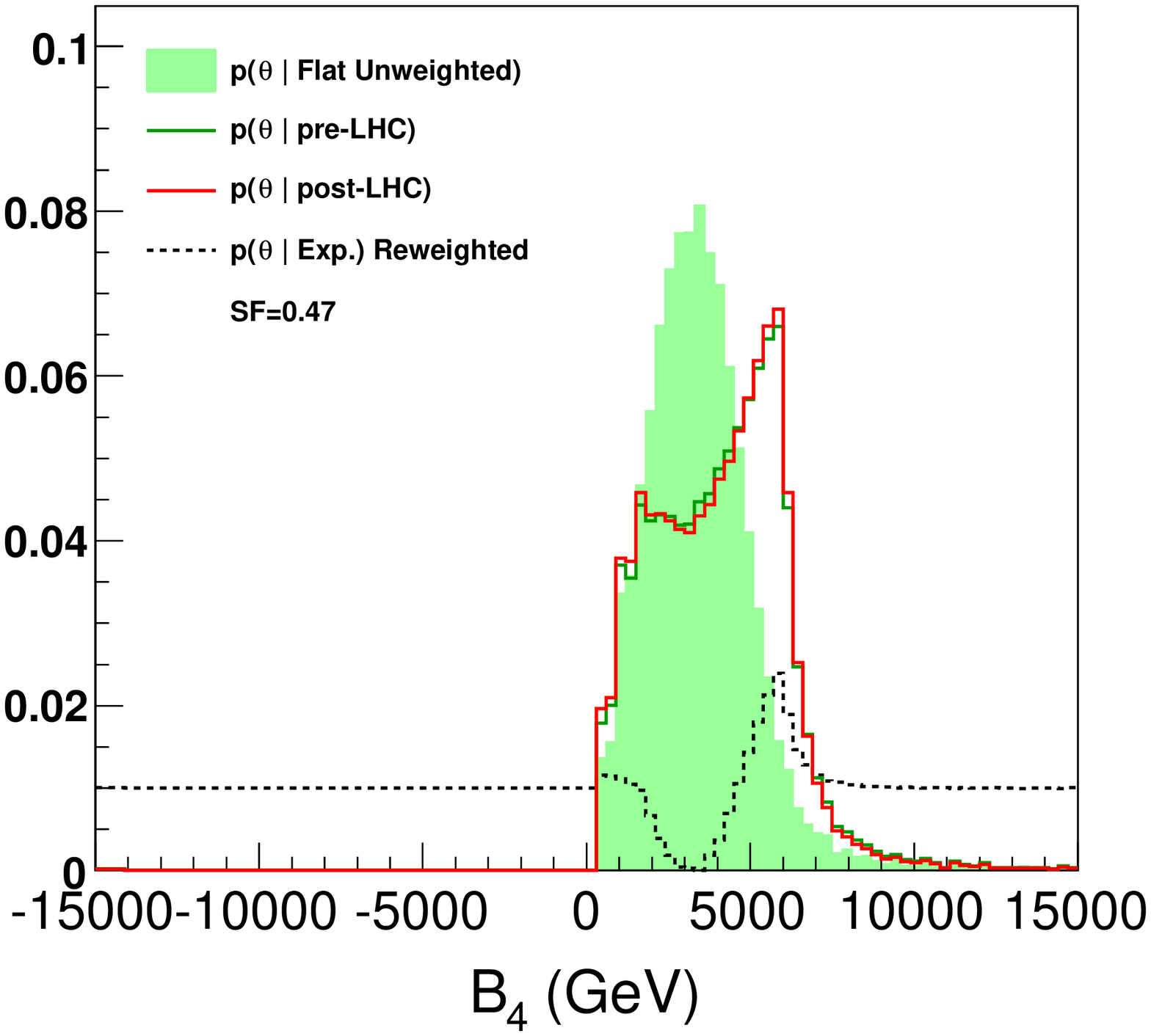}  &
\includegraphics[width=0.34\textwidth]{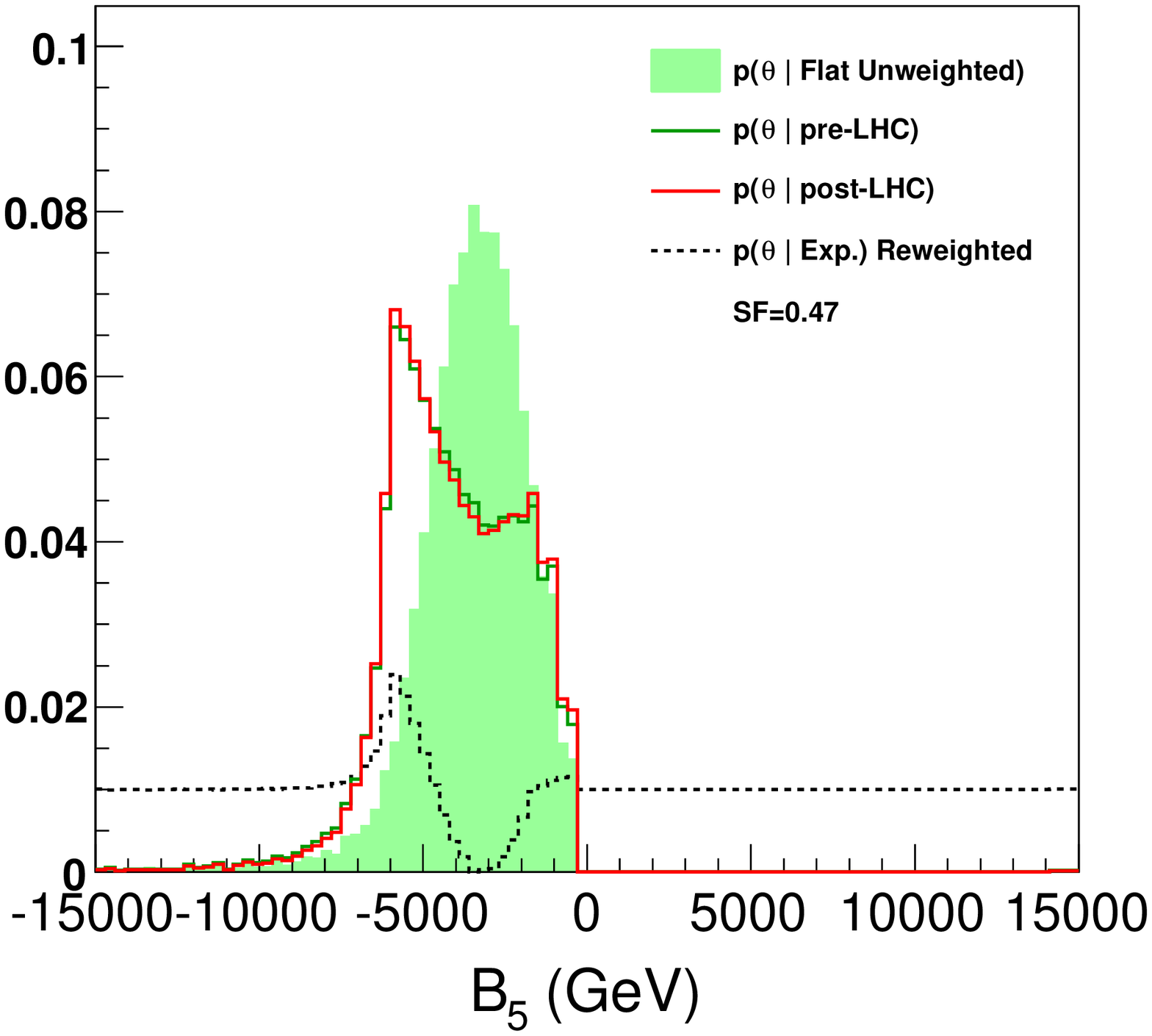}  &
\includegraphics[width=0.34\textwidth]{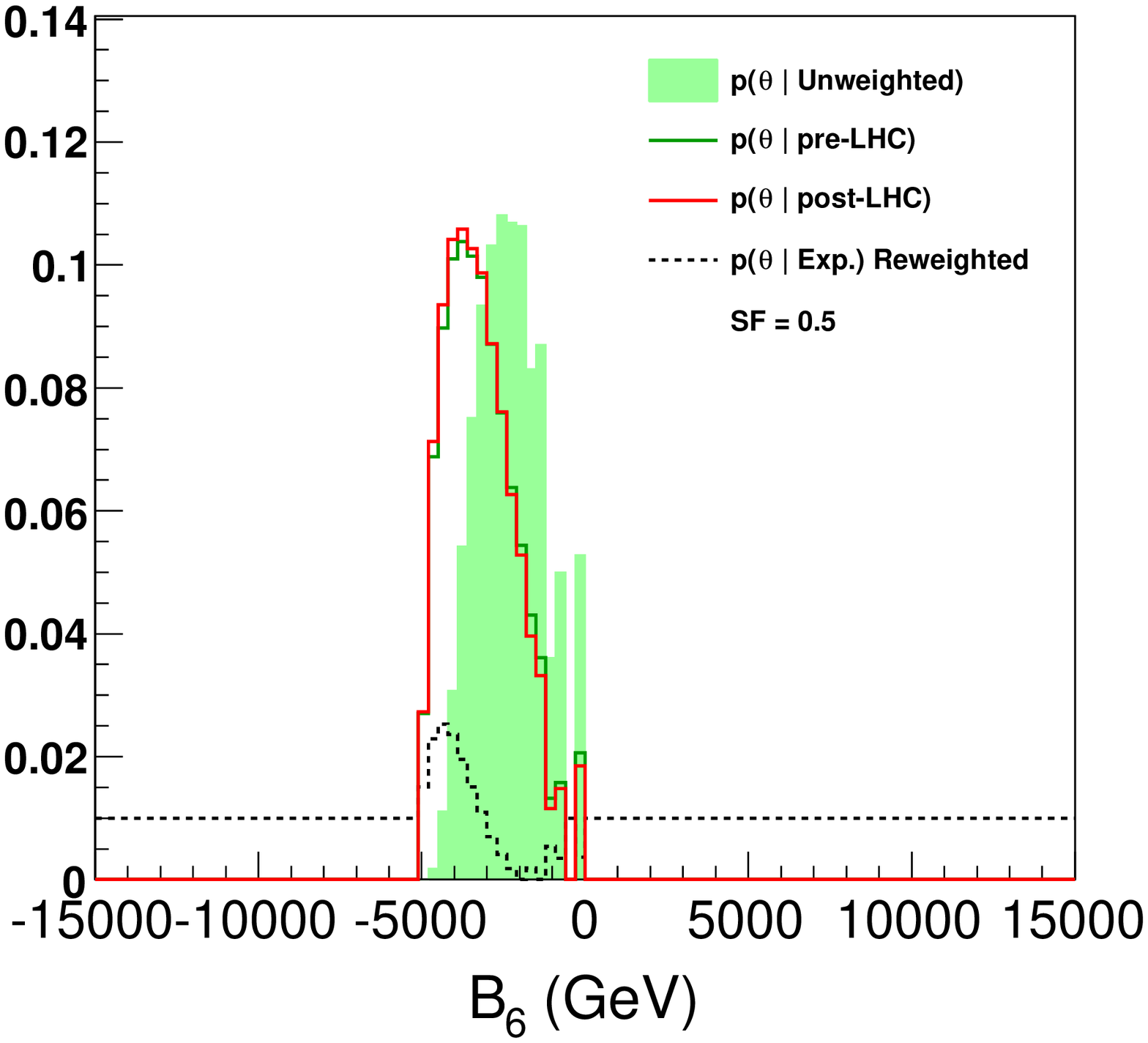}  \\
\includegraphics[width=0.34\textwidth]{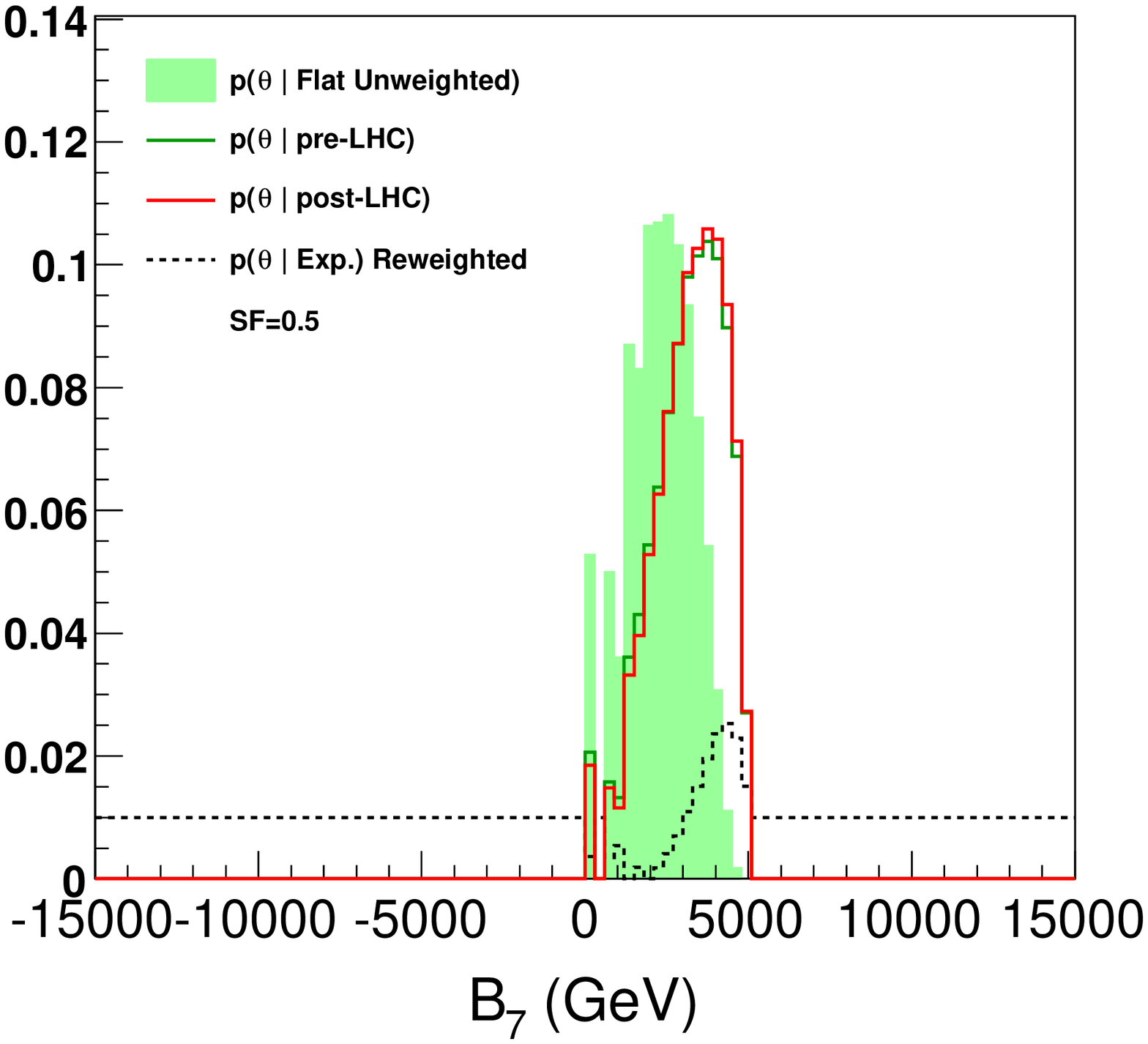}  &
\includegraphics[width=0.34\textwidth]{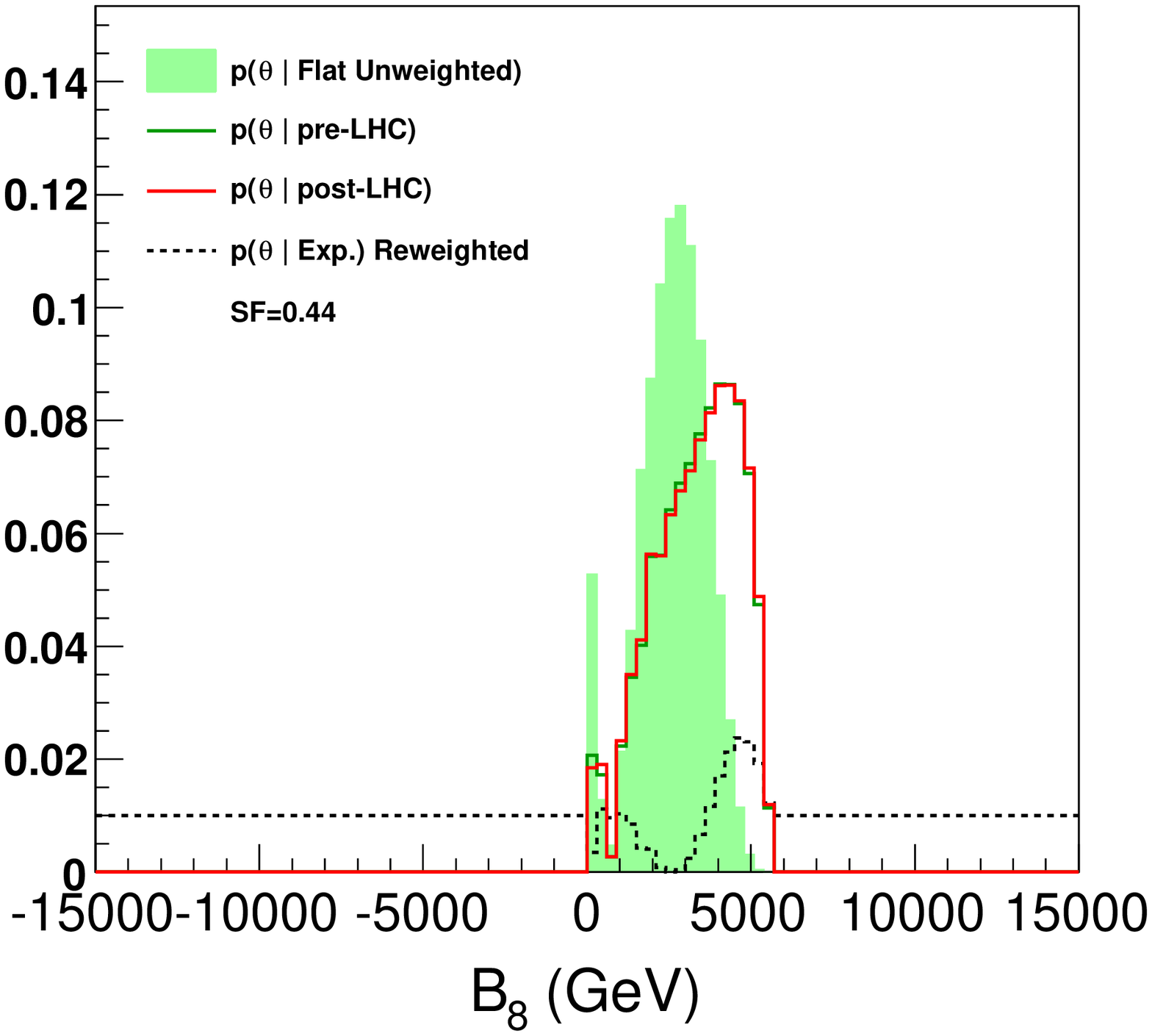}  &
\includegraphics[width=0.34\textwidth]{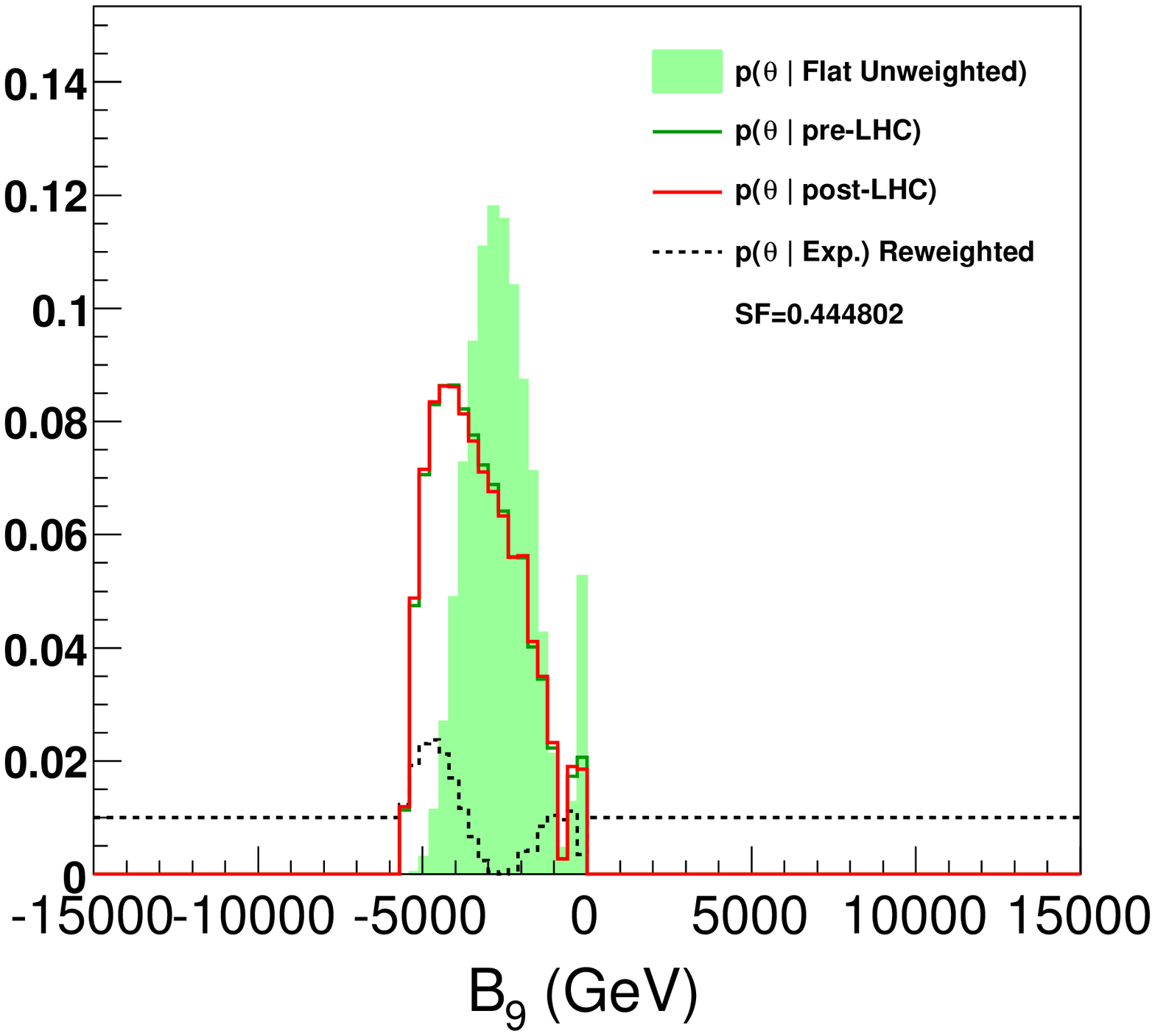}  \\
\end{tabular}
\end{center}
\caption{Distribution for the MGM parameter $B$ before and after the LHC constraints are added (green and red lines), flat distribution (shaded green) and subtracted probability distribution (dashed black line). The different sets are associated with different probability distributions listed in Eqs.~\ref{MGM_Bis}.}
\label{MGMParameters1}
\end{figure}

\begin{figure}
\begin{center}
\begin{tabular}{c c c}
\includegraphics[width=0.34\textwidth]{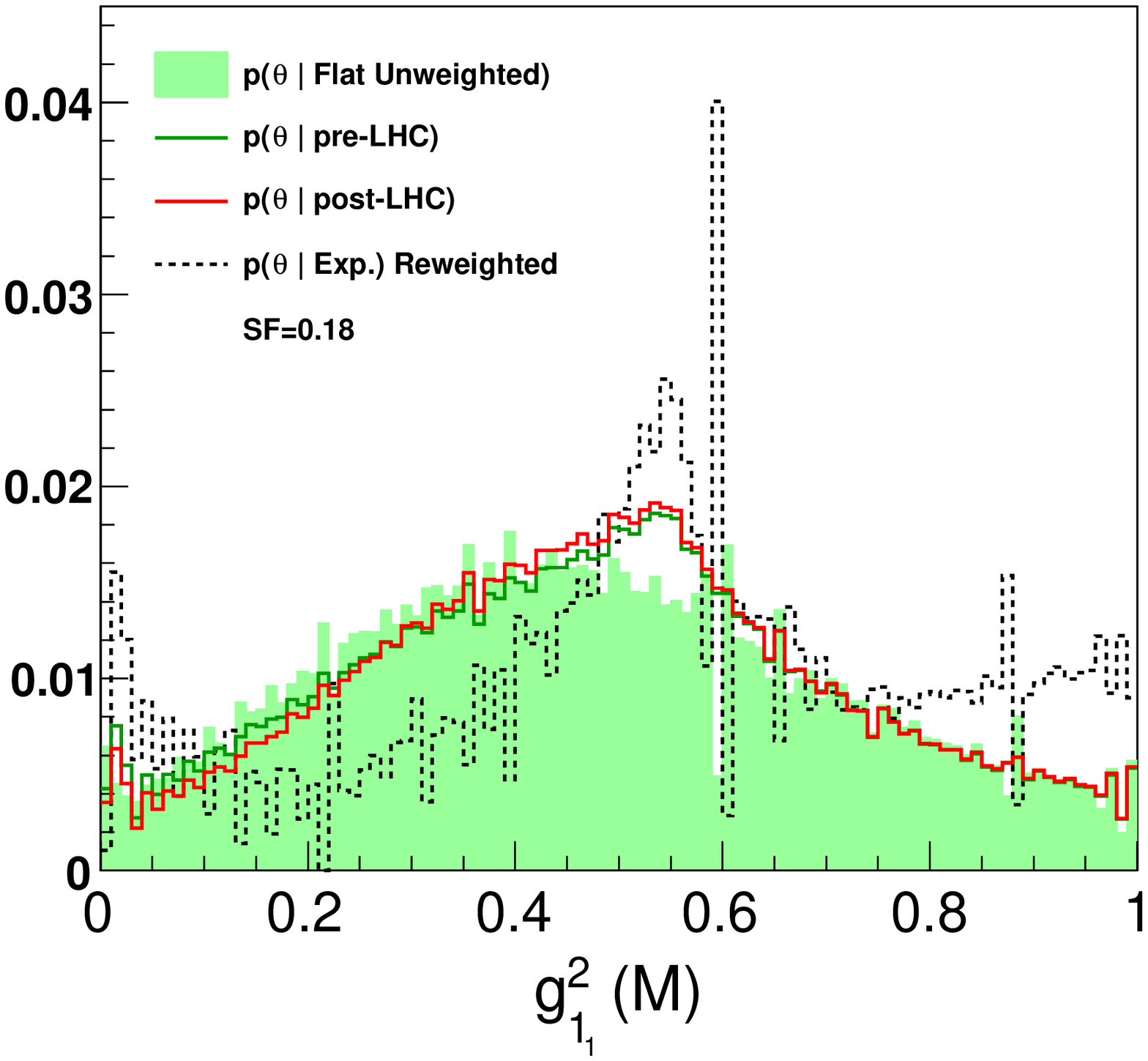}  &
\includegraphics[width=0.34\textwidth]{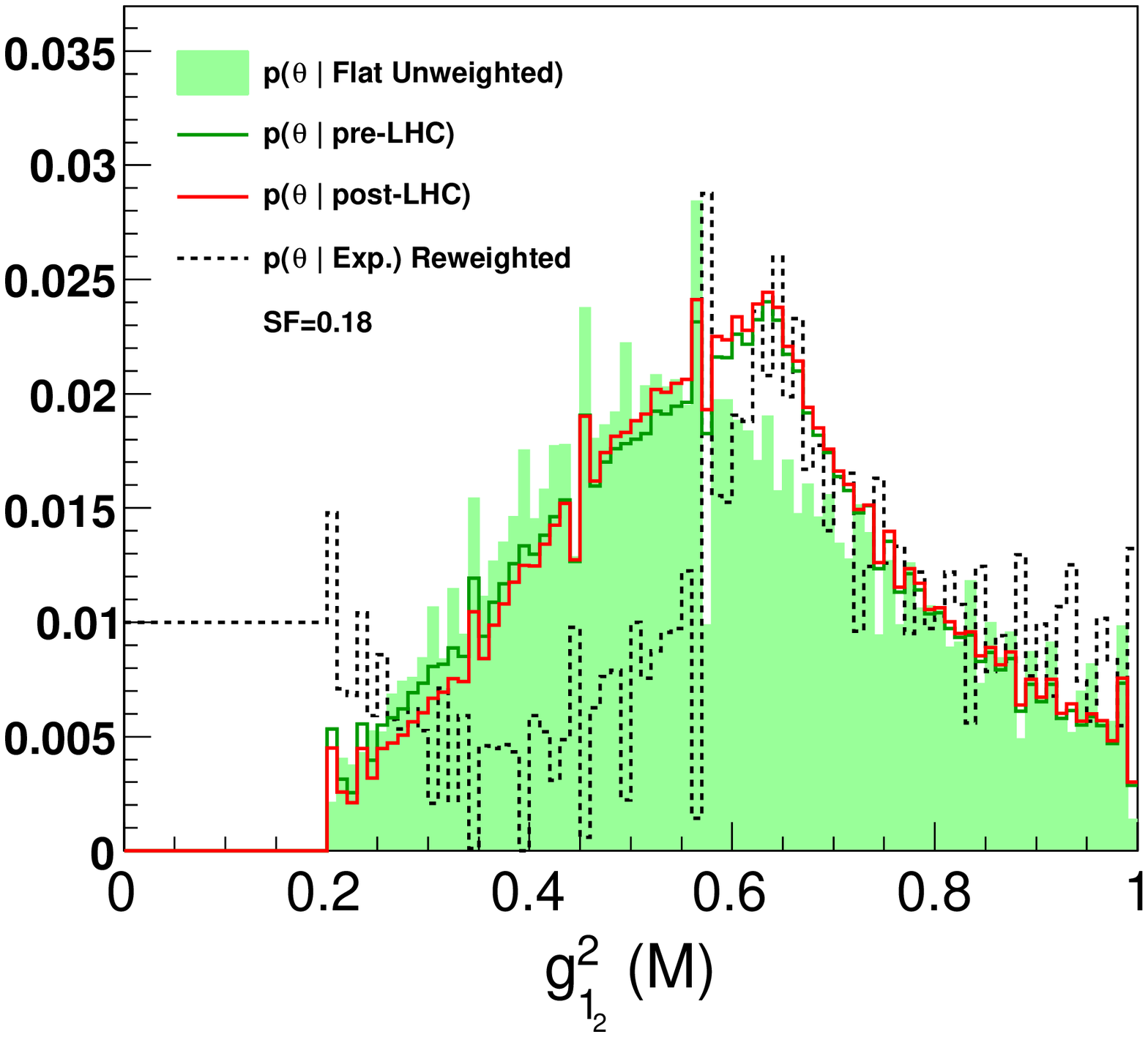}  &
\includegraphics[width=0.34\textwidth]{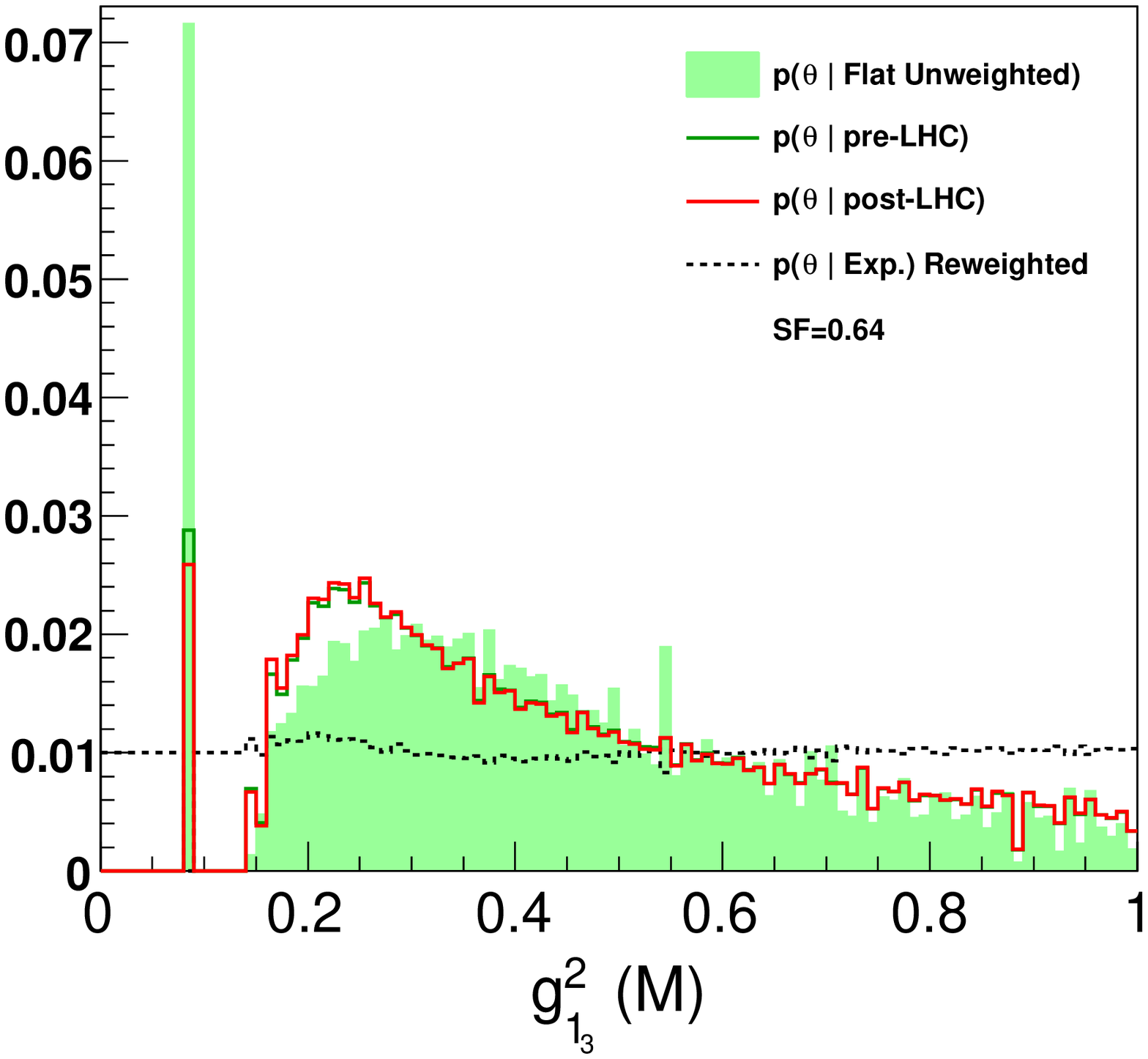}  \\
\includegraphics[width=0.34\textwidth]{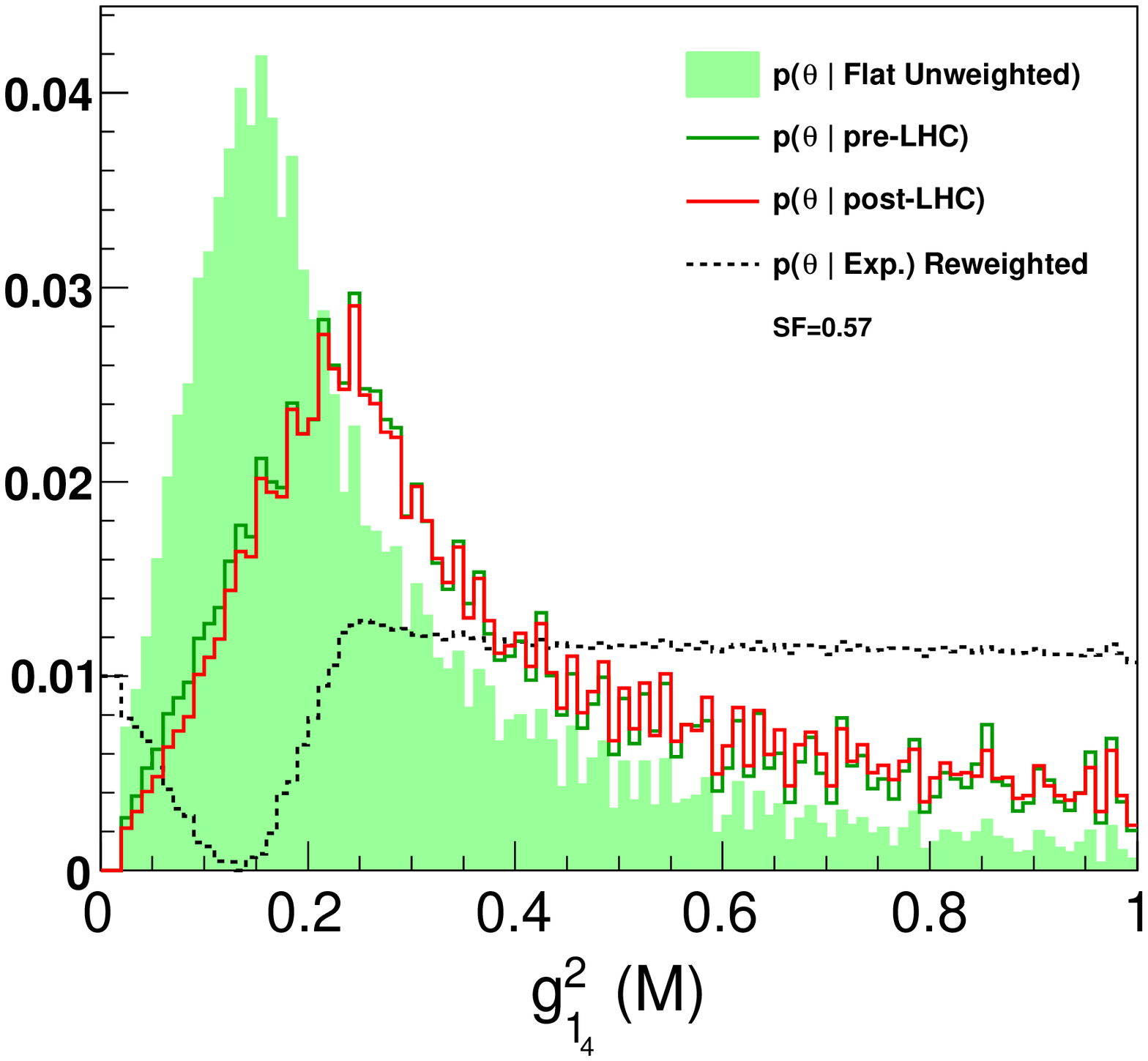}  &
\includegraphics[width=0.34\textwidth]{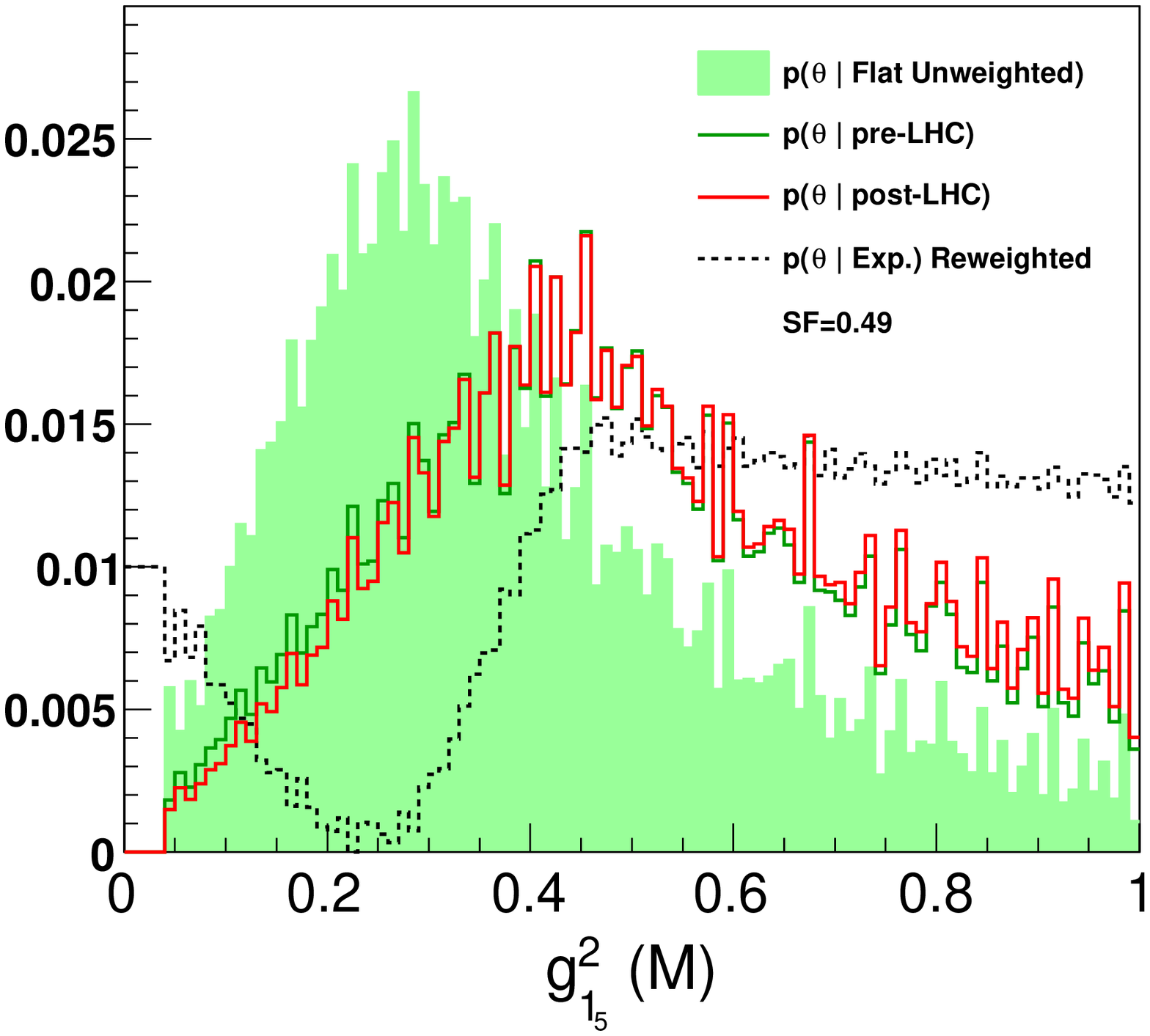}  &
\includegraphics[width=0.34\textwidth]{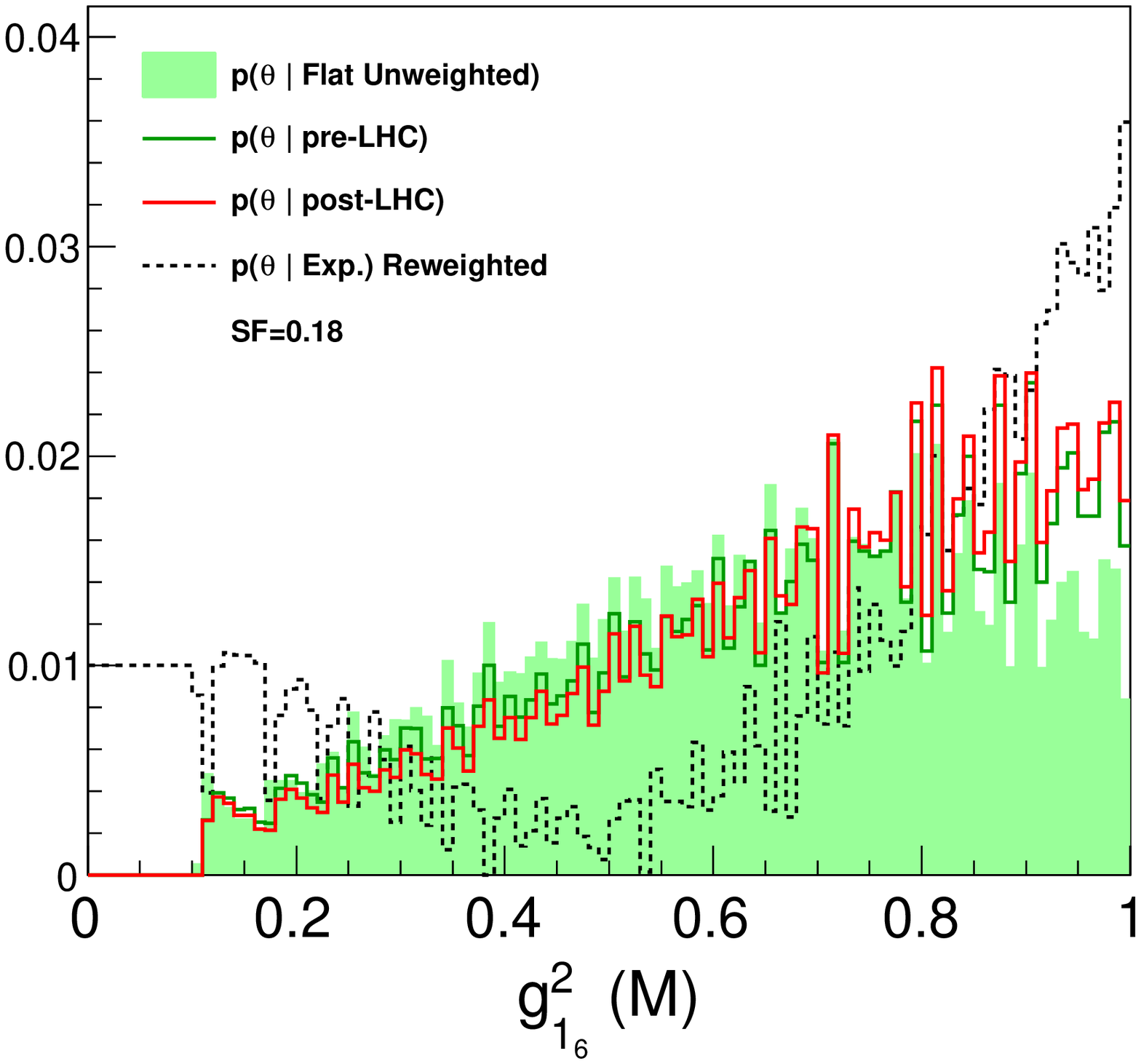}  \\
\includegraphics[width=0.34\textwidth]{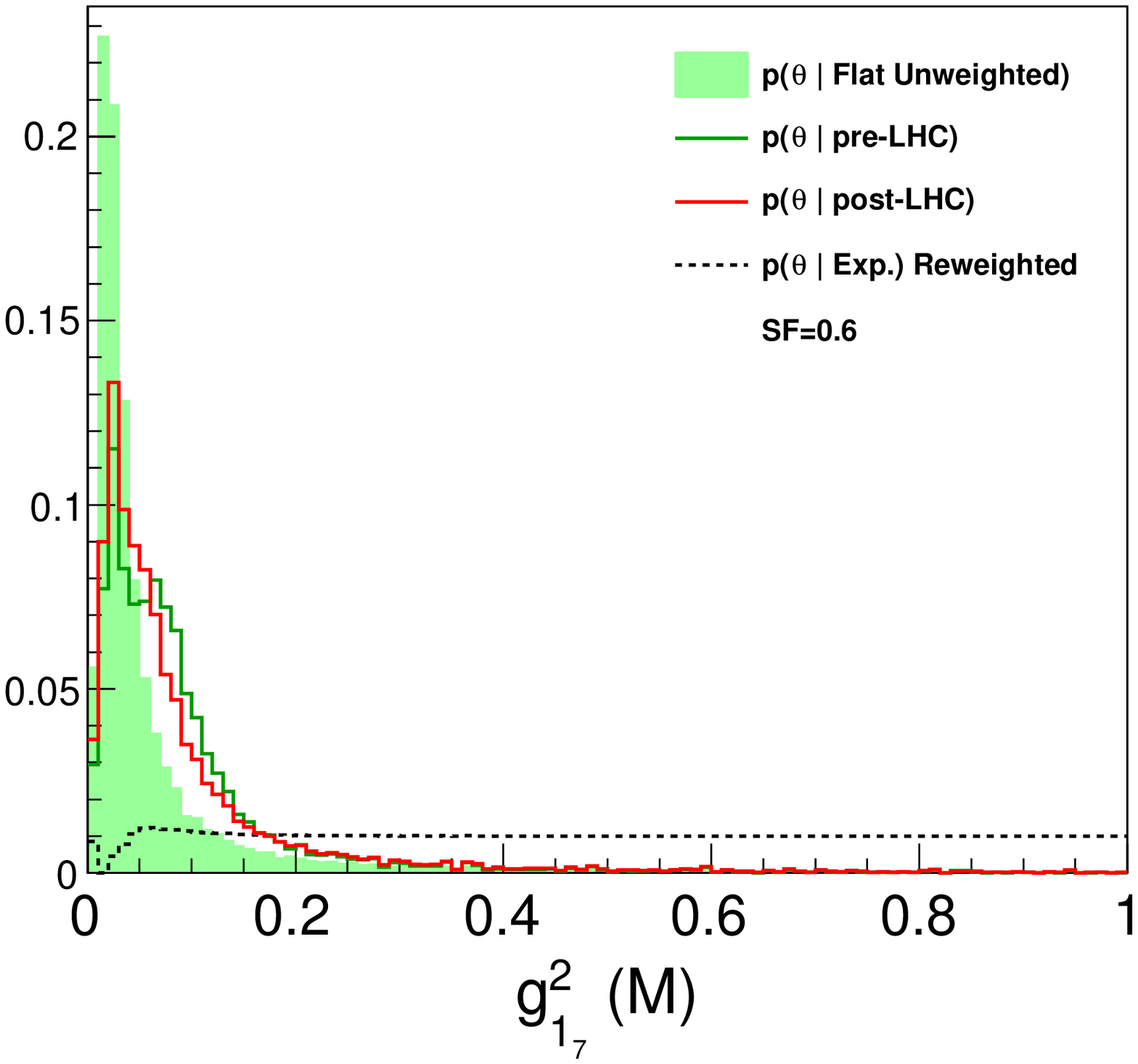}  &
\includegraphics[width=0.34\textwidth]{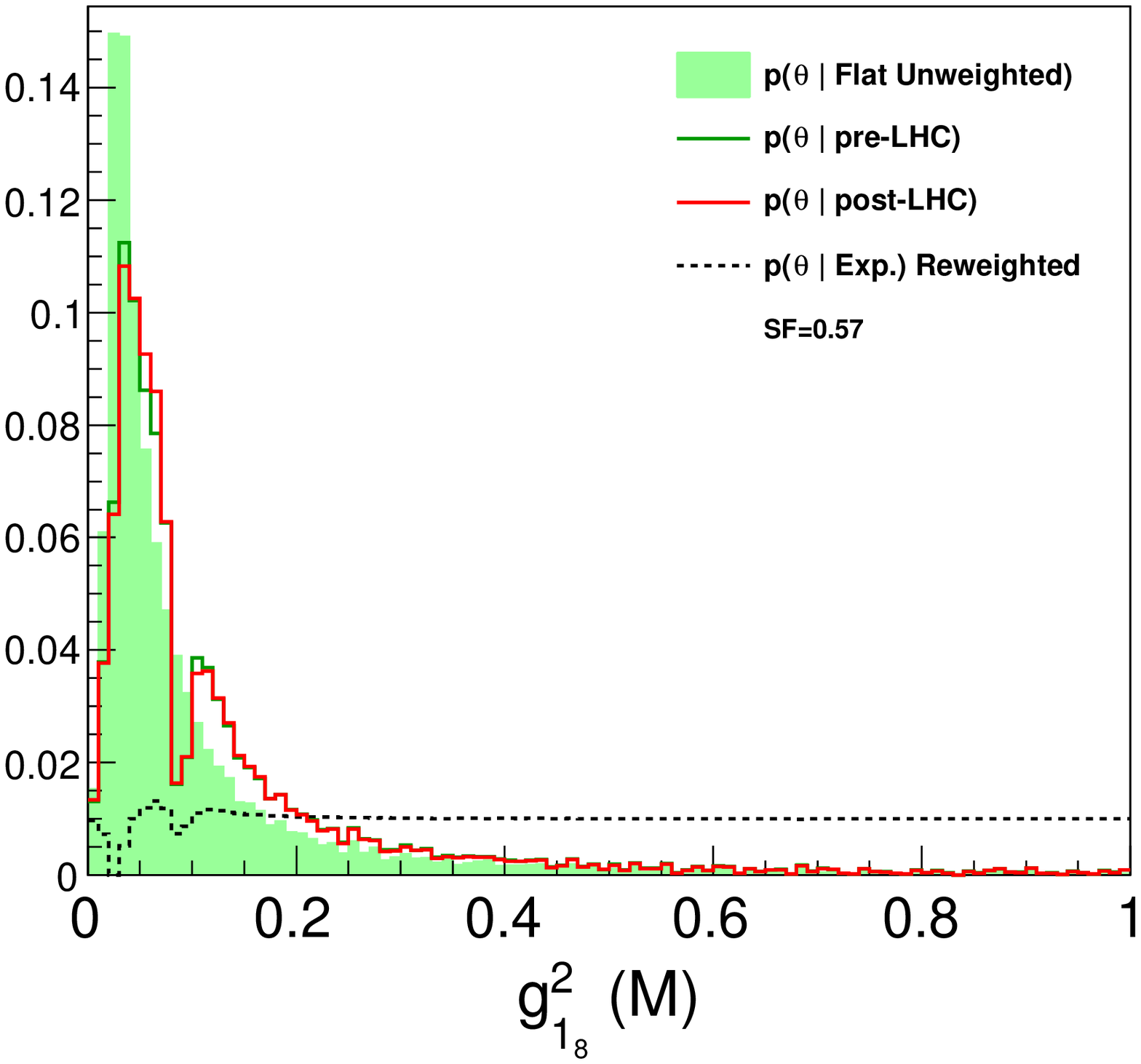}  &
\includegraphics[width=0.34\textwidth]{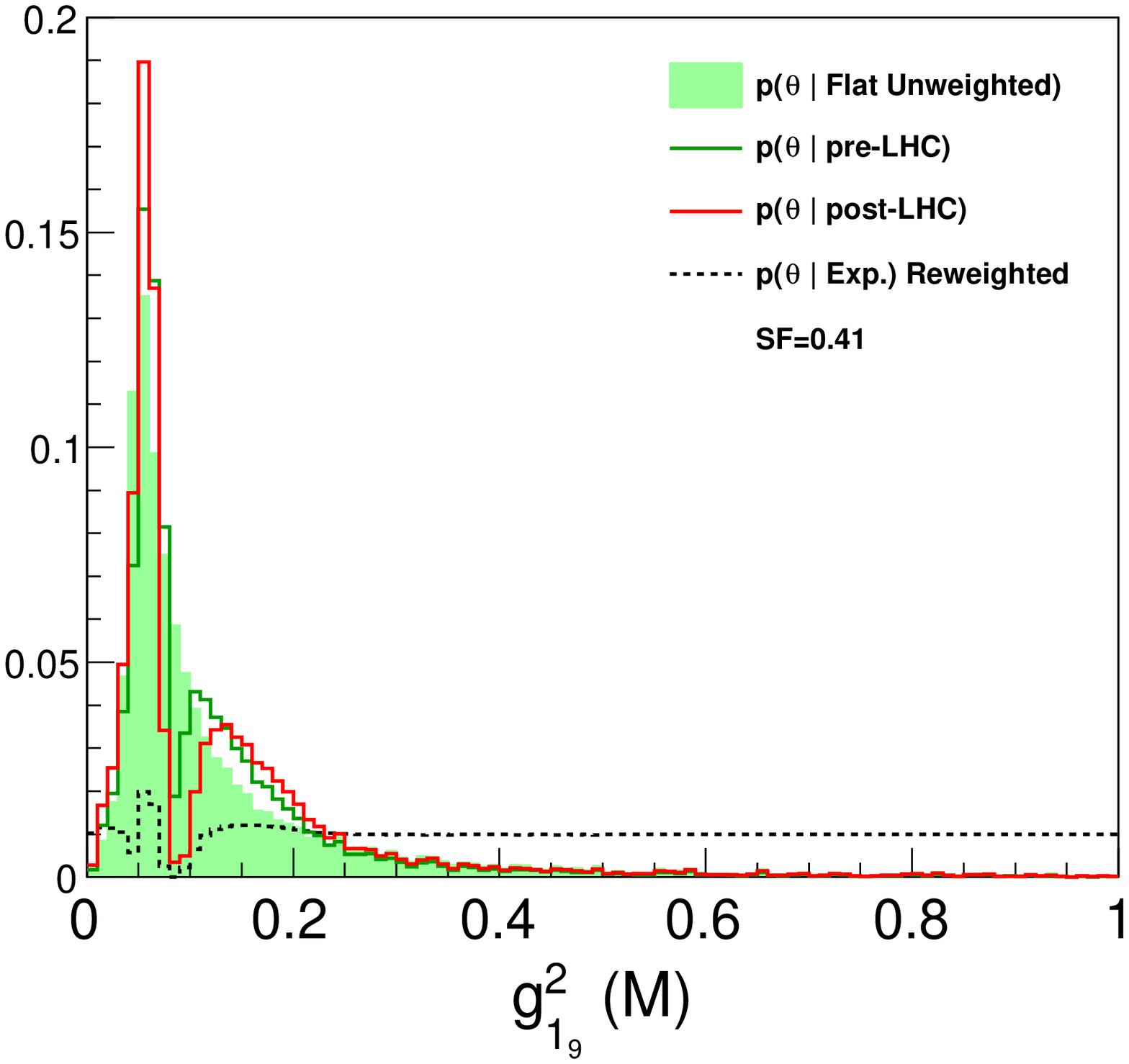}  \\
\end{tabular}
\end{center}
\caption{Distribution of the gauge coupling, $g_1^2$, at the messenger scale before and after the LHC constraints are added (green and red lines), flat distribution (shaded green) and subtracted probability distribution (dashed black line). The different sets are associated with different probability distributions given in Eqs.~\ref{MGM_gis}.}
\label{MGMParameters3}
\end{figure}

\begin{figure}
\begin{center}
\begin{tabular}{c c c}
\includegraphics[width=0.34\textwidth]{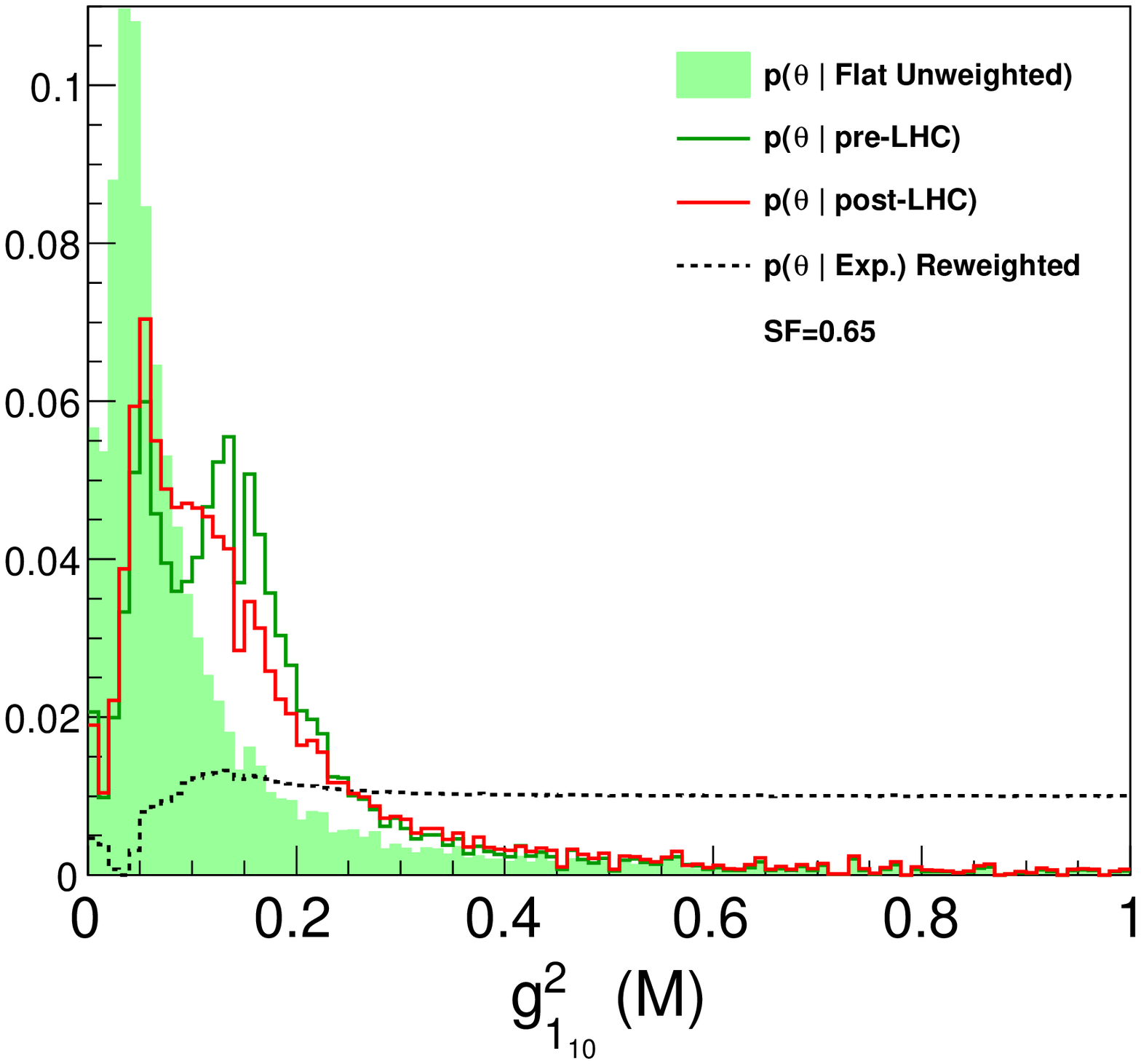}  &
\includegraphics[width=0.34\textwidth]{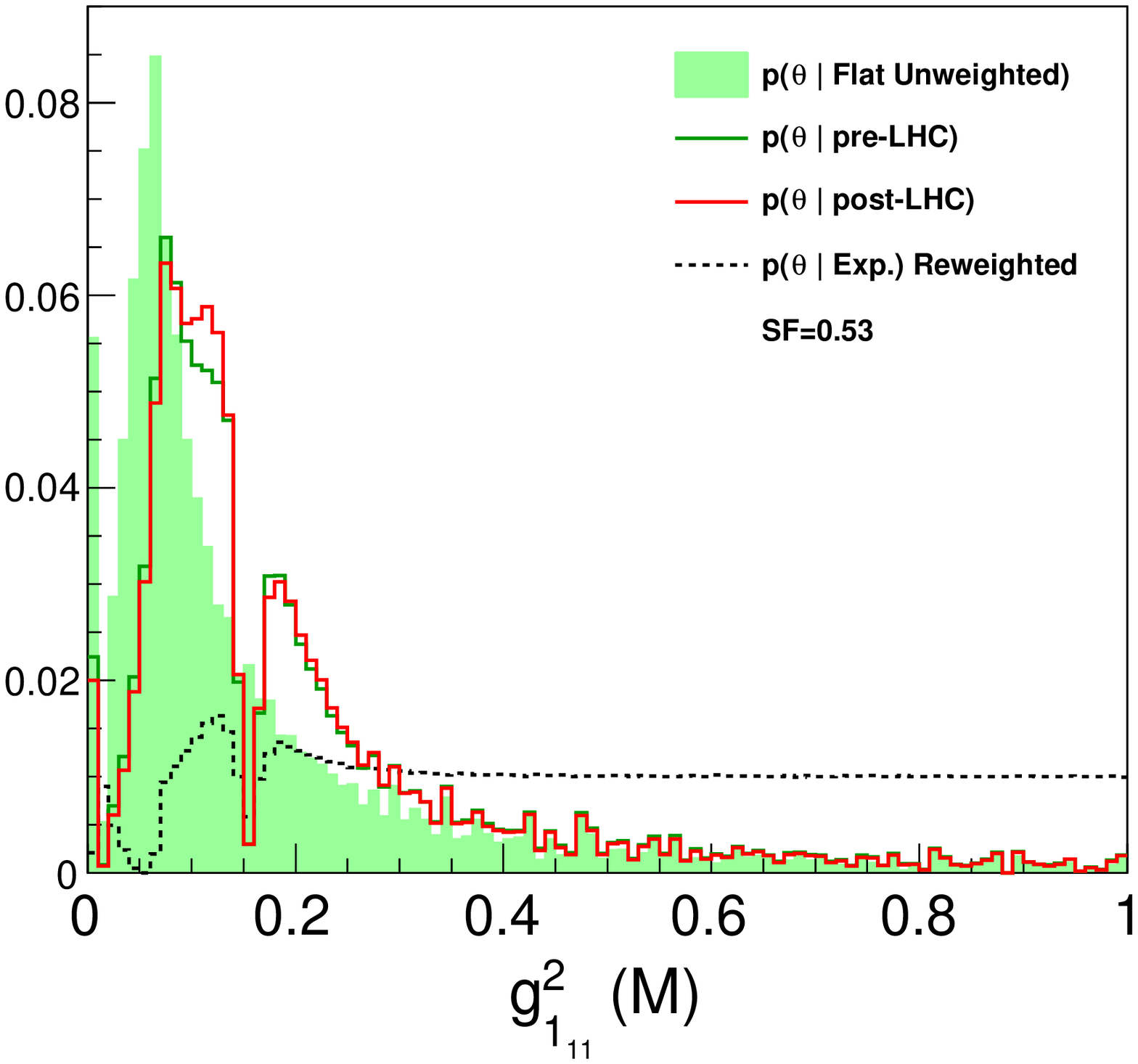}  &
\includegraphics[width=0.34\textwidth]{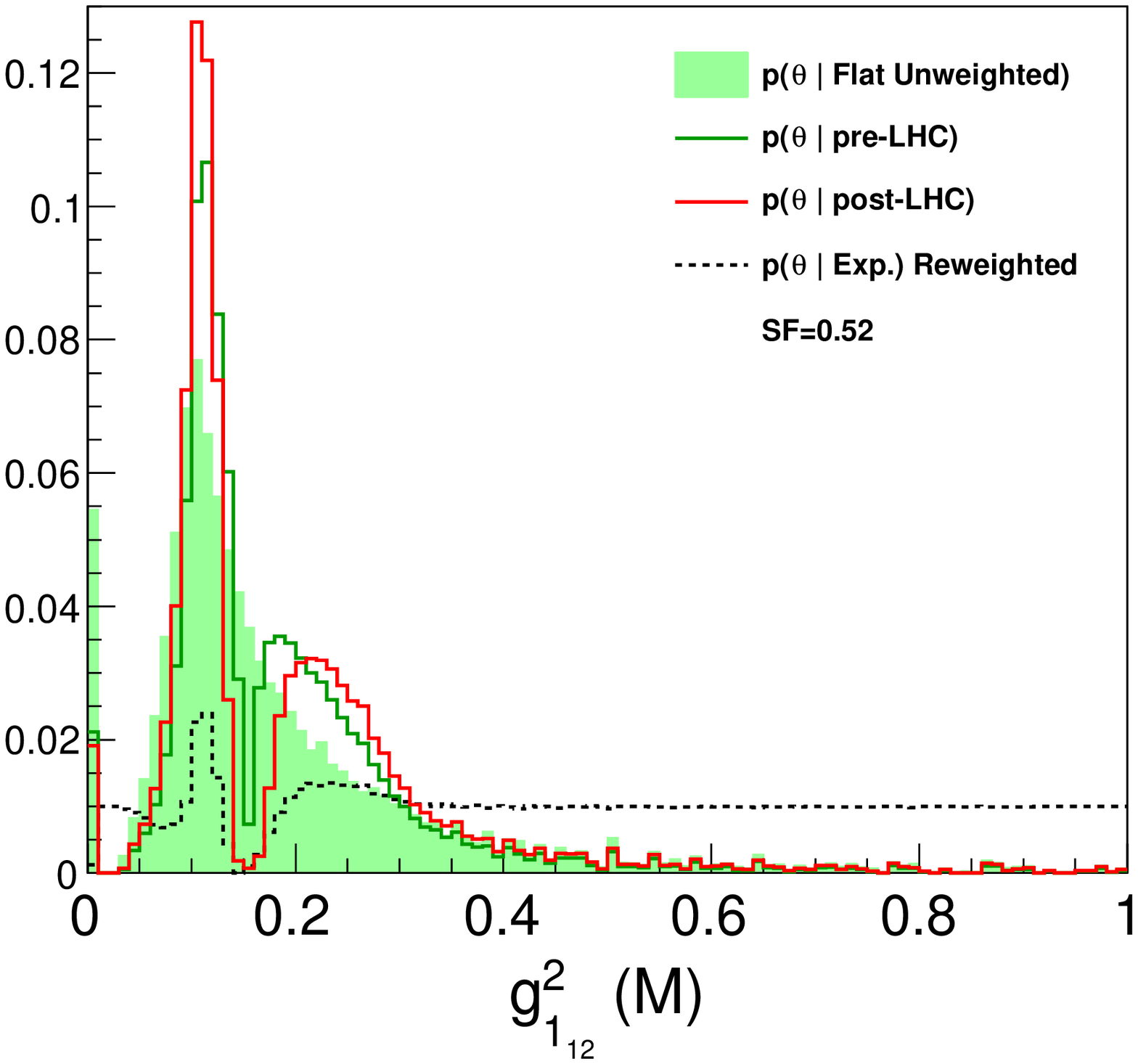}  \\
\end{tabular}
\end{center}
\caption{Distribution of the gauge coupling, $g_1^2$, at the messenger scale before and after the LHC constraints are added (green and red lines), flat distribution (shaded green) and subtracted probability distribution (dashed black line). The different sets are associated with different  probability distributions given in Eqs.~\ref{MGM_gis}.}
\label{MGMParameters4}
\end{figure}

The probabilities corresponding to each of these are plotted in Figs.~\ref{MGMParameters1}-\ref{MGMParameters4}.

Observe that, as is apparent from Eq.~(\ref{IMiB}) and Table~\ref{table.Inv}, MGM is associated with negative values of $I_{M_3}$ and positive values of $I_{M_{1,2}}$. As can be seen from Fig.~\ref{Invariants1}, these values of the RGIs are not the most likely ones consistent with the present constraints. However, wether a given model is likely or not is a very scan dependent question and hence we will not address that here. Instead, the probability distributions for the MGM parameters are computed for those configurations for which these conditions are fulfilled. The final distribution is obtained by multiplication of the independent probabilities of the 9 $B_i$ and 12 $g_{1_i}^2(M)$ solutions given in Eqs.~\ref{MGM_Bis}-\ref{MGM_gis}. The results are depicted in Fig.~\ref{MGMpardist}.

The messenger scale may be obtained from the value of the gauge coupling at this scale by using Eq.~(\ref{Munifgunif}), replacing $M_{\rm unif}$ by $M_{\rm mess}$. However, in contrast to the gaugino mass unification scale,  the messenger scale is always a physical scale and therefore expected to take values  between tens of TeV and the GUT scale, or equivalently,  gauge coupling values of $0.2 \simlt g_1^2(M_{\rm mess}) \simlt 0.5$. Fig.~(\ref{MGMpardist}) shows that values of the gauge couplings $g_1^2(M_{\rm mess}) \simgt 0.6$ tend to be preferred, which lie outside the physical region.  Considering only the physical range, values of the messenger scale close to the GUT scale are slightly preferred.  The most probable values of the parameter $B$ are about 1.25~TeV and 4.25~TeV.  Using the relation $M_i = g_i^2 B$  and the values of the gauge couplings at the weak scale, $M_{\rm mess} \simeq 1.25$~TeV would lead to a bino mass of the order of 250~GeV, a wino mass of about 500~GeV and a gluino mass of about 1.5~TeV. The larger value of $B$ would lead to gaugino masses 3.5 times heavier than these ones.

\section{Conclusions}

Supersymmetric extensions of the Standard Model provide a relationship between the weak scale and the scale of the supersymmetry-breaking parameters, rendering it stable under quantum corrections. In the MSSM, the SM-like Higgs particle is predicted to be light. The fast decoupling of the supersymmetric particles from the precision electroweak observables make the MSSM predictions consistent with those of the SM with a light Higgs, in full consistency with what current data seems to suggest. However, no direct hint of supersymmetric particles has been observed experimentally and hence no information of the structure and origin of the supersymmetry-breaking parameters is provided by current experiments, apart from perhaps the indirect hints provided by the anomalous magnetic moment and the Higgs mass range. Once additional information from direct searches  becomes available, a method to determine the structure of supersymmetry-breaking parameters at the messenger scale, as well at the messenger scale itself would be desirable. RGIs provide such a method, establishing a direct relationship between the observables at the weak scale and the messenger scale parameters.

In this article we have studied the probability distributions of a set of RGIs in the MSSM arising from symmetry arguments. The distributions are analyzed  at the TeV scale by making use of the constraints coming from flavor physics, LEP and Tevatron searches, Higgs physics and the anomalous magnetic moment of the muon, and separately from those, by constraints provided by the LHC. We have used a flat prior for the soft supersymmetry-breaking masses, using a pMSSM approach. The current constraints already provide interesting features in the probability distributions.

As an example of the application of the RGIs, we have used them to analyze the question of Gaugino Mass Unification and also the possible realization of General and Minimal Gauge Mediation. The methods described here are quite general and may be applied to analyze the ultraviolet properties of the MSSM parameters in other interesting supersymmetery-breaking scenarios.

We noticed that the scale of Gaugino Mass Unification is not necessarily identified with the messenger scale, but it can provide non-trivial information on the realization of minimal models of supersymmetry breaking.  GGM provides a well-motivated example of flavor independent, supersymmetry-breaking models. The probability distributions for the GGM parameters can be determined from those of the RGIs and present some interesting features as well. They also lead to information on possible non-universal Higgs mass parameters at the messenger scale. The determination of the messenger scale in GGM through RGIs demands the measurement of both the first and third generation fermion masses as well as the Higgs masses, and hence it is not practical at this moment. We also analyze the more simplistic subset of models  given by MGM. Since the entire model space of MGM is determined by only 2 parameters, we are able to extract information about the possible scale of SUSY particles as well as the messenger scale in this scenario.

It is clear that although the analysis we describe already has interesting features, the probability distributions of the RGIs will become particularly useful when the LHC starts revealing the presence of supersymmetric particles at the weak scale. In such a case, the probability distribution of the RGIs will become sharper and will start showing important features of the  supersymmetry-breaking mass parameters at the messenger scale.  Due to the higher cross sections for the production of supersymmetric particles, the higher luminosities and the higher energy reach, the 8~TeV run this year will lead to relevant constraints on the supersymmetric particle masses. It could also lead to the first hint of the presence of supersymmetry, beyond the indirect ones associated with Higgs search results. It will be therefore very interesting to repeat the analysis of the RGI distributions once the 2012 results are available. 

~\\
~\\
{\bf Acknowledgements}  We thank Sabine Kraml and Harrison Prosper for valuable discussions.
Fermilab is operated by Fermi Research Alliance, LLC, under contract DE-AC02-07CH11359 with the United States Department of Energy.
Work at ANL is supported in part by the U.S. Department of Energy~(DOE), Div.~of HEP, Contract DE-AC02-06CH11357.

\appendix 
\appendixpage 

\section{Probability Re-weighting and Re-scaling}

We are interested in a quantity which quantitative reflects the probability distributions of general functions of the masses, given the probability distributions for the masses themselves. The Markov Chain Monte Carlo (MCMC) method is used to scan over the pMSSM parameters in the range considered to be probed at the LHC. For each point corresponding to a model, a likelihood is computed, given certain experimental constraints. Since the MCMC technique scans the given parameter space along the isocontours of likelihood due to preLHC constraints listed in Table~\ref{tab:preLHCobs}, the ratio of the number of points scanned for any given value of a parameter to the total number of points gives the probability for that parameter value. This probability for a given point can then be re-weighted by the postLHC likelihoods to compute the current probabilities. We note however that the boundaries defining the pMSSM region scanned, introduce an artificial effect in the resulting probability distributions.  In the following, we will describe a method that can be used for eliminating this effect. In this method, we make the assumption that the LHC (as well as the pre-LHC) measurements will not be able to shed any light on the pMSSM parameter regions that are not scanned due to kinematic constraints, and assign a flat probability to these insensitive regions outside the scan boundary.

Let us consider a two dimensional probability distribution $p(x,y| O)$ of parameters $x$ and $y$ defined in a box where the variables $x$ and $y$ vary in the ranges $\{x_1, x_2\},\{y_1,y_2\}$, given some observables $O$. Assume that $x$ and $y$ have flat priors corresponding to the soft parameters that were scanned over in the MCMC. We are then interested in the probability distribution of some function, $\theta(x,y)$, given $O$:  $p(\theta(x,y)| O)$.  As explained in Section 2.2, the naive computation of this probability, especially using a flat prior $p_0^{f}(x, y) = constant$, will heavily reflect the size of the box alongside any other inherent probability distribution of this function.  The aim is to define a probability $p(\theta(x,y)|O)$ such that, if there is no condition on $\theta(x,y)$, then a flat distribution is obtained for $p_0(\theta(x,y))$. Any variations of this flatness should be something that reflects the actual variation of the probability due to the effect of $O$ rather than the effect of having a bounded box, as is the case in the example given in Section~\ref{RGIpMSSM}. 

Let us assume that the box contains $a$ bins in $x$ and $b$ bins in $y$. The flat distributions are defined such that in the absence of any additional condition:  

\begin{eqnarray}
p_0^{f}(x)&=&\frac{1}{a}\\
p_0^{f}(y)&=&\frac{1}{b}\\
p_0^{f}(x,y)&=&\frac{1}{a b}\\
p_0^{f}(\theta_i)&=& \sum_{{\rm All }\{x,y\} : \theta(x,y)=\theta_i} p_0^{f}(x,y)
\end{eqnarray}

Note that $p^{f}_0(\theta)$ is defined as the distribution that would be obtained for $\theta(x,y)$ if $x$ and $y$ have flat priors. This is the distribution that is referred to in the text as ``$p(\theta|\rm Flat Unweighted)$''. This distribution itself is generally not flat, but will have a distinct shape reflecting the boundary conditions of the original $x,y$ variables.  Analogously, the probability for $\theta(x,y)$ given $O$ is 
\begin{equation}
p(\theta_i|O)= \sum_{{\rm All }\{x,y\} : \theta(x,y)=\theta_i} p(x,y|O)\;,
\end{equation}
where this is referred to as  ``$p(\theta|\rm pre/post LHC)$'' in the text.
An easy  way to normalize this probability to obtain a flat distribution for the function $\theta(x,y)$ in the absence of non-trivial conditions is to weight each bin, $\theta_i$, by $1/p_0^{f}(\theta_i)$:
\begin{equation}
p^A(\theta_i|O) \propto \frac{p(\theta_i|O)}{p_0^{f}(\theta_i)}\;.
\end{equation}
The superscript $A$ denotes the fact that this effectively gives the average probability per unique $\{x,y\}$ combination for each $\theta_i$. However, this has the effect of washing out small effects on the probability distribution from $O$, when $\theta_i$ is such that a large number of unique combinations of $\{x,y\}$ contribute  to a given value of $\theta(x,y)$. 

We propose an alternative method. First, instead of taking the ratio we shall consider the difference: $p(\theta_i|O) - p_0^{f}(\theta_i)$.  Clearly this quantity is not always positive and cannot be identified with a probability distribution. It has, however, the property that it becomes positive whenever the probability of $\theta_i$ is enhanced by the observations $O$ and negative in the opposite case. We shall, hence, define a renormalized distribution $p^R(\theta_i|O)$ in the following way
\begin{eqnarray}
p_0^{f}(\theta_m)&=&Max\left[p_0^{f}(\theta_i)\right]\\
p^R(\theta_i|O)& \propto &p(\theta_i|O) + \left[p_0^{f}(\theta_m)-p_0^{f}(\theta_i)\right]\label{ReW},
\end{eqnarray}
which is always positive since $p(\theta_i|O)$ is positive and so is the quantity between brackets.  The above quantity, Eq.~(\ref{ReW}) has a clear interpretation : Let us first stress that, by definition,  $\theta_m$ is such that it has the largest number of unique combinations of $\{x,y\}$ contributing to it, with $x$ and $y$ varying with a flat distribution in the box.  Let's call $k_i$ the number of combinations corresponding to $\theta_i$.  Therefore,  all  $\theta_i$ have a smaller number than $\theta_m$, $k_i<k_m$. This is reflecting the fact that for $i\neq m$, the range of the original variables scanned over, $x$ and $y$, did not include all the combinations necessary to weight the $i$ bin of $\theta$ the same as $m$. We have made the argument that the values of $x$ and $y$ not scanned are ones that will not be affected by LHC measurements. Hence we propose that these combinations are given the same weight as  $p_0^f(\theta_i)/k_i=1/ab$. This leads, after proper normalization, to nothing more than the last term, between square brackets, in Eq.~(\ref{ReW}), and hence the quantity $p^R(\theta_i|O)$ reflects the actual probability distribution of $\theta_i$ given $O$, taking away the effect of the range of the original scan. For this quantity to represent a  probability distribution in the strict sense, it must be normalized to 1. Since $p(\theta_i|O)$ and $p_0^{f}(\theta_i)$ are quantities which are normalized to 1, assuming that the function  $\theta_i$ is evaluated in $l$ different bins, the normalization factor is nothing more than  $C=1/( p_0^{f}(\theta_m) l)$.  Hence the properly normalized probability distribution for $\theta$ is given by:
\begin{equation}
p^R(\theta_i|O)= \frac{1}{p_0^{f}(\theta_m) \; l}\left\{p(\theta_i|O) + \left[p_0^{f}(\theta_m)-p_0^{f}(\theta_i)\right]\right\}\label{ReWN}\;.
\end{equation}
We can see that this behaves the way we expect it to, by noting that when $O$ has not impacted the probability of $\theta$, i.e. $p(\theta_i|O)=p_0^{f}(\theta_i)$, $p^R(\theta_i|O)=1/l$, so we obtain a flat distribution. On the other hand, if the $p_0^{f}(\theta_i)$ is a constant, meaning that $\theta_i$ has a flat distribution in the same flat basis as the original variables $x$ and $y$, then $p_0^{f}(\theta_i) = p_0^{f}(\theta_m) = 1/l$ and we recovers $p(\theta_i|O)$ without any modification, as we should. 

 In order to emphasize the impact of the experimental constraints in a more clear way, however, we have gone a step further. Since we assumed that the probability outside the range we scanned is flat, the ratio of the difference of any two probabilities from flat, $(p^{R}(\theta_i|O)-1/l)/(p^{R}(\theta_j|O)-1/l)$, will remain invariant if we extended the range of the original scan, increasing the box size. Therefore,  this quantity is than also invariant under an overall rescaling of the differences with the flat probability. 
 
 Let us assume that there is a non-trivial impact of experiments on the RGI distributions, namely $p^R(\theta_i|O) \neq 1/l$ for at least one $i$.  Considering
\begin{equation}
p^R(\theta_n|O)=Min\left[p^R(\theta_i|O)\right]
\end{equation}
we define a scale factor, $SF$, such that the difference of this minimum with $1/l$ is scaled to be $1/l$ :
\begin{eqnarray}
SF \left( \frac{1}{l}-p^R(\theta_n|O) \right)&=&\frac{1}{l}\\
\implies SF^{-1}&=& 1-p^R(\theta_n|O) l .
\end{eqnarray}
We use the scale factor above to define a modified distribution
\begin{eqnarray}
p^{SS}(\theta_i|O)&=&\frac{1}{l}+SF\left[p^{R}(\theta_i|O)-\frac{1}{l}\right]\;,\label{pSS}\\
&=& \frac{1}{p_0^{f}(\theta_m) l} \left\{ p_0^{f}(\theta_m)+ SF \left[ p(\theta_i|O) - p_0^{f}(\theta_i) \right] \right\}\;.
\label{ModifiedRdistribution}
\end{eqnarray}
Once the scale factor $SF$ is given, it is easy to translate this modified distribution, Eq.~(\ref{ModifiedRdistribution}) to the original one, Eq.~(\ref{ReWN}).   The quantity $p^{SS}(\theta_i|O)$ has the virtue that when for a particular bin $p^{R}(\theta_i|O)=1/l$, meaning $O$ has had no impact on the $\theta_i$ probability, one obtains $p^{SS}(\theta_i|O)=1/l$. On the other hand when $p^R(\theta_i|O)=p^R(\theta_n|O)$, meaning when $O$ has maximally decreased the probability for that $\theta_i$,  $p^{SS}(\theta_i|O)=0$. 

 The fact that  $p^{SS}(\theta_i|O)$  will be invariant under a change in scan range of the original variables can be seen by inspecting Eq.~\ref{pSS} and noting that under a change of scan range,   $p^{SS}(\theta_i|O)= 1/l$ when $p^R(\theta_i|O)=1/l$  and by definition $p^{SS}(\theta_n|O)=0$.  
 
 Even though $p^{SS}(\theta_i|O)$ cannot be technically defined as a probability, it quantitatively reflects the actual impact of $O$ on the probability distribution of $\theta$ in a way which is independent of the artificial impact of scanning a finite region, and, as stressed above may be easily connected with $p^R(\theta_i|O)$, Eq.~(\ref{ReWN}).  We ran extensive numerical checks to make sure that this quantity indeed behaves in the expected manner. We have therefore used $p^{SS}(\theta_i|O)$ to represent the probability distribution of the RGIs, giving the associated scale factor $SF$ for every RGI distribution. In the text, in order to be more explicit about the meaning of these distributions, $p(\theta|O)$ was renamed $\lq\lq p(\theta|{\rm pre/post}$-${\rm LHC})$", while $p^{SS}(\theta|O)$ was renamed $\lq\lq p(\theta|Exp)$ Reweighted".

\section{pMSSM Parametrization}\label{sec:model}

The pMSSM, a 19-dimensional realization~\cite{Djouadi:1998di} of the R-parity conserving MSSM with parameters defined at the SUSY scale, $M_{\rm SUSY}=\sqrt{m_{\tilde t_1}m_{\tilde t_2}}$, employs only a few plausible assumptions motivated by experiment: there are no new CP phases, the sfermion mass matrices and trilinear couplings are flavor-diagonal, the first two generations of sfermions are degenerate and their trilinear couplings are negligible.
In addition, we assume that the lightest supersymmetric particle (LSP) is the lightest neutralino, $\tilde\chi^0_1$.
We thus arrive at a proxy for the MSSM characterized by 19 real, weak-scale, SUSY Lagrangian parameters:
\begin{itemize}
   \item 3 gaugino mass parameters $M_1$, $M_2$, and $M_3$;
   \item the ratio of the Higgs vevs, $\tan\beta=v_2/v_1$;
   \item the higgsino mass parameter, $\mu$, and
            the pseudo-scalar Higgs mass, $m_A$;
    \item 10 sfermion mass parameters $m_{\tilde{F}}$, where
         $\tilde{F} = \tilde{Q}_1, \tilde{U}_1, \tilde{D}_1,
                      \tilde{L}_1, \tilde{E}_1,
                      \tilde{Q}_3, \tilde{U}_3, \tilde{D}_3,
                      \tilde{L}_3, \tilde{E}_3$\\
(imposing $m_{\tilde{Q}_1}\equiv m_{\tilde{Q}_2}$,
           $m_{\tilde{L}_1}\equiv m_{\tilde{L}_2}$, etc.); and
   \item 3 trilinear couplings $A_t$, $A_b$ and $A_\tau$\,,
\end{itemize}
in addition to the SM parameters.

For each pMSSM point, 
{\tt SoftSUSY3.1.6}~\cite{Allanach:2001kg} was used to compute the SUSY spectrum,
{\tt SuperIsov3.0}~\cite{Mahmoudi:2008tp} was used to compute the low-energy constraints,
{\tt micrOMEGAs2.4}~\cite{Belanger:2001fz} was used for the SUSY mass limits, and
{\tt HiggsBounds2.0.0}~\cite{Bechtle:2011sb}  for the limit on the $h^0$ mass\footnote{In
evaluating the Higgs mass limit, a Gauss-distributed theoretical uncertainty of  $\sigma=1.5$~GeV  was applied to the $m_h$ computed with {\tt SoftSUSY},
cf.\ row 8 in Table~\ref{tab:preLHCobs} and row 12 in Table~\ref{tab:LHCobs}.}.
Moreover, 
{\tt SUSYHIT (SDECAY1.3b, HDECAY3.4)}~\cite{Djouadi:2006bz} was used to produce SUSY and Higgs decay tables, and
 {\tt micrOMEGAs2.4}~\cite{Belanger:2001fz} to compute the LSP relic density and direct detection cross sections.
The various codes were interfaced using the SUSY Les Houches Accord~\cite{Skands:2003cj}.

\section{Soft Mass Parameters and RGIs}

As mentioned in Section 2, one can make use of the RGIs and three independent masses to determine all other soft breaking masses. As an example, we write down 2 sets of solutions with different unknown masses. All the masses and gauge couplings are at the same scale. The gaugino masses in both cases are given by
\begin{equation}\label{gauginos}
M_i=I_{B_i}g_i^2\qquad i=1,2,3.
\end{equation}
We write the first set of solutions in terms of 3 third generation masses: $m_{Q_3}$, $m_{u_3}$ and $m_{e_3}$,
\begin{dmath}
m_{H_2}^2= \frac{D_{B_{13}}}{2}-\frac{D_Z}{2}-\frac{5 I_{M_1}}{66}+\frac{3I_{M_2}}{2}+\frac{4 I_{M_3}}{3}-D_{L_{13}}-\frac{247 D_{Y_{13H}}}{220}+\frac{D_{\chi_1}}{40}+\frac{3 I_{Y_\alpha} g_1^2}{22}+\frac{5}{66} I_{B_1}^2 g_1^4-\frac{3}{2} I_{B_2}^2 g_2^4-\frac{4}{3} I_{B_3}^2 g_3^4+\frac{3 m_{u_3}^2}{2},
\end{dmath}
\begin{dmath}
m_{H_d}^2= \frac{3 D_{B_{13}}}{2}-\frac{D_Z}{2}+\frac{2 I_{M_1}}{33}-3I_{M_2}+\frac{4 I_{M_3}}{3}-\frac{D_{L_{13}}}{2}-\frac{13 D_{Y_{13H}}}{44}+\frac{3 D_{\chi_1}}{8}-\frac{5 I_{Y_\alpha} g_1^2}{22}-\frac{2}{33} I_{B_1}^2 g_1^4+3 I_{B_2}^2 g_2^4-\frac{4}{3} I_{B_3}^2 g_3^4+\frac{m_{e_3}^2}{2}+3 m_{Q_3}^2-\frac{3 m_{u_3}^2}{2},
\end{dmath}
\begin{eqnarray}
m_{d_3}^2&=& D_{B_{13}}+\frac{I_{M_1}}{11}-3I_{M_2}-\frac{13 D_{Y_{13H}}}{165}+\frac{3 D_{\chi_1}}{10}-\frac{2 I_{Y_\alpha} g_1^2}{33}-\frac{1}{11} I_{B_1}^2 g_1^4+3 I_{B_2}^2 g_2^4+2 m_{Q_3}^2-m_{u_3}^2,\nonumber\\
\end{eqnarray}
\begin{dmath}
m_{Q_1}^2= \frac{1}{3960}\left(20 I_{M_1}+5940I_{M_2}-3520 I_{M_3}+78 D_{Y_{13H}}-627 D_{\chi_1}+60 I_{Y_\alpha} g_1^2-20 I_{B_1}^2 g_1^4-5940 I_{B_2}^2 g_2^4+3520 I_{B_3}^2 g_3^4\right),
\end{dmath}
\begin{dmath}
m_{L_3}^2= \frac{1}{220} \left(-10 I_{M_1}+330I_{M_2}-110 D_{L_{13}}-26 D_{Y_{13H}}-11 D_{\chi_1}-20 I_{Y_\alpha} g_1^2+10 I_{B_1}^2 g_1^4-330 I_{B_2}^2 g_2^4+110 m_{e_3}^2\right),
\end{dmath}
\begin{dmath}
m_{L_1}^2= \frac{1}{440} \left(20 I_{M_1}+660I_{M_2}-26 D_{Y_{13H}}-11 D_{\chi_1}-20 g_1^2 \left(I_{Y_\alpha}+I_{B_1}^2 g_1^2\right)-660 I_{B_2}^2 g_2^4\right),
\end{dmath}
\begin{dmath}
m_{d_1}^2= \frac{1}{1980}\left(40 I_{M_1}-1760 I_{M_3}+78 D_{Y_{13H}}+33 D_{\chi_1}+60 I_{Y_\alpha} g_1^2-40 I_{B_1}^2 g_1^4+1760 I_{B_3}^2 g_3^4\right),
\end{dmath}
\begin{dmath}
m_{u_1}^2= \frac{1}{990} \left(80 I_{M_1}-880 I_{M_3}-78 D_{Y_{13H}}-33 D_{\chi_1}-60 I_{Y_\alpha} g_1^2-80 I_{B_1}^2 g_1^4+880 I_{B_3}^2 g_3^4\right),
\end{dmath}
\begin{dmath}
m_{e_1}^2= \frac{1}{220} \left(40 I_{M_1}+26 D_{Y_{13H}}+11 D_{\chi_1}+20 I_{Y_\alpha} g_1^2-40 I_{B_1}^2 g_1^4\right)
\end{dmath}

Alternatively, the second set of solutions  is given in terms of the 2 soft masses for the Higgs, $m_{H_u}$ and $m_{H_d}$, and a third generation squark mass, $m_{Q_3}$:
\begin{dmath}
m_{u_3}^2= -\frac{D_{B_{13}}}{3}+\frac{D_Z}{3}+\frac{5 I_{M_1}}{99}-I_{M_2}-\frac{8 I_{M_3}}{9}+\frac{2 D_{L_{13}}}{3}+\frac{247 D_{Y_{13H}}}{330}-\frac{D_{\chi_1}}{60}-\frac{I_{Y_\alpha} g_1^2}{11}-\frac{5}{99} I_{B_1}^2 g_1^4+I_{B_2}^2 g_2^4+\frac{8}{9} I_{B_3}^2 g_3^4+\frac{2 m_{H_u}^2}{3},
\end{dmath}
\begin{dmath}
m_{e_3}^2= -4 D_{B_{13}}+2 D_Z+\frac{I_{M_1}}{33}+3I_{M_2}-\frac{16 I_{M_3}}{3}+3 D_{L_{13}}+\frac{156 D_{Y_{13H}}}{55}-\frac{4 D_{\chi_1}}{5}+\frac{2 I_{Y_\alpha} g_1^2}{11}-\frac{1}{33} I_{B_1}^2 g_1^4-3 I_{B_2}^2 g_2^4+\frac{16}{3} I_{B_3}^2 g_3^4+2 m_{H_d}^2+2 m_{H_u}^2-6 m_{Q_3}^2,
\end{dmath}
\begin{dmath}
m_{Q_1}^2= \frac{1}{3960}\left(20 I_{M_1}+5940I_{M_2}-3520 I_{M_3}+78 D_{Y_{13H}}-627 D_{\chi_1}+60 I_{Y_\alpha} g_1^2-20 I_{B_1}^2 g_1^4-5940 I_{B_2}^2 g_2^4+3520 I_{B_3}^2 g_3^4\right),
\end{dmath}
\begin{dmath}
m_{d_3}^2= \frac{4 D_{B_{13}}}{3}-\frac{D_Z}{3}+\frac{4 I_{M_1}}{99}-2I_{M_2}+\frac{8 I_{M_3}}{9}-\frac{2 D_{L_{13}}}{3}-\frac{91 D_{Y_{13H}}}{110}+\frac{19 D_{\chi_1}}{60}+\frac{I_{Y_\alpha} g_1^2}{33}-\frac{4}{99} I_{B_1}^2 g_1^4+2 I_{B_2}^2 g_2^4-\frac{8}{9} I_{B_3}^2 g_3^4-\frac{2 m_{H_u}^2}{3}+2 m_{Q_3}^2,
\end{dmath}
\begin{dmath}
m_{L_3}^2= -2 D_{B_{13}}+D_Z-\frac{I_{M_1}}{33}+3I_{M_2}-\frac{8 I_{M_3}}{3}+D_{L_{13}}+\frac{13 D_{Y_{13H}}}{10}-\frac{9 D_{\chi_1}}{20}+\frac{1}{33} I_{B_1}^2 g_1^4-3 I_{B_2}^2 g_2^4+\frac{8}{3} I_{B_3}^2 g_3^4+m_{H_d}^2+m_{H_u}^2-3 m_{Q_3}^2,
\end{dmath}
\begin{dmath}
m_{L_1}^2= \frac{1}{440} \left(20 I_{M_1}+660I_{M_2}-26 D_{Y_{13H}}-11 D_{\chi_1}-20 g_1^2 \left(I_{Y_\alpha}+I_{B_1}^2 g_1^2\right)-660 I_{B_2}^2 g_2^4\right),
\end{dmath}
\begin{dmath}
m_{d_1}^2= \frac{1}{1980}\left(40 I_{M_1}-1760 I_{M_3}+78 D_{Y_{13H}}+33 D_{\chi_1}+60 I_{Y_\alpha} g_1^2-40 I_{B_1}^2 g_1^4+1760 I_{B_3}^2 g_3^4\right),
\end{dmath}
\begin{dmath}
m_{u_1}^2= \frac{1}{990} \left(80 I_{M_1}-880 I_{M_3}-78 D_{Y_{13H}}-33 D_{\chi_1}-60 I_{Y_\alpha} g_1^2-80 I_{B_1}^2 g_1^4+880 I_{B_3}^2 g_3^4\right),
\end{dmath}
\begin{dmath}
m_{e_1}^2= \frac{1}{220} \left(40 I_{M_1}+26 D_{Y_{13H}}+11 D_{\chi_1}+20 I_{Y_\alpha} g_1^2-40 I_{B_1}^2 g_1^4\right)
\end{dmath}

\section{Flavor-Blind Models}

The most immediate consequence of flavor-blindness is the vanishing of $D_{B_{13}}$ and $D_{L_{13}}$. Therefore these invariants provide us with a direct test of the flavor-independent hypothesis with a minimal set of measurements. More precisely, they allow this hypothesis to be ruled out: measuring $D_{B_{13}}\neq 0$ or $D_{L_{13}}\neq 0$ at the low scale implies high-scale family non-universality; however, as noted in Ref.~\cite{Kane:2006hd}, measuring $D_{B_{13}}=0$ and $D_{L_{13}}=0$ at the low scale does not necessarily indicate high-scale universality. 

Current experimental data from flavor physics strongly motivates a flavor-universal mediation mechanism for SUSY-breaking. Accordingly, if $D_{B_{13}}$ and $D_{L_{13}}$ are found to vanish, it is reasonable to proceed a step further and attempt to extract constraints on the high-scale values of the flavor-blind MSSM soft parameters from the RGIs.

The 7 scalar and 3 gaugino soft mass parameters in the flavor-blind MSSM can be expressed uniquely in terms of the 10 invariants $D_{\chi_1}$ through $I_{M_3}$ listed in Table~\ref{table.Inv}. These are listed in Eqs.~(\ref{gauginos}) and (\ref{mL1})-(\ref{flaveq2}). Note that these relations depend on the 3 gauge couplings and further all couplings and soft parameters are assumed to be given at the messenger scale: 
\begin{align}
m_{\tilde{L}}^2 &= -\frac{1}{440} \big(26 D_{Y_{13H}}+11 D_{\chi_1}+20 \big(\big(g_1^4I_{B_1}^2 +33 g_2^4 I_{B_2}^2\big)-\big(I_{M_1}+33 I_{M_2}\big)+g_1^2I_{Y\alpha} \big)\big)\;,\label{mL1}\\
m_{H_d}^2 &=  m_{\tilde{L}}^2 - \frac{1}{2}D_Z \; ,\\
m_{H_u}^2 &= m_{\tilde{L}}^2 - \frac{1}{2}D_Z - \frac{13}{11} D_{Y_{13H}} + \frac{g_1^2}{11}  I_{Y\alpha} \; ,\\
m_{\tilde{e}}^2 &= \frac{1}{220} \big(26 D_{Y_{13H}}+11 D_{\chi_1}-20 \big(2\big( g_1^4I_{B_1}^2  - I_{M_1}\big)-g_1^2I_{Y\alpha}\big)\big)\;,\\
m_{\tilde{u}}^2 &= -\frac{1}{990} \big(78 D_{Y_{13H}}+33 D_{\chi_1}+20 \big(4\big(\big(g_1^4 I_{B_1}^2 -11 g_3^4 I_{B_3}^2\big) -\big( I_{M_1}-11 I_{M_3}\big)\big)+3 g_1^2I_{Y\alpha}\big)\big)\;,
\end{align}
\begin{eqnarray}
m_{\tilde{d}}^2 &= &\frac{1}{1980}\big(78 D_{Y_{13H}}+33 D_{\chi_1}-20 \big(2 \big(\big(g_1^4I_{B_1}^2 -44 g_3^4 I_{B_3}^2\big) -\big(I_{M_1}-44 I_{M_3}\big)\big)-3 g_1^2I_{Y\alpha}\big)\big)\;,\nonumber\\
&&\\
m_{\tilde{Q}_1}^2 &=& \frac{1}{3960}\big( 78 D_{Y_{13H}}-627 D_{\chi_1}\label{mQ1}\nonumber\\
&&-20 \big(\big(g_1^4 I_{B_1}^2 +297 g_2^4 I_{B_2}^2-176 g_3^4 I_{B_3}^2\big)
 -\big(I_{M_1}+297 I_{M_2}-176 I_{M_3}\big)-3 g_1^2I_{Y\alpha}\big)\big)\; .\label{flaveq2}
\end{eqnarray}

Using the invariants $I_{g_2}$ and $I_{g_3}$ these may be expressed entirely in terms of $g_1$. Equivalently, one can reduce the degrees of freedom at the high scale to a single parameter, which can be taken to be the value of that scale. In particular this permits tests of more restrictive flavor-universal models such as mSUGRA, taking $g_1$ at the GUT scale.

\end{document}